\documentclass[11pt, a4paper]{article}

\usepackage{amsmath}
\usepackage{amsfonts}
\usepackage[dvipdfmx]{graphicx}
\usepackage[dvipdfmx]{color}

\usepackage{algorithm}
\usepackage{algorithmic}

\usepackage[a4paper]{geometry}

\begin{document}








%
%



%





\pagestyle{empty}

\newgeometry{left=0.5cm, right=0.5cm}
\addtolength{\topmargin}{2.6cm}

{\center

	{\Huge Canonical approach}

	\vspace{1em}

	{\LARGE --- Investigation of finite density QCD phase transition ---}

	\vspace{10cm}

	{\huge Shotaro Oka}
	
	\vspace{1em}
	
	{\Large Rikkyo University}

	\vspace{2.4cm}

	{\Large A thesis submitted for the degree of}

	\vspace{1em}
	
	{\Large \textit{Doctor of Philosophy}}

	\vspace{1em}

	{\Large January 2018}

}

\newpage

\pagestyle{empty}

\addtolength{\topmargin}{-2.6cm}
\restoregeometry
\addtolength{\leftmargin}{0.3cm}
\addtolength{\rightmargin}{0.3cm}








%
%



\begin{abstract}

QCD is the fundamental theory which describes the dynamics of quarks and gluons.
If we understand the dynamics at finite density and temperature,
i.e. QCD phase diagram and equation of states,
we can progress many studies such as the studies of unstable nucleus, nuclear fusion,
early universe and neutron stars.
The study of QCD phase diagram is very interesting,
but we have not understood it well for a long time.
This is because we face a problem in this thesis at finite density.
The problem is called the sign problem.
It causes a decrease of the calculation accuracy.
That is why we can not calculate physical quantities with high accuracy at finite chemical potential.

In this thesis, we try to \textit{beat} the sign problem using the canonical approach
of finite density lattice QCD.
Although it is known that the canonical approach has several numerical problems,
we can reduce them and calculate thermodynamic observables well at finite density.
Concretely, in order to reduce the computation cost of the fermion determinant,
we use the winding number expansion,
in order to enhance the calculation accuracy of physical quantities,
we adopt the multi precision calculation to our program based on the canonical approach.
In this thesis, we will see how to improve the canonical approach
and a result of thermodynamic observables which is related with the QCD phase transition
at finite density.

Our result shows that we do not observe the peak which represents the confinement--deconfinement
phase transition in baryon number susceptibility.
Therefore, we do not see the QCD phase transition yet.
However, in this thesis, we find that canonical approach can explore the QCD phase diagram
beyond $\mu_B/T = 3$ ($\mu_B$ is the baryon chemical potential).
That is, we explored the QCD phase structure
beyond the validity range of Taylor expansion and reweighting method.
This opens a bright window of study of QCD phase diagram at finite density.
With our improvement, canonical approach has the possibility for investigation of thermodynamic observables
at any chemical potential.

\end{abstract}  

\newpage

\tableofcontents

\vspace{2em}
* means the author's works.

\begin{flushleft}
\vspace{2em}

\hrulefill

\vspace{1em}

Parts of this thesis have been published in the following journal articles.
\end{flushleft}

\begin{itemize}
	\item
		A. Nakamura, S. Oka and Y. Taniguchi, 
		``QCD phase transition at real chemical potential with canonical approach,"
		JHEP 02 (2016) 054.

	\item
		R. Fukuda, A. Nakamura and S. Oka, ``Canonical approach to finite density QCD
		with multiple precision computation,"
		Phys. Rev. D 93, 094508 (2016).
		
	\item
		S. Oka, ``Exploring finite density QCD phase transition with	canonical approach
		---Power of multiple precision computation---,"
		PoS (LATTICE2015) 067 (2016).

	\item
		R. Fukuda, A. Nakamura, S. Oka, S. Sakai, A. Suzuki and Y. Taniguchi, 
		``Beating the sign problem in finite density lattice QCD,"
		PoS (LATTICE2015) 208 (2016).

	\item
		R. Fukuda, A. Nakamura and S. Oka,
		``Validity range of canonical approach to finite density QCD,"
		PoS (LATTICE2015) 167 (2016).

\end{itemize}

\newpage

\clearpage
\pagestyle{plain}
\setcounter{page}{1}

\section{Introduction}

\subsection{Property of quarks}

Our world is constructed by the matter such as quarks and leptons.
We can predict their behavior using several gauge theories,
that is Weinberg--Salam theory for electroweak interaction
and quantum chromodynamics (QCD, Yang--Mills theory for SU(3)) for strong interaction.\\

It is well known that, for electroweak interaction,
experimental results have been consistent with predictions based on
the perturbative gauge theory with high accuracy.
On the other hand, for strong interaction,
we can not quantitively analyze most of the QCD phase diagram on a temperature--density
plane because perturbative methods do not work in the region.

\begin{figure}[h]
	\centering
 	\includegraphics[width=0.85\hsize, clip]{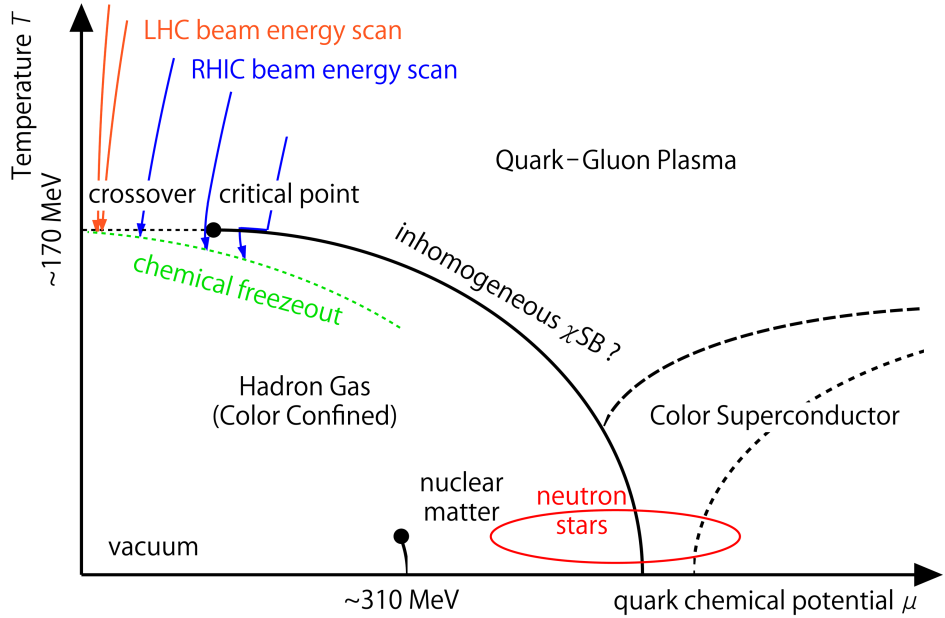}
	\caption{Schematic QCD phase diagram}
	\label{phase_diagram}
\end{figure}

However, we have expected that QCD has a rich phase structure
at finite temperature and finite density.
This is because qualitative analysis has given the clues which strongly support the expectation.
For example, they suggest that quarks are confined and form hadronic bound states
at a low temperature and a low density.
Since our world is located on this region,
our bodies, this paper, the earth and the other planets are made of hadronic matters.

At high temperature, $T > 10^2$ MeV ($= 10^{12}$ K)\cite{Fukushima}, quarks are disentangled from hadrons.
This means that quarks can travel almost freely even though they interact with gluons.
This phase is called quark--gluon plasma (QGP).
We consider that the origin of our world was in this phase
because our universe began with high temperature.
Therefore, the study of QGP has been one of the most important subjects
to reveal the early universe for ages.

\begin{figure}[h]
	\centering
 	\includegraphics[width=0.8\hsize, clip]{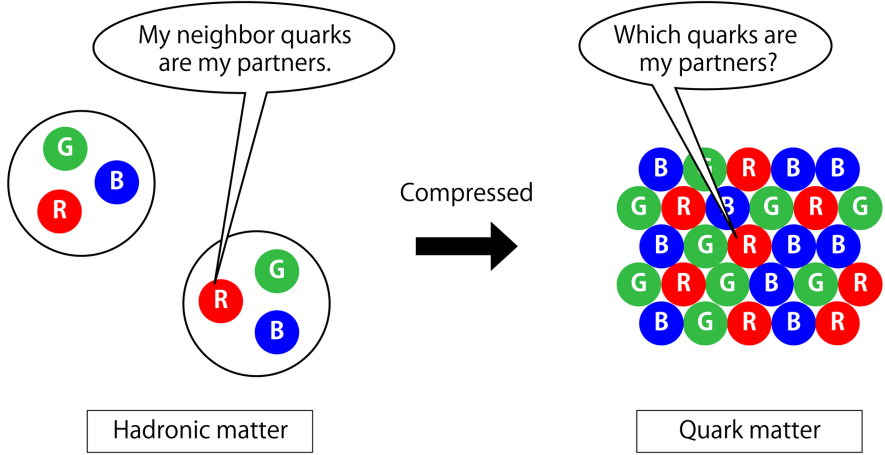}
	\caption{Difference of hadronic matters and quark matters.
			R, G and B are the colors of quarks.}
	\label{quark_matter}
\end{figure}

At high baryon density, $\mu_B > 10^3$ MeV ($= 10^{18}$ kg/m$^3$)\cite{Fukushima, QM},
quarks can not distinguish the boundary of baryons (see Fig.~\ref{quark_matter}).
Therefore, baryons dissolve into a degenerate Fermi system of quarks;
it is called the quark matter.
Such an exotic system may exist in the central core of the neutron stars.
\\

We have mainly obtained a lot of interesting information about the quark matter
by experiments, model calculations and lattice QCD simulations.
Among these, lattice QCD simulations play very important roles.
This is because it is the first--principles calculation based on the discretized theory of QCD.
In other words, it can reliably discuss nature of the world managed by strong interaction.

\subsection{QCD}

QCD describes the dynamics of quarks and gluons.
A quark has a color charge which is one of quantum numbers.
It is classified into three states of red, green and blue.
As an electric charge is origin of the electromagnetic interaction,
color charge causes the strong interaction for quarks.
Then, the interaction is propagated by gluons.

Let $\psi^a$ be a fermion field and $A_\mu = A^\alpha_\mu T^\alpha$ be a gauge field.
Here, $a$ is a color index; it describes a type of quarks
($a = 1, 2, 3$ correspond to red, green and blue, respectively.)
$T^\alpha$ ($\alpha = 1, \cdots, 8$) are the generators of SU(3).
Gluon field $A_\mu$ is a matrix which interacts with fermions and gluons.
Therefore, we write it as
\begin{equation}
	A_\mu = A^{a b}_\mu(x,y,z,t) = A^\alpha_\mu(x,y,z,t) T^{\alpha, a b}  \  .
\end{equation}

Then, QCD Lagrangian is written as follows,
\begin{align}
	L &= -\frac{1}{4} G_{\mu \nu}^\alpha G^{\alpha \mu \nu} + \bar{\psi}^a (i \gamma^\mu D_\mu - m) \psi^a  \  ,  \label{QCD_lagrangian}  \\
	G^\alpha_{\mu \nu} &= \partial_\mu A^\alpha_\nu - \partial_\nu A^\alpha_\mu + g f^{\alpha \beta \gamma} A^\beta_\mu A^\gamma_\nu  \  ,  \\
	D_\mu \psi^a &= \partial_\mu \psi^a - i g A^\alpha_\mu T^{\alpha, a b} \psi^b  \  ,
\end{align}
where $G^\alpha_{\mu \nu}$ is the strength of the gauge field,
$D_\mu$ is the covariant derivative and $g$ is the coupling constant.
$f^{\alpha \beta \gamma}$ are the structure constant of SU(3).
$f^{\alpha \beta \gamma}$ and $T^\alpha$ are written as follows using the Gell--Mann matrices $\lambda^\alpha$,
\begin{align}
	T^\alpha &= \frac{\lambda^\alpha}{2}  \  ,  \\
	\left[ T^\alpha, T^\beta \right] &= i f^{\alpha \beta \gamma} T^\gamma  \  .
\end{align}

Here, since gluons have color indices (that is, gluons are not commutative), it can interact with itself.
We can see this from the first term of Eq.~(\ref{QCD_lagrangian}),
\begin{align}
	G_{\mu \nu}^\alpha G^{\alpha \mu \nu}
		&= \left( \partial_\mu A^\alpha_\nu - \partial_\nu A^\alpha_\mu + g f^{\alpha \beta \gamma} A^\beta_\mu A^\gamma_\nu \right)  \notag  \\
			&\hspace{3em} \times \left( \partial^\mu A^{\alpha \, \nu} - \partial^\nu A^{\alpha \, \mu} + g f^{\alpha \delta \epsilon} A^{\delta \, \mu} A^{\epsilon \, \nu} \right)  \\
		&= \partial_\mu A^\alpha_\nu \partial^\mu A^{\alpha \, \nu}
			- \partial_\mu A^\alpha_\nu \partial^\nu A^{\alpha \, \mu}
			+ \underline{g \partial_\mu A^\alpha_\nu f^{\alpha \delta \epsilon} A^{\delta \, \mu} A^{\epsilon \, \nu}}  \notag  \\
		&\phantom{=} - \partial_\nu A^\alpha_\mu \partial^\mu A^{\alpha \, \nu}
			+ \partial_\nu A^\alpha_\mu \partial^\nu A^{\alpha \, \mu}
			- \underline{g \partial_\nu A^\alpha_\mu f^{\alpha \delta \epsilon} A^{\delta \, \mu} A^{\epsilon \, \nu}}  \notag  \\
		&\phantom{=} + \underline{g f^{\alpha \beta \gamma} A^\beta_\mu A^\gamma_\nu \partial^\mu A^{\alpha \, \nu}}
			- \underline{g f^{\alpha \beta \gamma} A^\beta_\mu A^\gamma_\nu \partial^\nu A^{\alpha \, \mu}}  \notag  \\
		&\phantom{=} + \underline{g^2 f^{\alpha \beta \gamma} A^\beta_\mu A^\gamma_\nu f^{\alpha \delta \epsilon} A^{\delta \, \mu} A^{\epsilon \, \nu}}  \label{gluon_int}  \  .
\end{align}
The underlined terms of Eq.~(\ref{gluon_int}) describe the interactions between gluons,
and the other terms describe the motions of gluons.
In the QED, photons do not interact with itself
because the structure constant of U(1) gauge theory satisfies $f^{\alpha \beta \gamma} = 0$.

This characteristic property affects the dynamics of quarks significantly.
For example, quarks (color charges) emit or absorb the gluons
as well as electric charges do for the photons.
Then, although photons are emitted radially,
gluons are emitted to only one direction because of gluon--gluon interaction.
That is, emitted gluons form a flux tube,
and they are necessarily absorbed by the other color charge.
We may consider that such a flux tube has a binding energy in order to converge the gluons.
Because the energy may become larger with increasing the distance,
the flux is stretched between two charges in the shortest way,
and a quark can not appear alone since it needs infinite energy to be free.

\begin{figure}[h]
	\centering
 	\includegraphics[width=0.8\hsize, clip]{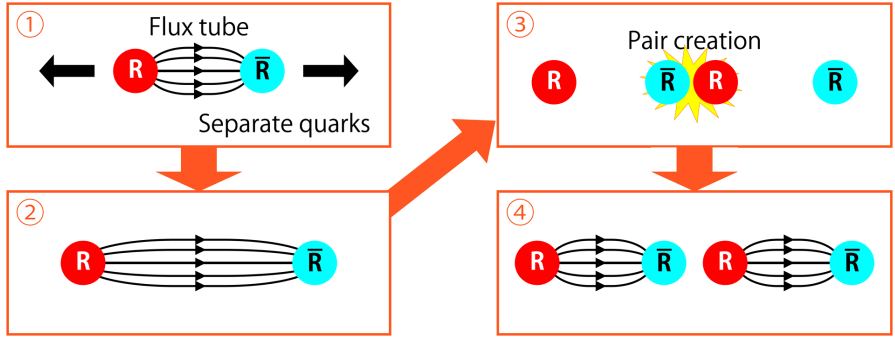}
	\caption{Schematic drawing of gluon flux tube}
	\label{flux_tube}
\end{figure}

In fact, if we try to take away a quark from a baryon,
a pair creation of quarks occurs in the flux tube,
we finally get one meson and one new baryon.
(See also Fig.~\ref{flux_tube}; it is a meson case.)
This phenomenon is called the confinement of quarks.\\

QCD also has other interesting properties,
for instance, the asymptotic freedom is a significant feature of QCD.

Coupling constant $g$ of QCD is not a constant.
In practice, it depends on energy scale $\mu$ which is used for a renormalization,
e.g. the momentum of external lines of a Feynmann diagram.
The behavior of $g(\mu)$ is given by a calculation of the beta function.
For QCD, It is known that beta function $B$ associates coupling constant $g(\mu)$
with energy scale $\mu$ as follows,
\begin{align}
	B(g(\mu)) &= \mu \frac{d g(\mu)}{d \mu} = -B_0 g^3 - B_1 g^5 + \cdots  \  ,  \\
	B_0 &= \frac{1}{(4 \pi)^2} \left( \frac{11}{3} N_c - \frac{2}{3} N_f \right)  \  ,  \\
	B_1 &= \frac{1}{(4 \pi)^2} \left( \frac{34}{3} {N_c}^2 - \frac{10}{3} N_c N_f - \frac{{N_c}^2 - 1}{N_c} N_f \right)  \  ,
\end{align}
using the two--loop perturbation theory\cite{beta_function}.
Here, $N_c = 3$ is the number of colors and $N_f$ is the number of flavors.
From these, we find the following relation,
\begin{equation}
	\log(\mu) = \frac{1}{2 B_0 g^2} + \frac{B_1}{2 {B_0}^2} \log(g^2) - \frac{B_1}{2 B_0} \log(B_0 + B_1 g^2) + (\textrm{Const.})  \  .
\end{equation}
This equation indicates that $\mu$ becomes large when $g$ is small.
In other words, decreasing distance scale $a = 1/\mu$ makes coupling constant $g$ small.
It means that quarks behave as asymptotic free particles in a high energy collision
although they are confined at low energy scale.
This behavior indicates the presence of QGP at high temperature,
and asks us what happens between hadron gas phase and QGP phase;
these two phases are respectively ruled by different particles and physics.

\begin{figure}
	\centering
 	\includegraphics[width=\hsize, clip]{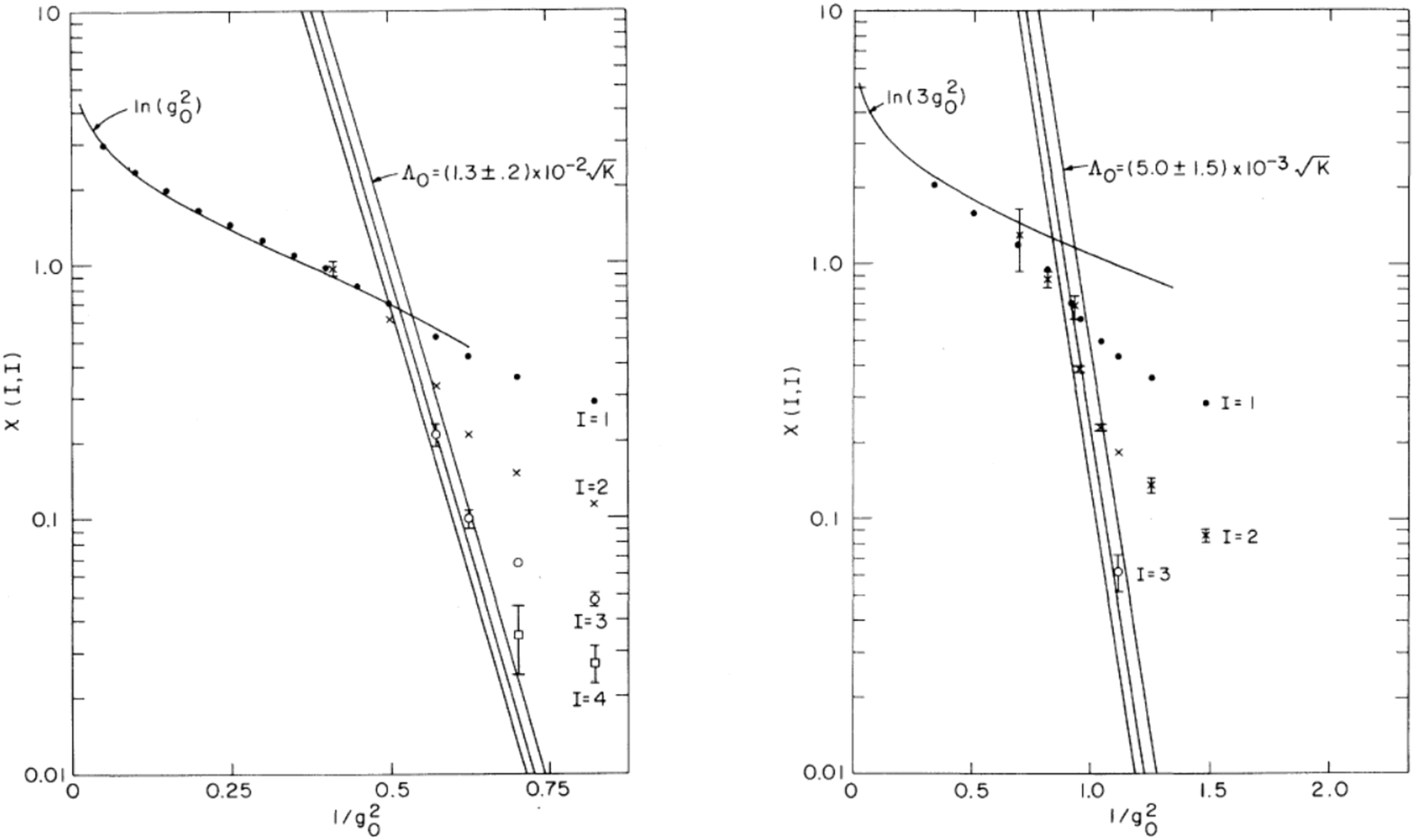}
	\caption{Running coupling constant dependence on string tension.
		The left panel and right panel show the results for SU(2) and SU(3), respectively.
		Vertical axis denotes the quantities $\chi = a^2 \sigma$ which is related with the beta function,
		and horizontal axis denotes the running coupling constant $\beta = 4/g^2$(left panel)
		or $6/g^2$ (right panel).
		This figure is taken from Ref.~\cite{Creutz-2}.}
	\label{fig_Creutz}
\end{figure}

The asymptotic freedom was found by a collider experiment of SLAC\cite{SLAC-exp, SLAC-th} in 1968,
which was explained theoretically by D.J. Gross and F. Wilczek\cite{Gross},
and H.D. Politzer\cite{Politzer} in 1973.
In addition, it is confirmed that the results of string tension\footnote{
	We can consider the gluon flux tube which bonds two quarks as a string.
	Then, introducing the string tension, we can explain the confinement of quarks as
	a string model.
	The asymptotic freedom is explained as the string tension becomes small
	at short distance scale.
} of a lattice QCD simulation
reproduce the theoretical expectation with high accuracy as Fig.~\ref{fig_Creutz}.
That is, the non--perturbative result supports the perturbative exception
and shows the behavior of running coupling constant at middle region
where the perturbation theory does not work;
it is a valuable result.
This simulation was performed by M. Creutz\cite{Creutz-1, Creutz-2} in 1980.


\subsection{Finite temperature and density QCD}

When we discuss actual QCD systems, we use methods of statistical mechanics and thermodynamics.
QCD is a quantum theory, that is, we can apply the path integral formulation to it.
Thus, we can discuss the system of QCD statistically.
Thermodynamic variables, e.g. temperature and chemical potential can be adopted into QCD
by the following way.

In order to introduce temperature to QCD,
we consider a partition function,
\begin{equation}
	Z = \textrm{Tr} \left( e^{-\beta ( \hat{H} - \mu \hat{N} )} \right)  \  .  \label{QCD_Z}
\end{equation}
where $\beta = 1/T$ is an inverse temperature, $\hat{H}$ is the Hamiltonian operator
and $\hat{N}$ is the number operator.
Let $\left| \phi \right\rangle$ be a state of the system on the Hilbert space.
Then, we can rewrite the trace as a summation of physical states,
\begin{equation}
	Z = \sum_\phi \left\langle \phi(x) \left| e^{-\beta ( \hat{H} - \mu \hat{N} )} \right| \phi(x) \right\rangle  \  .  \label{Z_trace}
\end{equation}
For simplicity, we will neglect fermions and $\mu \hat{N}$ for a moment.
Then, we notice that the operator of Eq.~(\ref{Z_trace}), $\exp(-\beta \hat{H})$,
corresponds to the Euclidean time evolution operator of quantum mechanics, $\exp(-\hat{H} t)$.
We therefore can consider that, essentially, $\beta$ is the elapsed time of states.

Now, states are represented by gauge field $A_\mu$.
Taking the Coulomb gauge, we can write the partition function as
\begin{equation}
	Z = \sum_{\mathbf{A}} \left\langle \mathbf{A}(\beta, \mathbf{x}) \left| e^{-\beta \hat{H}} \right| \mathbf{A}(0, \mathbf{x}) \right\rangle  \  ,
\end{equation}
where 3--dimensional vector $\mathbf{A}$ means the spatial components of $A_\mu$.
In this equation, the final state of the bracket is not equivalent to the initial state of it
unlike Eq.~(\ref{Z_trace}).
In order to remove this inconsistency, we require the periodic boundary condition along time direction,
\begin{equation}
	\mathbf{A}(\beta, \mathbf{x}) = \mathbf{A}(0, \mathbf{x})  \  .
\end{equation}
Requiring this condition, we can construct the finite temperature QCD.
(See also Ref.~\cite{finite_temperature_QCD}.)\\

In order to introduce density to QCD,
we need to consider how to include chemical potential $\mu$ in it.
Because we need fermions when we discuss the density of the system,
$\mu$ should be included in the fermionic part of the Lagrangian,
especially in the Dirac operator.

Then, we can get a hint from Eq.~(\ref{Z_trace}).
Using the path integral formalism, we can write the partition function
with chemical potential $\mu$ as follows,
\begin{equation}
	Z = \int DA D\psi D\bar{\psi} \exp \left( -\int_{0}^{\beta} d\tau \int d^3x \left[ -\frac{1}{4} F^2 + \bar{\psi} (\gamma D - m) \psi - \mu \psi^\dagger \psi \right] \right) \, ,
\end{equation}
where $\tau = i t$ is the Euclidean time coordinate with the periodic boundary condition.
(Now, we omit the indices.)
Here, we use the fact that the number operator is written as $\hat{N} = \int d^3x \psi^\dagger \psi$
and $\hat{N}$ is constant of time.

However, it is complicated because $\psi^\dagger$ appears.
We can simplify this equation via the change of variables
\begin{equation}
	A_\nu \rightarrow A_\nu - \frac{i}{g} \left( \partial_\nu e^{\mu \tau} \right) e^{-\mu \tau}  \  .  \label{trans_mu}
\end{equation}
The partition function now becomes
\begin{equation}
	Z = \int DA D\psi D\bar{\psi} \exp \left( -\int_{0}^{\beta} d\tau \int d^3x \left[ -\frac{1}{4} F^2 + \bar{\psi} (\gamma D - m) \psi \right] \right)  \  .
\end{equation}
That is, the chemical potential appears as the imaginary part of temporal component of gauge field.

A. Roberge and N. Weiss considered above discussion
with pure imaginary chemical potential $\mu = i \mu_I$\cite{RW}.
Then, because $e^{i \mu_I}$ is periodic, $Z(\mu_I)$ has the periodicity,
\begin{equation}
	Z(\mu_I) = Z\left( \mu_I + \frac{2 \pi k}{N} \right)  \  ,  \label{RW_period}
\end{equation}
for any integer $k$.
This result is proved for SU(N) theory.
($N=3$ is QCD case.)
Thus, the partition function $Z(\mu)$ of QCD has $Z_{3}$ symmetry
which is the center of SU(3) at imaginary chemical potential.
This is called Roberge--Weiss (RW) symmetry.
In addition, a phase transition which RW symmetry causes is called RW phase transition.\\

Using above methods, we can investigate QCD phase diagram at finite temperature and density.
Ones may perform it numerically with lattice QCD simulations and/or several models
because it is very hard to investigate QCD theoretically with non--parturbative methods.

However, we have not been able to study it well,
although we use the numerical methods.
For example, the QCD critical end point have been searched for a long time.
It is known that, in $T$--$\mu$ plane, the critical point is located at $T_c \approx 150$ MeV
from zero density lattice QCD simulations\footnote{
	The critical point is located at \textit{finite density}.
	This means that we can not use usual order parameters such as the Polyakov loop and chiral condensate
	because the center symmetry of SU(3) and chiral symmetry is broken by the massive fermions.
	Thus, when we discuss the location of critical point,
	we use the critical temperature at zero density alternatively
	or \textit{pseudo} critical temperature $T_{pc}$ which is approximately estimated at finite density
	by the Polyakov loop or chiral condensate.
	In this thesis, we write the both of critical and pseudo critical temperature as $T_c$.
}
\cite{Fukushima}.
But, critical density (chemical potential) $\mu^c_B$ is not known well.

Z. Fodor and S.D. Katz studied it in 2004 for $2+1$ flavor lattice QCD\cite{Fodor-Katz},
then they gave $\mu^c_B = 320(40)$ MeV.
On the other hand, S. Datta, R.V. Gavai and S. Gupta studied it in 2012 for two flavor lattice QCD\cite{Taylor_critical}.
They then gave $\mu^c_B = 243(20)$ MeV.
Moreover, O. Scavenius, A. Mocsy, I.N. Mishustin and D.H. Rischke introduced
the phase diagrams of $\sigma$--model and NJL model which are effective models of QCD, in 2000\cite{models}.
From this paper, critical point of these models are at $\mu^c_B = 220$ MeV and $330$ MeV, respectively.

As we saw, the location of the critical point depends on models.
In addition, at a finite density, there is the sign problem which is discussed in section below.
This problem causes the decease of calculation accuracy.
That is why lattice QCD simulations (and several models) can not estimate physical quantities well
in this region of finite chemical potential.

Although the sign problem is a severe difficulty,
ones have been studying how to avoid this problem.
As one of such methods, the canonical approach was developed.
This method has potential to avoid the problem,
but the other several numerical problems are left in the method.\\

In this thesis, we try to completely avoid the effect of the sign problem
and investigate the QCD phase diagram at finite density by solving the problems of canonical approach.

\newpage

\section{Basic definitions of finite density lattice QCD}

\subsection{Lattice variables and simple lattice actions}

Let us begin to consider the lattice QCD.
In this theory, space--time coordinates are discretized as lattice.
We put fermions and gluons on the lattice.
More precisely, fermions are on the sites of the lattice,
and gluons are on the links of the lattice because they have a direction as Lorentz index.
\begin{figure}[h]
	\centering
 	\includegraphics[width=0.75\hsize, clip]{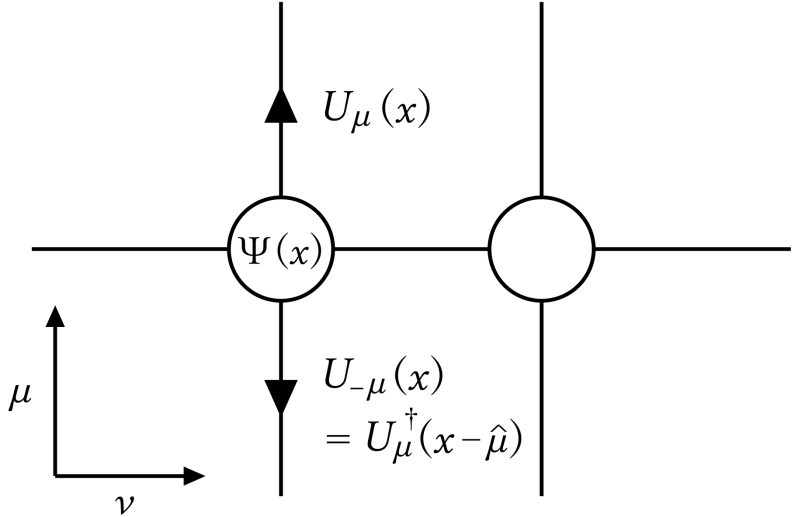}
	\caption{Schematic sketch of lattice variables}
	\label{lattice}
\end{figure}

Let $a$ be the lattice spacing.
Because $A_\mu$ is not covariant under the SU(3) gauge transformation,
we alternatively use the following covariant quantity,
\begin{equation}
	U_\mu (x) = U (x \rightarrow x+\hat{\mu}a) := e^{i a g A_\mu(x+\hat{\mu}/2)} \in \textrm{SU(3)}  \  ,
\end{equation}
for a lattice variable.
Here, $\hat{\mu}$ is the unit vector along $\mu$ direction.
$U_\mu$ are called the link variables.

Since they are unitary matrices, we find the following relation,
\begin{equation}
	U_\mu(x) U_{-\mu}(x+\hat{\mu}a) = 1 = U_\mu(x) U^\dagger_\mu(x)  \  ,
\end{equation}
thus $U_{-\mu}(x+\hat{\mu}a) = U^\dagger_\mu(x)$.\\

We can write the gauge action by gauge invariant quantities which are constructed by $U_\mu$.
For example, the trace of closed loops of $U_\mu$ is gauge invariant.
The simplest loop, that is a unit square on the lattice,
\begin{equation}
	\textrm{tr} U_{\mu}(x) U_{\nu}(x+\hat{\mu}) U_{\mu}^{\dagger}(x+\hat{\nu}) U_{\nu}^{\dagger}(x)  \  ,
\end{equation}
is called the plaquette.
\begin{figure}
	\centering
 	\includegraphics[width=0.6\hsize, clip]{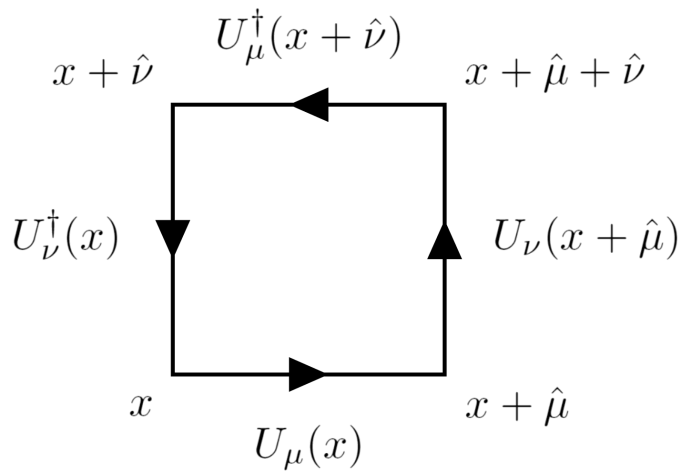}
	\caption{Schematic drawing of plaquette}
	\label{plaq}
\end{figure}
It is invariant under the following gauge transformation,
\begin{align}
	\Psi(x) &\rightarrow \Psi^\prime(x) = e^{-i \theta^\alpha(x) T^\alpha} \Psi(x) =: \Omega(x) \Psi(x)  \  ,  \\
	A_\mu(x) &\rightarrow A^\prime_\mu(x) = \Omega(x) A^\alpha_\mu T^\alpha \Omega^\dagger(x) - \frac{i}{g} [\partial_\mu \Omega(x)] \Omega^\dagger(x)  \  ,
\end{align}
with $\Omega(x) \in$ SU(3). That is,
\begin{equation}
	U_\mu(x) \rightarrow U^\prime_\mu(x) = \Omega(x) U_\mu(x) \Omega^\dagger(x+\hat{\mu}a)  \  .
\end{equation}
Using the plaquette, we can write the gauge action as
\begin{equation}
	S_G = \frac{\beta}{3} \sum_{\mu \not= \nu} \textrm{tr} U_{\mu}(x) U_{\nu}(x+\hat{\mu}) U_{\mu}^{\dagger}(x+\hat{\nu}) U_{\nu}^{\dagger}(x)  \  .
\end{equation}
Now, we use $\beta$ as the effective coupling constant. Note that it is not the inverse temperature.
This action is called Wilson gauge action.

It is difficult to construct the fermion action.
This is because fermions get extra degrees of freedom
when we discretize the Dirac operator naively.
This problem is known as the fermion doubling problem.

To reduce the problem, K.G. Wilson considered the following action,
\begin{align}
	S_F &= a^4 \sum_{x, x^\prime} \bar{\Psi}(x) \Delta(x, x^\prime) \Psi(x^\prime)  \  ,  \\
	\Delta(x, x^\prime) &= \delta_{x, x^\prime}
		- \kappa \sum_{i=1}^{4} \Bigl( \bigl( r - \gamma_i \bigr) U_i (x) \delta_{x^\prime, x+\hat{i}}
                                     + \bigl( r + \gamma_i \bigr) U_i^\dagger (x^\prime) \delta_{x^\prime, x-\hat{i}} \Bigr)  \,  ,  \label{Wilson_fermion}
\end{align}
where $\kappa$ is the hopping parameter.
$\kappa$ rules the movement of quarks, which depends on the quark mass $m_q$\cite{kappa} as
\begin{equation}
	\kappa = \frac{1}{8 + 2 m_q a}  \  .
\end{equation}
The parameter $r$ ranges from zero to one.
Nonzero $r$ gives the mass which diverges at $a = 0$ to the extra fermions.
We usually set $r = 1$.
This term is called Wilson term.
Although Wilson terms break the chiral symmetry on the lattice
because they introduce extra masses into the theory,
the symmetry is recovered when we take the continuum limit $a \rightarrow 0$.
Such fermion action is called Wilson fermion action.

\subsection{Partition function and observables}


In order to calculate physical quantities, let us firstly consider
a representation of the grand partition function.
We can construct it using path integral formulation,
\begin{equation}
	Z(\mu) = \int DU D\Psi D\bar{\Psi} e^{-S_G} e^{-S_F(\mu)}  \ .  \label{def_Zmu}
\end{equation}
Then $U$ is link variables, $\Psi$ and $\bar{\Psi}$ are fermions and anti--fermions
on a lattice respectively.
$S_G$ and $S_F$ are a gauge action and a fermion action respectively.
We will see them in detail below. \\

For the gauge part $S_G$, we use the improved Wilson gauge action,
\begin{equation}
	S_G = \frac{\beta}{3} \sum_{\mu \not= \nu} \left[ c_0 \mathrm{tr} P_{\mu \nu}^{1 \times 1} + c_1 \mathrm{tr} P_{\mu \nu}^{1 \times 2} \right]  \ ,  \label{S_G}
\end{equation}
where $\beta = 6/g^2$ is the effective coupling constant ($g$ is a bare coupling constant),
$P_{\mu \nu}^{1 \times 1}$ is the usual plaquette
and $P_{\mu \nu}^{1 \times 2}$ is the rectangle plaquette:
\begin{align}
	P_{\mu \nu}^{1 \times 1} (x) &= U_{\mu}(x) U_{\nu}(x+\hat{\mu}) U_{\mu}^{\dagger}(x+\hat{\nu}) U_{\nu}^{\dagger}(x)  \\
	P_{\mu \nu}^{1 \times 2} (x) &= U_{\mu}(x) U_{\mu}(x+\hat{\mu}) U_{\nu}(x+2\hat{\mu}) \notag \\
		&\hspace{2em}  U_{\mu}^{\dagger}(x+\hat{\mu}+\hat{\nu}) U_{\mu}^{\dagger}(x+\hat{\nu}) U_{\nu}^{\dagger}(x) \ .
\end{align}
We set the coefficients in Eq.~(\ref{S_G}) as $c_1 = -0.331$ and $c_0 = 1 - 8 c_1$ as Ref.~\cite{Iwasaki}.
It is called Iwasaki gauge action. 
\begin{figure}
	\centering
 	\includegraphics[width=0.8\hsize, clip]{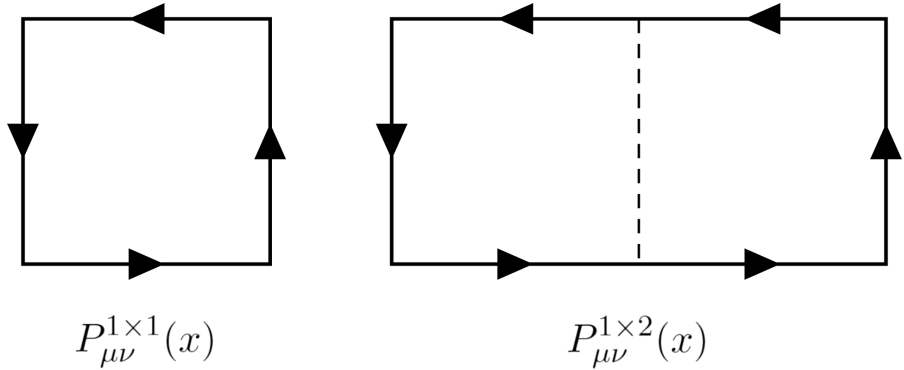}
	\caption{Plaquette and rectangle plaquette.
		The left panel and right panel is the usual (unit square) plaquette
		and rectangle plaquette, respectively.}
	\label{plaq_2}
\end{figure}

This action was suggested by Y. Iwasaki in 1983.
The aim of this action is that
the numerical errors from lattice spacing $a$ are reduced when we calculate the action.
In other word, the lattice action approaches the original continuous action rapidly
when we take the continuum limit $a \rightarrow 0$.\\

The fermion part $S_F$ is written as 
\begin{equation}
	S_F(\mu) = a^4 \sum_{x, x^\prime} \bar{\Psi}(x) \Delta(\mu, x, x^\prime) \Psi(x^\prime)
\end{equation}
where $\Delta(\mu)$ is a ferimon matrix.
Now, we adopt the improved Wilson fermions with chemical potential $\mu$ for future reference,
\begin{align}
	\Delta(\mu, x, x^\prime) &= \delta_{x, x^\prime}
		- \kappa \sum_{i=1}^{3} \Bigl( \bigl( 1 - \gamma_i \bigr) U_i (x) \delta_{x^\prime, x+\hat{i}}
                                             + \bigl( 1 + \gamma_i \bigr) U_i^\dagger (x^\prime) \delta_{x^\prime, x-\hat{i}} \Bigr)  \notag \\
		&\hspace{1em}  - \kappa \Bigl( e^{+\mu a} \bigl( 1 - \gamma_4 \bigr) U_4 (x) \delta_{x^\prime, x+\hat{4}}
                              + e^{-\mu a} \bigl( 1 + \gamma_4 \bigr) U_4^\dagger (x^\prime) \delta_{x^\prime, x-\hat{4}} \Bigr)  \notag \\
        &\hspace{1em}  - \kappa C_{SW} \delta_{x, x^\prime} \sum_{\mu < \nu} \sigma_{\mu \nu} F_{\mu \nu} (x)  \ ,  \label{def_delta}
\end{align}  
where the second line which includes $e^{\pm \mu a}$ denotes the quarks hopping along time direction
(that is, $\hat{4}$ represents the unit vector along time direction),
\begin{align}
	F_{\mu \nu} = P_{\mu \nu}^{1 \times 1} (x) + P_{\nu -\mu}^{1 \times 1} (x) + P_{-\mu -\nu}^{1 \times 1} (x) + P_{-\nu \mu}^{1 \times 1} (x)  \  ,
\end{align}
and $\sigma_{\mu \nu} = i [\gamma_\mu, \gamma_\nu] / 2$.
\begin{figure}
	\centering
 	\includegraphics[width=0.65\hsize, clip]{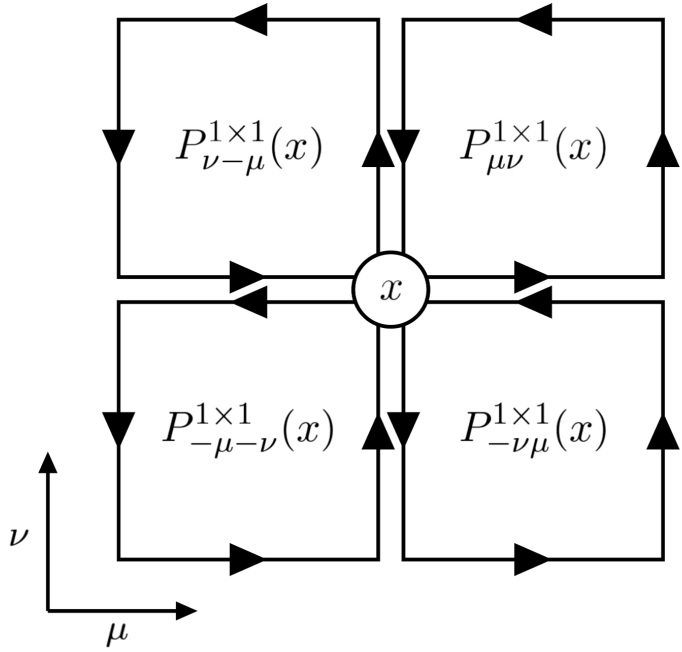}
	\caption{Schematic drawing of clover term.}
	\label{plaq_3}
\end{figure}

Then, we introduce $e^{\mu a}$ to the terms of $U_4$ and $U^\dagger_4$
because of Eq.~(\ref{trans_mu}). 
This means that fermions get the coefficient $e^{\mu a}$ when they travel
along $t$--coordinate.
Since the inverse temperature is related with the boundary conditions of $\psi$ and $A_\mu$,
if fermions wind around the lattice along the time direction,
they get $e^{\mu a N_t} = e^{\mu / T}$.

The term of $C_{SW}$ is called the clover term.
This is the correction term of Wilson fermions,
which includes the field strength on the lattice $F_{\mu \nu}$.
The term reduces the discretization error of the action.
By a result of one--loop perturbation theory\cite{clover_fermions},
we set $C_{SW} = (1 - 0.8412/\beta)^{-3/4}$. (This $\beta$ is the coupling constant.)


Note that we can integrate out $e^{-S_F}$
because it forms the Gaussian integral of the Grassmann number $\Psi$.
(See Appx.~\ref{Grassmann_int}.)
As a result, Eq.~(\ref{def_Zmu}) is rewritten as follows.
\begin{equation}
	Z(\mu) = \int DU \left[ \det \Delta(\mu) \right] e^{-S_G}  \  .
\end{equation}
It is obtained for one flavor fermion case.
If we want to consider many flavor fermions,
we can use the following equation, 
\begin{equation}
	Z(\mu) = \int DU {\left[ \det \Delta(\mu) \right]}^{N_f} e^{-S_G}  \  ,
\end{equation}
where $N_f$ is the number of fermions. \\


Using above definitions, we can calculate the expectation values of thermodynamic observables,
\begin{equation}
	\left\langle O(\mu) \right\rangle = \frac{1}{Z(\mu)} \int DU O(\mu) {\left[ \det \Delta(\mu) \right]}^{N_f} e^{-S_G}  \ .  \label{def_O}
\end{equation}
Here, operater $O(\mu)$, fermion matrix $\Delta(\mu)$ and gauge action $S_G$
depend on link variables $U$.
Therefore, in a practical caluculation,
we calculate them for each configuration (a set of $\{ U \}$) and average their product.

\newpage

\section{Basic methods of finite density lattice QCD}

\subsection{Monte Carlo method}

Here, link variables $U$ have many indices, specifically
\begin{equation}
	U = U^{a b}_{\mu}(x,y,z,t)  \  ,
\end{equation}
where $a$ and $b$ are the color indices, $\mu$ is the Lorentz index,
and $(x,y,z,t)$ are the space--time indices.

"$DU$" means the integration with all of the combinations of $U$.  
Now, $a, b = 1, 2, 3$ and $\mu = 1, 2, 3, 4$, and
we discretize the coordinates $x,y,z,t = 1, \cdots, 10$ for instance.
Then, $DU$ represents $3.6 \times 10^5$ times integration.

Moreover, we have to perform $\int D U^{a b}_{\mu}(x,y,z,t)$ using the division quadrature.
That is,
if we write the component of gauge field with fixed $a$, $b$, $\mu$ and $(x,y,z,t)$
as $U^{a b}_{\mu}(x,y,z,t)$ $ =: u$,
\begin{align}
	\int d \left(U^{a b}_{\mu}(x,y,z,t)\right) f[U] &= \int d u f(u)  \notag \\
		&\approx \sum_{k=1}^{N} f(u_k) \Delta u  \  ,
\end{align}
where $u_k$ ($k \in \{ 1, 2, \cdots N \}$) is a sampled $u$.
Therefore, if $N=10$, we should perform $10^{3.6 \times 10^5}$ operations in this case.
Hence, it is difficult to calculate Eq.~(\ref{def_O}) naively. \\

We then can use the Monte Carlo method in order to reduce
the computational cost of the integration.
In this method, we do not use the sequence number,
and alternatively use the random number as $U$.
Here, we consider the probability density function $p$ of the random number as 
\begin{equation}
	p \propto {\left[ \det \Delta(\mu) \right]}^{N_f} e^{-S_G}  \  ,
\end{equation}
in this case.
We are going to see it in detail below.

\subsection{Improved Monte Carlo methods}

\subsubsection{Importance sampling method}

When we generate the gauge configurations using the Monte Carlo method,
we naively use the uniform random numbers.
However, sometimes it is not an efficient method because it generates
not only statistically important configurations
but also unimportant others with the same probability.
Therefore, we have to improve it in order to pick only important ones from them.\\

Importance sampling method is one of the methods which enhances the statistic accuracy.
In this method, we adopt alternative probability density function
and use it as a filter.
For instance, we consider a simple Monte Carlo case,
\begin{equation}
	E[O] = \int O(x) f(x) dx  \  ,  \label{MC_method}
\end{equation}
where $E[O]$ is the expectation value of a function $O(x)$,
$f(x)$ is the probability density function,
$x$ is the integration variable.
We calculate this in practice,
\begin{equation}
	E[O] = \lim_{m \rightarrow \infty} \frac{1}{m} \sum_{j=1}^{m} O(x_m)  \label{MC_method_d}
\end{equation}
where $x_m$ are the random numbers which conforms to the distribution $f(x)$.
Then, if we want to change the probability density $f(x)$
to another convenient function $f^\prime(x)$,
we can rewrite Eq.~(\ref{MC_method}) as
\begin{equation}
	E[O] = \int O(x) \frac{f(x)}{f^\prime(x)} f^\prime(x) dx  \  .
\end{equation}
Thus, Eq.~(\ref{MC_method_d}) can be rewritten,
\begin{equation}
	E[O] = \lim_{m \rightarrow \infty} \frac{1}{m} \sum_{j=1}^{m} O(x^\prime_m) \frac{f(x^\prime_m)}{f^\prime(x^\prime_m)}  \  ,
\end{equation}
where $x^\prime_m$ are the random numbers which conforms to the distribution $f^\prime(x)$,
$f(x) / f^\prime(x)$ plays the role of the weight function.

Using such a procedure, if $f(x)$ is intricate or it is not useful function,
we can use any useful functions through $f^\prime(x)$.
Naively, we use the uniform distribution as $f^\prime(x)$,
but we usually use more suitable function, e.g. Gauss distribution in order to
enhance the accuracy of the integration.

\subsubsection{Metropolis method}

To perform the Monte Carlo integration, we have to generate the set of
integration variable $x$ which conforms to the distribution $f(x)$ or $f^\prime(x)$.
In this thesis, we use the Metropolis method for this purpose.

Let $Q(x|y)$ be a conditional probability density function.
It generates new sample $x$ from previous sample $y$.
$Q(x|y)$ is also called the proposal density or jumping distribution.
Now, we require that $Q$ has to satisfy the detailed balance condition,
\begin{equation}
	Q(x|y) = Q(y|x)  \  .
\end{equation}

Metropolis algorithm is constructed as follows.
\begin{enumerate}
	\item Initialization: choose an arbitrary point $x_0$ as an initial sample and
		an arbitrary probability density $Q(x|y)$.

	\item Iteration: for each $t$,
	\begin{enumerate}
		\item Generate new sample $x^\prime$ from $Q(x^\prime|x_t)$.

		\item Calculate acceptance ratio $\alpha = \frac{f(x^\prime)}{f(x_t)}$.
			Note that it is just the weight function.
		
		\item If $\alpha \geq 1$, $x^\prime$ is more likely than $x_t$,
			then it is accepted for next sample $x_{t+1}$;
			if $\alpha < 1$, $x^\prime$ is accepted according to probability $\alpha$;
			when $x^\prime$ is rejected, set $x_{t+1} = x_t$.
	\end{enumerate}
\end{enumerate}

We can generate the gauge configurations using this algorithm,
but it has a demerit
--- the generated samples are correlated.
This means that, when we use these samples,
we have to extract every $n$--th samples and only use them
($n$ is determined by the autocorrelation between adjacent samples.
we do not discuss it in detail.)
Thus, this algorithm is not efficient.
Usually, an improved method is alternatively used;
we will see it in the following section.

\subsubsection{Hybrid Monte Carlo method}

Hybrid Monte Carlo (HMC) method is one of the Markov chain Monte Carlo (MCMC) methods.
It is constructed by the combination of hybrid molecular dynamics (HMD) and the Metropolis method,
and mainly used in the two flavor lattice QCD simulations.

\begin{enumerate}
	\item Initialization:
	\begin{enumerate}
		\item Choose $\phi$ as the integration variable.
		\item Generate the conjugation momentum $\pi$ with
			probability $P(\pi) \propto e^{-\pi^2/2}$ (Gauss distribution).
	\end{enumerate}
	
	\item Iterations (HMD part):
		evolve $\phi$ and $\pi$ through the following equations,
		\begin{equation}
			\left\{
			\begin{array}{l}  \displaystyle
				H[\phi, \pi] = -\frac{1}{2} \pi^2 + S(\phi)  \\  \displaystyle
				\dot{\phi} = \frac{\partial H[\phi, \pi]}{\partial \pi} = \pi  \\  \displaystyle
				\dot{\pi} = -\frac{\partial H[\phi, \pi]}{\partial \phi} = -\frac{\partial S(\phi)}{\partial \phi}
			\end{array}
			\right.
		\end{equation}
		where $H[\phi, \pi]$ is the Hamiltonian and
		$S(\phi)$ is the action for $\phi$ (it is given).

	\item Metropolis test: using the Metropolis method,
		accept or reject $\phi$ and $\pi$ according to the following probability,
		\begin{align}
			&P( \{ \phi_\textrm{init}, \pi_\textrm{init} \} \rightarrow \{ \phi_\textrm{new}, \pi_\textrm{new} \} ) = \min \left\{ 1, e^{-\Delta H} \right\}  \  ,  \\
			&\Delta H = H(\phi_\textrm{new}, \pi_\textrm{new}) - H(\phi_\textrm{init}, \pi_\textrm{init})  \  .
		\end{align}

	\item Repeat 2 and 3.
\end{enumerate}


\section{Sign problem}


\subsection{Sign problem}

Basically, using above definitions and methods,
we can start the calculation of finite density lattice QCD.
However, $\det \Delta$ becomes a complex number when $\mu$ is nonzero.
We can see this using the definition of $\Delta$ and the property of the determinant.
First, we observe the following relation by a property of a determinant:
\begin{equation}
	{\left[ \det \Delta(\mu) \right]}^\star = \det \Delta^\dagger (\mu)  \ .
\end{equation}
Then, $\Delta^\dagger$ appears,
and we can calculate it using Eq.~(\ref{def_delta}).
Here, we use the following equations,
\begin{align}
	{\left( 1 - \gamma_i \right)}^\dagger &= 1 - \gamma_i  \hspace{2em}  \left( i = 1,2,3 \right) \notag \\
		&= \gamma_5 \left( 1 + \gamma_i \right) \gamma_5  \\
	{e^{\mu} \left( 1 - \gamma_4 \right)}^\dagger &= e^{\mu^\star} \left( 1 - \gamma_4 \right)  \notag  \\
		&= e^{\mu^\star} \gamma_5 \left( 1 + \gamma_4 \right) \gamma_5  \  .
\end{align}
Note that we now use the Euclidean gamma matrix $\{ \gamma_i, \gamma_j \} = 2 \delta_{i j}$.
Thus, we find
\begin{align}
	{\left[ \det \Delta(\mu) \right]}^\star &= \det \left( \gamma_5 \Delta(-\mu^\star) \gamma_5 \right)  \\
		&= \det \left( \gamma_5 \right) \, \det \left( \Delta(-\mu^\star) \right) \, \det \left( \gamma_5 \right)  \\
		&= \det \Delta(-\mu^\star)  \  .
\end{align}
$\det \Delta(\mu)$ is a complex number if $\mathrm{Re}(\mu) \not= 0$.

When we calculate thermodynamic observables in the finite density region ($\mathrm{Re}(\mu) \not= 0$),
Monte Carlo method does not work well
because the probability ${\left[ \det \Delta(\mu) \right]}^{N_f} e^{-S_G}$ is complex.
Such a problem is called the sign problem\cite{sign1, sign2}.
This problem was found in 1985\cite{sign_orig1, sign_orig2}.
Although researchers have studied it about 30 years, it is not solved yet.


\subsection{Reweighting technique}


The reweighting technique\cite{reweighting-1} may be the most famous method
which is made in order to avoid the sign problem.
In this method, we rewrite fermion matrix as
\begin{equation}
	\det \Delta(\mu) = \frac{\det \Delta(\mu)}{\det \Delta(\mu_0)} \det \Delta(\mu_0)  \  .
\end{equation}
We usually set $\mu_0 = 0$.
Then, partition function becomes
\begin{align}
	Z(\mu) &= \int DU {\left[ \frac{\det \Delta(\mu)}{\det \Delta(\mu_0)} \right]}^{N_f} {\left[ \det \Delta(\mu_0) \right]}^{N_f} e^{-S_G}  \notag  \\
		&= Z(\mu_0) \cdot \frac{1}{Z(\mu_0)} \int DU {\left[ \frac{\det \Delta(\mu)}{\det \Delta(\mu_0)} \right]}^{N_f} {\left[ \det \Delta(\mu_0) \right]}^{N_f} e^{-S_G}  \notag  \\
		&= Z(\mu_0) \cdot \left\langle {\left[ \frac{\det \Delta(\mu)}{\det \Delta(\mu_0)} \right]}^{N_f} \right\rangle_{\mu_0}  \  ,
\end{align}
where $\langle \cdot \rangle_{\mu_0}$ means the expectation value
on the gauge configurations which are generated at $\mu_0$.
Similarly, the observables is written as
\begin{align}
	\left\langle O(\mu) \right\rangle_{\mu} &= \frac{1}{Z(\mu)} \int DU O(\mu) {\left[ \frac{\det \Delta(\mu)}{\det \Delta(\mu_0)} \right]}^{N_f} {\left[ \det \Delta(\mu_0) \right]}^{N_f} e^{-S_G}  \notag  \\
		&= \frac{1}{Z(\mu)} Z(\mu_0) \cdot \frac{1}{Z(\mu_0)} \int DU O(\mu) {\left[ \frac{\det \Delta(\mu)}{\det \Delta(\mu_0)} \right]}^{N_f} {\left[ \det \Delta(\mu_0) \right]}^{N_f} e^{-S_G}  \notag  \\
		&= \left. \left\langle O(\mu) {\left[ \frac{\det \Delta(\mu)}{\det \Delta(\mu_0)} \right]}^{N_f} \right\rangle_{\mu_0} \middle/ \left\langle {\left[ \frac{\det \Delta(\mu)}{\det \Delta(\mu_0)} \right]}^{N_f} \right\rangle_{\mu_0} \right.  \  .
\end{align}
At $\mu_0 = 0$ or pure imaginary, we can calculate $\langle \cdot \rangle_{\mu_0}$
because the determinant of fermion matrix is real.
In other words, we consider the complex probability of the Monte Carlo integration
as a part of the observables.

Note that this method is just one of the importance sampling method.\\

This method seems to work out, but it does not work well in practice.
The reason is as follows.
When we vary $\mu$, the distribution of $\det \Delta(\mu)$ is varied drastically.
Therefore the distributions do not overlap and we can not sample the gauge configuration well.
This problem is called the overlap problem.\\

Multi--parameter reweighting (MPR) technique is the expansion method of original reweighting technique
which is made in order to reduce the effect of overlap problem.
In this method, we reweight $e^{-S_G}$ too.
Concretely, 
\begin{align}
		{\left[ \det \Delta(\mu) \right]}^{N_f} e^{-S_G(\beta)} = {\left[ \frac{\det \Delta(\mu)}{\det \Delta(\mu_0)} \right]}^{N_f} \frac{e^{-S_G(\beta)}}{e^{-S_G(\beta_0)}} \cdot {\left[ \det \Delta(\mu_0) \right]}^{N_f} e^{-S_G(\beta_0)}  \  ,
\end{align}
where $\beta$ is the effective coupling constant (see Eq.~(\ref{S_G}), the definition of $S_G$).
Using this technique, we can enhance accuracy in part at high temperature\cite{reweighting-2}, $T > T_c$,
but the overlap problem still remains in the other part\cite{NN_eos}.

\subsection{Taylor expansion method}


We calculate the thermodynamic observables via the partition function $Z(\mu)$.
For example, the pressure is given as
\begin{align}
	\frac{\Delta p(\mu, T)}{T^4} &:= \frac{p(\mu, T) - p(0, T)}{T^4}  \notag  \\
		&= \frac{1}{V T^3} \left[ \log(Z(\mu, T)) - \log(Z(0, T)) \right]  \notag  \\
		&= \frac{{N_t}^3}{N_x N_y N_z} \log \frac{Z(\mu, T)}{Z(0, T)}  \  .
\end{align}
where $p$ is the pressure and $V = N_x a \times N_y a \times N_z a$ is the spatial volume of
the system.
Then, we can expand it in powers of $\mu/T$ around $\mu = 0$,
\begin{equation}
	\frac{\Delta p(\mu, T)}{T^4} = \sum_{n=2,4,6,\cdots}^{\infty} c_n(T) {\left(\frac{\mu}{T}\right)}^n  \  .  \label{pre}
\end{equation}
Here, the even powers of $\mu/T$ only survive because the QCD partition function has the
symmetry $Z(\mu) = Z(-\mu)$.
We can calculate $c_n$ via $Z(\mu)$ as follows,
\begin{align}
	c_n &= \frac{1}{n !} \left. \frac{\partial^n}{\partial (\mu/T)^n} \frac{\Delta p(\mu, T)}{T^4} \right|_{\mu=0}  \notag  \\
		&= \frac{1}{n !} \frac{{N_t}^3}{N_x N_y N_z} \left. \frac{\partial^n}{\partial (\mu/T)^n} \log Z(\mu, T) \right|_{\mu=0}  \  .
\end{align}
Because $c_n$ does not depend on $\mu$, we can calculate $\Delta p /T^4$ at any $\mu$ using Eq.~(\ref{pre})
once we calculate $c_n$.

In order to use such a method,
we have to calculate $Z(\mu)$ at several values of $\mu$.
Although it is difficult to perform the calculation at finite $\mu$ because of the sign problem,
it is easy to perform near $\mu=0$.
Thus, we calculate $c_n$ around $\mu=0$ and extrapolate the observables to finite $\mu$.
Such method is called Taylor expansion method,
which is employed in order to calculate thermodynamic observables at any $\mu$.\\

However, in general, although this method works out well at $\mu/T \leq 1$,
it may not work well at $\mu/T > 1$ because the numerical errors become severe.
In fact, the studies of Taylor expansion method\cite{Taylor_critical, Taylor-1, Taylor-2, Taylor-3}
show the results of equation of state for $\mu/T \leq 1$.
(Several studies\cite{WHOT} show it for $\mu/T \leq 1.2$.)
Here, K. Nagata and A. Nakamura note that Taylor expansion and MPR methods
work out at the same region of $\mu$\cite{NN_eos}.
They concluded that they are consistent methods and the overlap problem of MPR
and the truncation error of Taylor expansion method are negligible.
That is, we may not explore QCD phase diagram beyond $\mu/T=1$ using these methods.

We need to develop another method in order to explore the QCD phase diagram.

\subsection{Canonical approach}



Canonical approach is one method to avoid the sign problem.
It was suggested by A. Hasenfratz and D. Toussaint\cite{canonical} in 1992.

They focused on the relation between grand partition function $Z(\mu)$
and canonical partition function $Z_C(n)$.
In the statistical mechanics,
it is well known that we can write $Z(\mu)$ as a polynomial of fugacity,
\begin{align}
	Z(\mu) &= \mathrm{Tr} \left( e^{-(\hat{H} - \mu \hat{N}) / T} \right)  \notag \\
		&= \sum_{n=-\infty}^{\infty} \left\langle n \left| e^{-\hat{H}/T} \right| n \right\rangle e^{n \mu/T}  \notag \\
		&=: \sum_{n=-\infty}^{\infty} Z_C(n) \left(e^{\mu/T}\right)^n  \label{fuga} \  ,
\end{align}
with a thermodynamical limit.
Here, $|n\rangle$ is the eigenvectors of the number operator $\hat{N}$
($\hat{N} |n\rangle = n |n\rangle$) and $Z_C(n)$ is the canonical partition function
with net quark number $n$.
In the second line, we use the fact that the number operator does not depend on time, that is $[H, N] = 0$.
Equation~(\ref{fuga}) is called the fugacity expansion.

Hasenfratz and Toussaint noticed that $Z(\mu)$ is just the Laplace transform of $Z_C(n)$.
They consider that, if we use pure imaginary chemical potential $\mu = i\mu_I$ ($\mu_I$ is real),
we get $Z(\mu)$ as the inverse Fourier transform of $Z_C(n)$,
\begin{equation}
	Z(\mu = i\mu_I) = \sum_{n=-\infty}^{\infty} Z_C(n) e^{i (\mu_I/T) n}  \  .
\end{equation}
Therefore, we also get $Z_C(n)$ as the Fourier transform of $Z(\mu)$,
\begin{equation}
	Z_C(n) = \frac{1}{2 \pi} \int d\left( \frac{\mu_I}{T} \right) Z(\mu=i\mu_I) e^{-i(\mu_I/T)n}  \  .  \label{fourier}
\end{equation}

Note that, for pure imaginary chemical potential,
$\det \Delta$ is real and we can calculate $Z(i\mu_I)$ and $Z_C(n)$ without the sign problem.
Moreover, if we get $Z_C(n)$, we can calculate $Z(\mu)$ with ANY chemical potential using Eq.~(\ref{fuga}).
Thus, adopting such a procedure, we can calculate $Z(\mu)$ without the sign problem.
This method is called the canonical approach.

In this thesis, we use the canonical approach in order to avoid the sign problem.

\newpage

\section{Review for study of QCD phase structure with canonical approach}

\label{review}

\subsection{Experiment data and canonical partition function}

The canonical partition function $Z_C(n)$ is related with the number of events
of the collider experiments.
This is because $Z_C(n)$ represents the existence probability of the system
which $n$ particles are in.
That is, we can get the following equation from Eq.~(\ref{fuga}),
\begin{equation}
	1 = \sum_{n} \frac{Z_C(n) (e^{\mu/T})^n}{Z(\mu)}  \  .
\end{equation}
From this equation, we can consider $Z_C(n) (e^{\mu/T})^n / Z(\mu)$
as the existence probability.


Here, let us consider the ratio $Z_C(n) / Z_C(0)$.
This ratio is the normalized canonical partition function.
We will use it when we discuss the existence probability because
\begin{equation}
	\left. \frac{Z_C(n) \left(e^{\mu/T}\right)^n}{Z(\mu)} \middle/ \frac{Z_C(0) \left(e^{\mu/T}\right)^0}{Z(\mu)} \right. = \frac{Z_C(n)}{Z_C(0)} \left(e^{\mu/T}\right)^n  \  .
\end{equation}
Thus, we can cancel $Z(\mu)$ using the ratio.\\

\begin{figure}
	\centering
 	\includegraphics[width=0.8\hsize, clip]{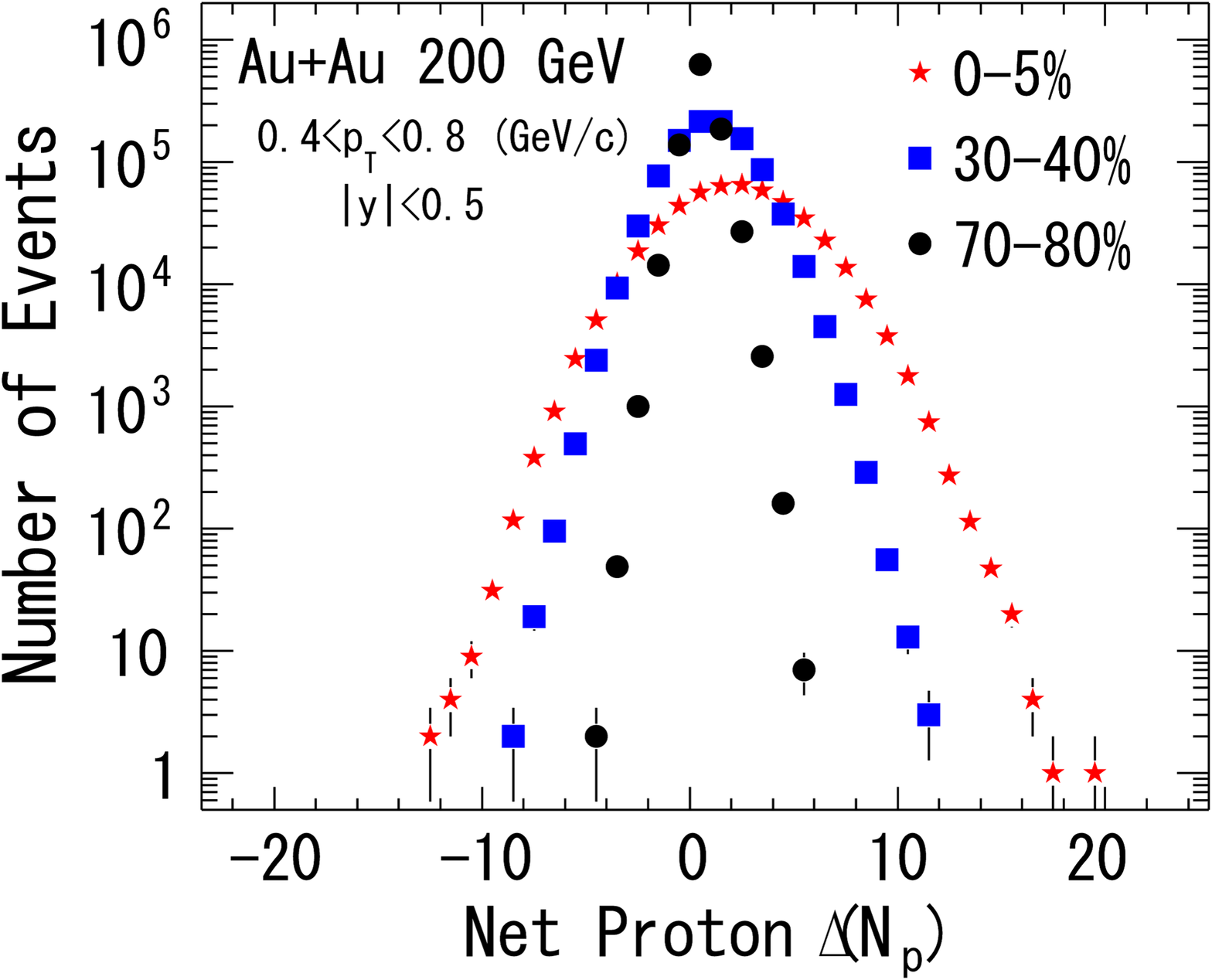}
	\caption{Net proton dependence on number of events in Au+Au collision at RHIC.
		This figure is taken from Ref.~\cite{RHIC}.}
	\label{Zn_exp_RHIC}
\end{figure}

We can see normalized existence probability $Z_C(n) \left(e^{\mu/T}\right)^n / Z_C(0)$ in collider experiments.
For example, Fig.~\ref{Zn_exp_RHIC} shows the net proton dependence on the number of events
in the Au+Au collision at RHIC.

Let $N_\textrm{event}(n_p)$ is the number of event with the net proton number $n_p = 3 n$
and $N_\textrm{total}$ is the total number of events.
Then, normalized number of events $N_\textrm{event}(n_p) / N_\textrm{event}(0)$ is given as follows,
\begin{align}
	\frac{N_\textrm{event}(n_p)}{N_\textrm{event}(0)} &= \frac{ N_\textrm{total} \cdot \left.Z_C(n_p) \left(e^{\mu/T}\right)^{n_p} \middle/ Z(\mu)\right. }{ N_\textrm{total} \cdot \left.Z_C(0) \left(e^{\mu/T}\right)^0 \middle/ Z(\mu)\right. }  \notag  \\
		&= \frac{Z_C(n_p)}{Z_C(0)} \left(e^{\mu/T}\right)^{n_p}  \  .  \label{Zn_from_exp_data}
\end{align}
Note that $Z_C(n)$ satisfies
\begin{equation}
	Z_C(n) = Z_C(-n)  \  .  \label{sym_Zn}
\end{equation}
This is because the QCD partition function has the symmetry of the exchange
of perticles for anti--particles (C--symmetry).

\begin{figure}
	\centering
 	\includegraphics[width=0.8\hsize, clip]{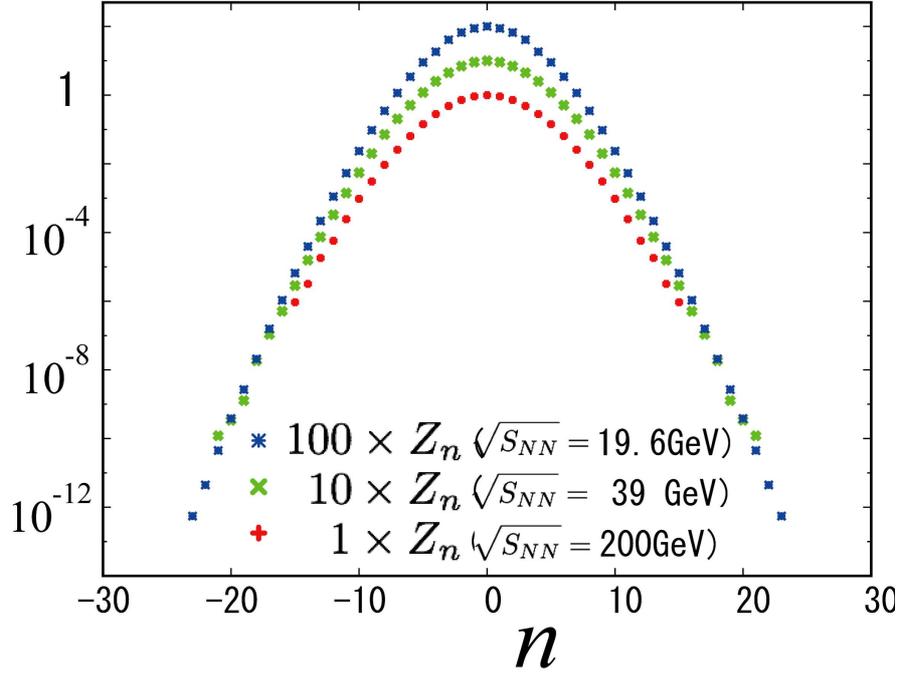}
	\caption{Proton number dependence on normalization canonical partition function from RHIC data
		with $\sqrt{s_{NN}} = 19.6$, $39$ and $200$ GeV.
		Vertical axis represents normalized canonical partition function $Z_C(n_p)/Z_C(0)$
		and horizon axis represents net proton number $n_q$.
		This figure is taken from Ref.~\cite{RHIC_fit}.}
	\label{Zn_fit_RHIC}
\end{figure}

Using Eqs.~(\ref{Zn_from_exp_data}) and (\ref{sym_Zn}),
we can calculate $Z_C(n) / Z_C(0)$ from RHIC data.
Figure~\ref{Zn_fit_RHIC} shows the net proton number dependence on 
normalized canonical partition function from RHIC data.
From this figure, we see that $Z_C(n)$ becomes smaller rapidly with increasing $|n|$
and it is raised with increasing the collision energy $\sqrt{s_{NN}}$.

\subsection{Lattice results for QCD phase structure with canonical approach}

In this section, we focus on the lattice QCD results with canonical approach at finite density.
We will not introduce the results for SU(2) or results at zero density.

\begin{figure}[h]
	\centering
 	\includegraphics[width=0.9\hsize, clip]{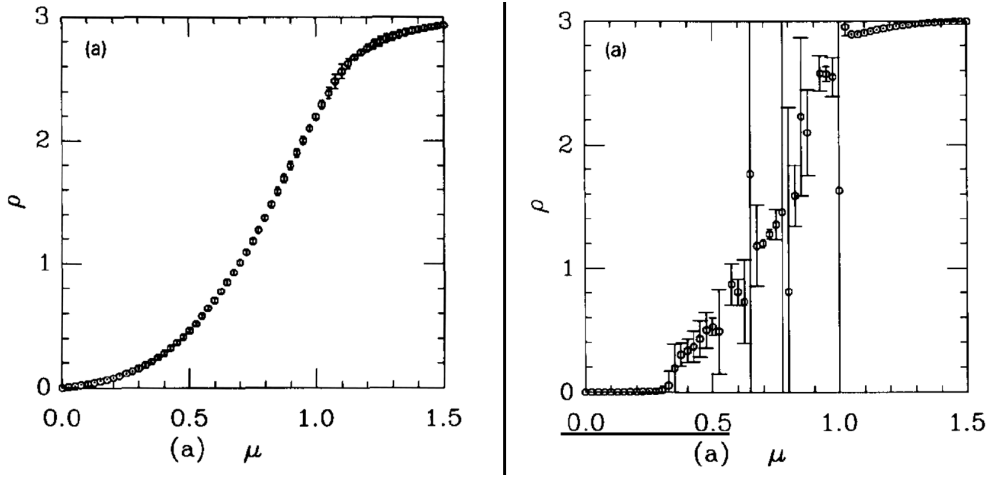}
	\caption{Number density as a function of chemical potential 
		by Hasenfratz and Toussaint in Ref.~\cite{canonical}.
		The left and right panel is calculated at high and low temperature, respectively.}
	\label{HT_graph}
\end{figure}

Because the canonical approach requires large computation resources,
we could not perform full QCD simulation with canonical approach well in 1990's.
In the first study with canonical approach\cite{canonical},
Hasenfratz and Toussaint performed the calculation of number density
on $2^4$ and $4^4$ lattice. See Fig.~\ref{HT_graph}.
Their results tell us that, at low temperature, we can not perform this calculation well,
although at high temperature, we may perform it well.
\\

\begin{figure}
	\centering
 	\includegraphics[width=0.6\hsize, clip]{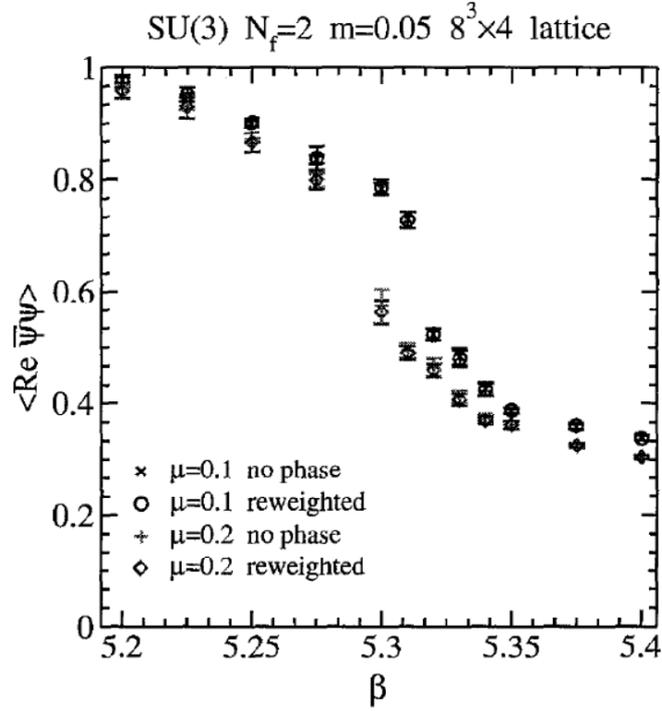}
	\caption{Chiral condensate as a function of the effective coupling constant
		at several finite density
		by Sasai, Nakamura and Takaishi in Ref.~\cite{SNT}.}
	\label{SNT_graph}
\end{figure}

In 2004, Y. Sasai, A. Nakamura and T. Takaishi investigated
the crossover phase transition of QCD at finite density\cite{SNT}.
In this study, they used $8^3 \times 4$ lattice.
They calculated the chiral condensate as Fig.~\ref{SNT_graph}.
From this figure, we see the crossover phase transition as the jump of
the chiral condensate.
Thus, we can conclude that the canonical approach can be used in the study of
phase transition.
Note that, if $T_c \approx 200$ MeV, they studied at $\mu \approx 100$ MeV.

\begin{figure}
	\centering
 	\includegraphics[width=0.7\hsize, clip]{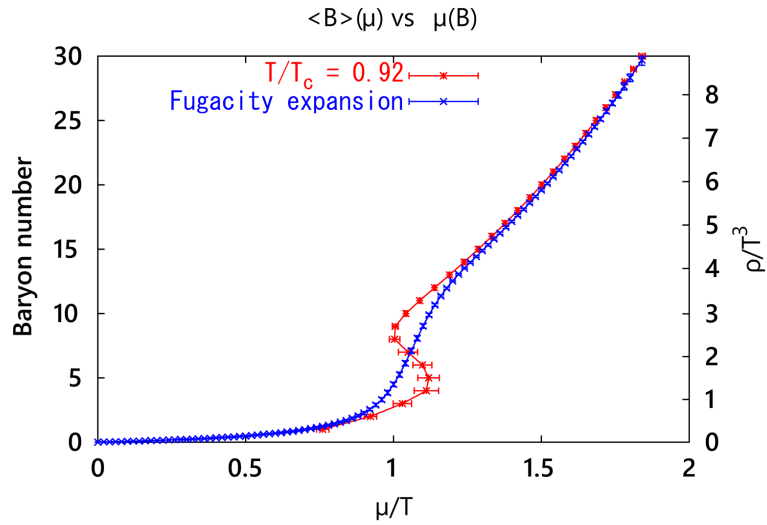}
	\caption{Baryon number as a function of $\mu/T$ by de Forcrand and Kratochvila
	in Ref.~\cite{Forcrand-1}. Blue points correspond to their results.}
	\label{Forcrand_graph-1}
\end{figure}

\begin{figure}
	\centering
 	\includegraphics[width=0.9\hsize, clip]{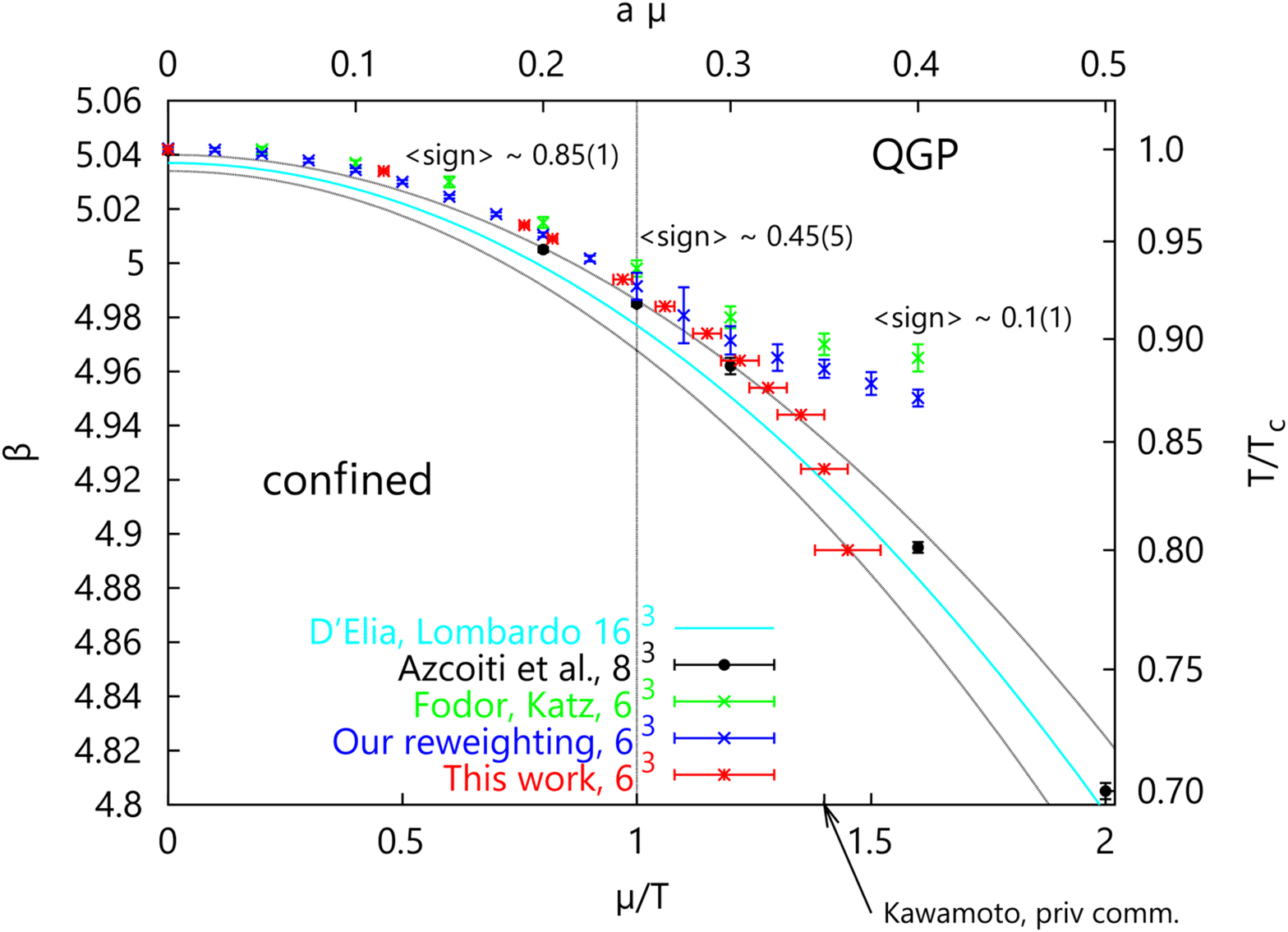}
	\caption{Phase diagram in the $T$--$\mu$ plane by de Forcrand and Kratochvila
		in Ref.~\cite{Forcrand-1}. Red points correspond to their results.}
	\label{Forcrand_graph-2}
\end{figure}

Ph. de Forcrand and S. Kratochvila studied the QCD phase diagram
via baryon number density 
with canonical approach in 2005\cite{Forcrand-1, Forcrand-2}.
They then used $6^3 \times 4$ lattice.
Their results are shown in Figs.~\ref{Forcrand_graph-1} and \ref{Forcrand_graph-2}.
In this study, they calculated the baryon number beyond $\mu/T = 1$.
It seems that, using the canonical approach,  we can explore the QCD phase diagram at high density.
However, they note that their results are affected by the finite size effect strongly.
Therefore, we have to use larger lattice in order to investigate the QCD phase transition.\\

The lattice QCD group of Kentucky ($\chi$QCD collaboration)
performed several important studies for canonical approach.
For example, they developed the winding number expansion for the fermion matrix
in Ref.~\cite{WNE} in 2008. This is introduced below in detail.
Using this method, we can perform the calculation of the fermion determinant faster
than the other methods.\\

\begin{figure}
	\centering
 	\includegraphics[width=0.6\hsize, clip]{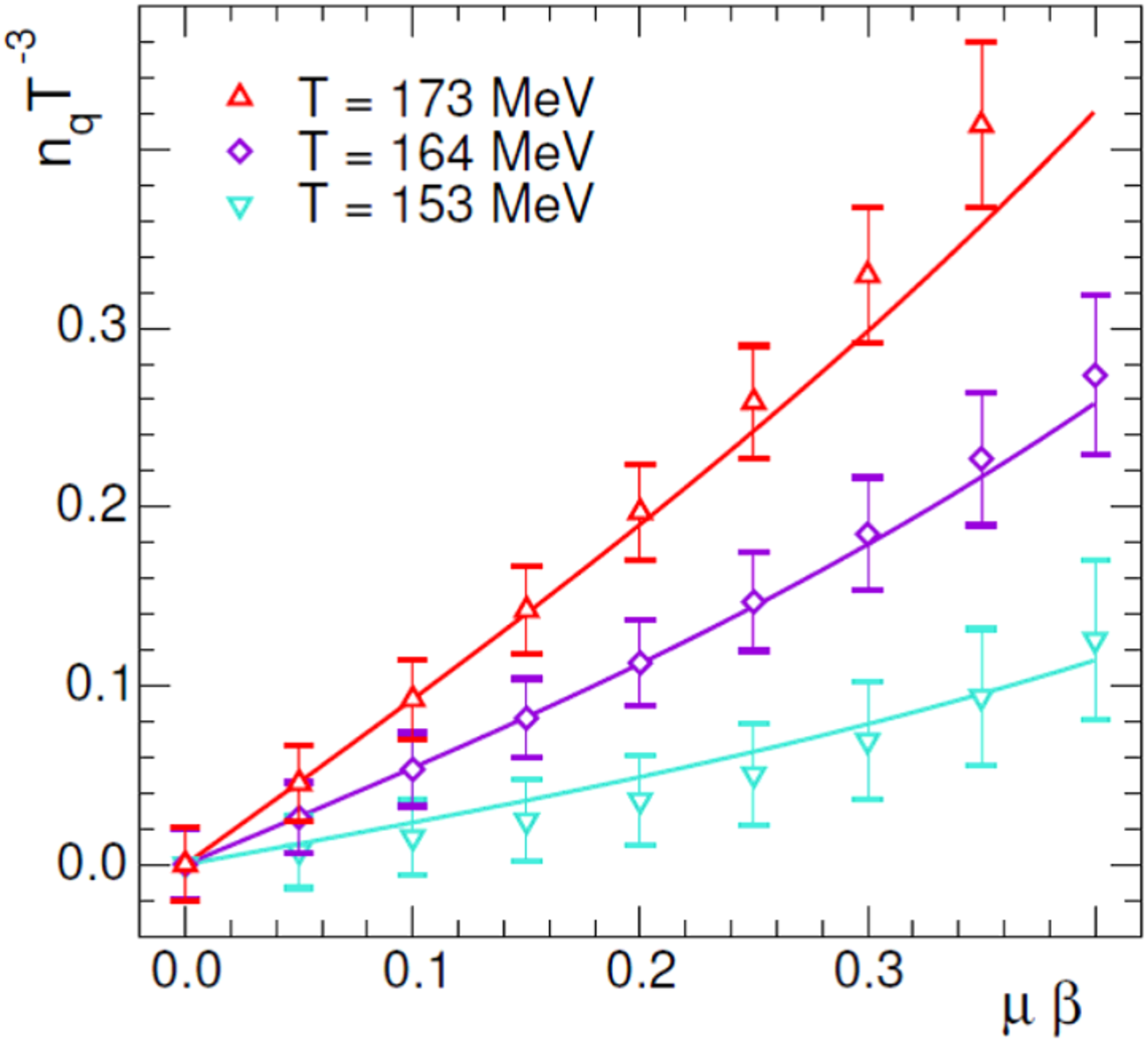}
	\caption{Comparison between the number density from canonical approach
		and Taylor expansion method.
		This figure was made by Danzer and Gattringer in Ref.~\cite{Gatt}.}
	\label{Gatt_graph}
\end{figure}

J. Danzer and C. Gattringer performed the calculation of number density
and number susceptibility at finite chemical potential with $8^3 \times 4$ lattice
in 2011\cite{Gatt}.
Then, they compare the results of canonical approach with
the results of Taylor expansion method as Fig.~\ref{Gatt_graph} and Tab.~\ref{Gatt_table}.
\begin{table}
	\centering
 	\includegraphics[width=0.8\hsize, clip]{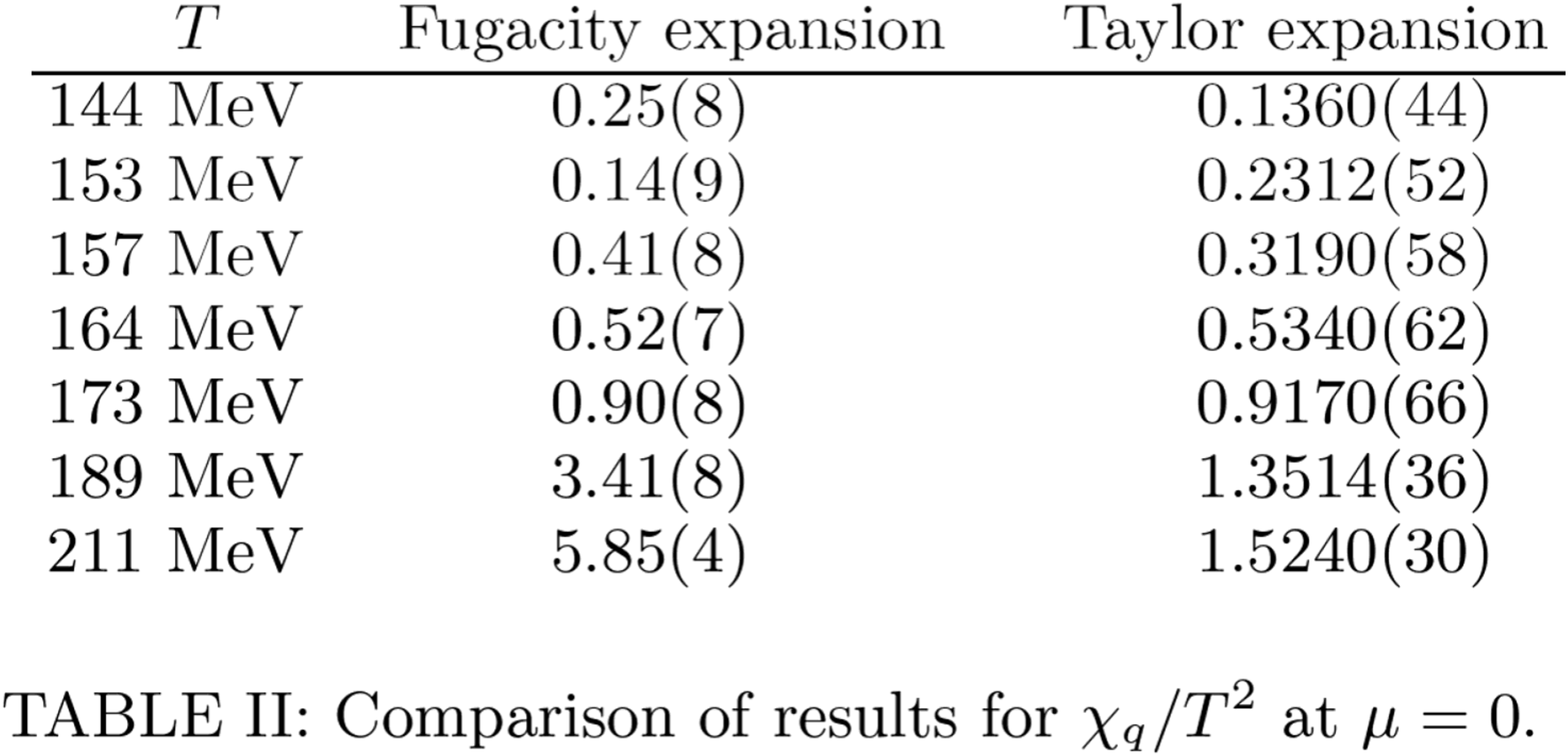}
	\caption{Comparison between the number susceptibility from canonical approach
		(fugacity expansion)
		and Taylor expansion method.
		This table was made by Danzer and Gattringer in Ref.~\cite{Gatt}.}
	\label{Gatt_table}
\end{table}
From Fig.~\ref{Gatt_graph}, we see that the results of canonical approach and Taylor expansion
method are consistent.
However, from Tab.~\ref{Gatt_table},
we see the difference between two methods.

Danzer and Gattringer noted that, at finite volume, fugacity series is a finite Laurent series
in the fugacity parameter $e^{\mu/T}$,
while the Taylor series is an infinite series even at this volume.
Since the fugacity series and Taylor series have different properties,
they suggested that we have to choose a better method (fugacity expansion or
Taylor expansion) for our purposes.

\newpage

\section{Calculation methods of fermion determinant}

Before discussing the problems of canonical approach,
we discuss the computing cost of the fermion determinant.


Since the fermion matrix $\Delta(\mu)$ has many degrees of freedom,
its rank becomes very large.
Concretely, $\mathrm{rank} \, \Delta = N_x N_y N_z N_t \times 4 \times 3$
($N_x$ is a number of points of x coodinate, and so on.)
and if we choose $N_x = N_y = N_z = N_t = 10$, $\mathrm{rank} \, \Delta \approx 10^5$.
Moreover, we have to calculate it for EACH configuration.

\subsection{Reduction formula}


We firstly introduce the reduction formula of Wilson fermion matrix.
It was suggested by Nagata and Nakamura\cite{NN-reduction} in 2010.
(See also Ref.\cite{reduction-2}.)
Note that the reduction formula for staggered fermion have been suggested since 1986
\cite{canonical, reduction_stag}.
Nagata and Nakamura noticed that Wilson fermions form a sparse band matrix.
It can be resolved into the diagonal part and its neighbor parts.

For future reference, we divide the determinant of Wilson fermions into three terms,
\begin{equation}
	\Delta = B - 2 z^{-1} \kappa r_{-} V - 2 z \kappa r_{+} V^\dagger  \  .
\end{equation}
Here, $r_{\pm} = (1 \pm \gamma_4)/2$ is the projection operators, $z = e^{-\mu a}$ and
\begin{align}
	B(x, x^\prime) &:= \delta_{x, x^\prime}
		- \kappa \sum_{i=1}^{3} \Bigl( \bigl( 1 - \gamma_i \bigr) U_i (x) \delta_{x^\prime, x+\hat{i}}
			+ \bigl( 1 + \gamma_i \bigr) U_i^\dagger (x^\prime) \delta_{x^\prime, x-\hat{i}} \Bigr)  \notag \\
		&\hspace{1em}  - \kappa C_{SW} \delta_{x, x^\prime} \sum_{\mu < \nu} \sigma_{\mu \nu} F_{\mu \nu} (x)  \  ,  \\
	V(x, x^\prime) &:= U_4 (x) \delta_{x^\prime, x+\hat{4}}  \  ,  \\
	V^\dagger(x, x^\prime) &:= U_4^\dagger (x) \delta_{x^\prime, x-\hat{4}}  \  .
\end{align}
These matrices are also given as follows in a time--plane block matrix form,
\begin{equation}
	B = 
	\begin{array}{cc}
		& t^\prime=1 \hspace{3em} \cdots \hspace{3em} t^\prime=N_t  \\
		\begin{array}{c}
			t=1 \\
			t=2 \\
			t=3 \\
			\vdots \\
			t=N_t
		\end{array}
		& \left(
		\begin{array}{cccccc}
			B_1 & 0   & 0   & \cdots & 0 & 0  \\
			0   & B_2 & 0   & \cdots & 0 & 0  \\
			0   & 0   & B_3 & \cdots & 0 & 0  \\
			\vdots & \vdots & \vdots & \ddots & \vdots & \vdots  \\
			0   & 0   & 0 & \cdots & 0 & B_{N_t}  
		\end{array}
		\right)  \  ,
	\end{array}
\end{equation}

\begin{equation}
	V = \footnotesize \left(
		\begin{array}{ccccccc}
			0          & U_4(t=1) & 0        & \cdots   & 0            & 0  \\
			0          & 0        & U_4(t=2) & \cdots   & 0            & 0  \\
			\vdots     & \vdots   & \vdots   & \ddots   & \vdots       & \vdots  \\
			0          & 0        & \cdots   & \cdots   & U_4(t=N_t-2) & 0  \\
			0          & 0        & \cdots   & \cdots   & 0            & U_4(t=N_t-1)  \\
			U_4(t=N_t) & 0        & \cdots   & \cdots   & 0            & 0  \\
		\end{array}
		\right) \normalsize  \  .
\end{equation}

In order to cancel $V^\dagger$, we multiply $\Delta$ and
\begin{equation}
	P := r_{-} + z^{-1} r_{+} V  \  .
\end{equation}
As a result, 
\begin{equation}
	\Delta P = (B r_{-} - 2 \kappa r_{+}) + (B r_{+} - 2 \kappa r_{-}) z^{-1} V  \  ,
\end{equation}
Here, $(B r_{-} - 2 \kappa r_{+})$ is the diagonal part
and $(B r_{+} - 2 \kappa r_{-}) z^{-1} V$ is the other part.
As the time--plane block matrix, it is given as 
\begin{equation}
	\Delta P =
		\left(
		\begin{array}{ccccc}
			\alpha_1 & \beta_1 z^{-1} &&&  \\
			& \alpha_2 & \beta_2 z^{-1} &&  \\
			&& \ddots & \ddots &  \\
			&&& \alpha_{N_t-1} & \beta_{N_t-1} z^{-1}  \\
			\beta_{N_t} z^{-1} &&&& \alpha_{N_t}
		\end{array}
		\right)  \  ,
\end{equation}
where,
\begin{align}
	\alpha_i &= \alpha^{a b, \mu \nu}(\vec{x},\vec{y},t_i)  \notag  \\
		&= B^{a b, \mu \sigma}(\vec{x},\vec{y},t_i) r_{-}^{\sigma \nu}
			- 2 \kappa r_{+}^{\mu \nu} \delta^{a b} \delta(\vec{x}-\vec{y})  \  ,  \\
	\beta_i &= \beta^{a b, \mu \nu}(\vec{x},\vec{y},t_i)  \notag  \\
		&= B^{a c, \mu \sigma}(\vec{x},\vec{y},t_i) r_{+}^{\sigma \nu} U_4^{c b}(\vec{y},t_i)
			- 2 \kappa r_{-}^{\mu \nu} \delta(\vec{x}-\vec{y}) U_4^{c b}(\vec{y},t_i)  \  .
\end{align}
Here, $a, b$ is color indices

Using these equations, we can write the determinant of $\Delta P$,
\begin{equation}
	\det(\Delta P) = \left( \prod_{i=1}^{N_t} \det(\alpha_i) \right) \det \left( 1 + z^{-N_t} Q \right)  \  ,
\end{equation}
where $Q := (\alpha_1^{-1} \beta_1) \cdots (\alpha_{N_t}^{-1} \beta_{N_t})$.
Since $\det(P) = z^{-N/2}$ and $N = N_x N_y N_z N_t \times 3 \times 4$, we get
\begin{align}
	\det \Delta &= z^{N/2} \left( \prod_{i=1}^{N_t} \det(\alpha_i) \right) \det \left( 1 + z^{-N_t} Q \right)  \notag  \\
		&= z^{-N/2} \left( \prod_{i=1}^{N_t} \det(\alpha_i) \right) \det \left( z^{N_t} + Q \right)  \  .
\end{align}
Then, we use a property of determinant $\det(A B) = \det(A) \det(B)$,
and $\det(cA) = c^{N} \det(A)$ with a scalar $c$ and the rank $N$.\\

We are going to see that this method is related with the canonical partition function.

Let $\lambda_n$ be the eigenvalues of $Q$.
Then, the determinant of the reduced matrix is written as
\begin{equation}
	\det(z^{N_t} + Q) = \prod_{n=1}^{N_\textrm{red}} (\lambda_n + z^{N_t})  \  ,
\end{equation}
where $N_\textrm{red}$ is the rank of the reduced matrices.
We can expand it as
\begin{equation}
	z^{-N/2} \prod_{n=1}^{N_\textrm{red}} (\lambda_n + z^{N_t})
		= \sum_{n=-N_\textrm{red}/2}^{N_\textrm{red}/2} c_n {\left( z^{N_t} \right)}^n  \  .
\end{equation}
Here, using $z^{N_t} = e^{\mu a N_t} = e^{\mu/T}$, we finally obtain the following relation,
\begin{equation}
	\det \Delta(\mu) = \sum_{n=-N_\textrm{red}/2}^{N_\textrm{red}/2} C_n {\left( e^{\mu/T} \right)}^n  \  .
\end{equation}
We can calculate $C_n$ numerically.
Nagata and Nakamura noticed later that $C_n$ is just the canonical partition function $Z_C(n)$.
Thus, we can calculate $Z_C(n)$ with the eigenvalues of reduction matrix.\\

As seen above, we can dissolve the determinant of $\Delta$ into one of $\alpha_i$ and $\beta_i$.
Then, we reduce the rank of matrix, from $\textrm{rank} \Delta = N_x N_y N_z N_t \times 4 \times 3$
to $\textrm{rank} \alpha_i = \textrm{rank} \beta_i = N_x N_y N_z \times 4 \times 3 = (\textrm{rank} \Delta)/N_t$.
The time complexity of a determinant calculation depending on the rank $N_\textrm{rank}$
is $O(N_\textrm{rank}^3)$ in a typical algorithm.
In the reduction formula, we repeat the calculation of $\det \alpha_i$ and $\det \beta_i$ $N_t$ times.
Therefore, the complexity of its algorithm is $O \left( (N_x N_y N_z \times 4 \times 3)^3 \times N_t \right)$.
On the other hand, the complexity of naive method is $O \left( (N_x N_y N_z N_t \times 4 \times 3)^3 \right)$.
If we set $N_x = N_y = N_z = N_t = 10$, it is reduced from $O(10^{15})$ to $O(10^{13})$.
Then, the running time is decreased to $1/100$, roughly.

When we use the order function $O(\cdot)$, we usually neglect the constants.
In following sections, we only consider the dependence on $N_x, N_y, N_z$ and $N_t$
because the physics of system depends on them.

Although the reduction formula also seems efficient, calculation cost is still large.
Especially, when we calculate with large lattice,
the time complexity becomes to diverge with $O \left( (N_x N_y N_z)^3 \right)$.
Therefore, we need to reduce it.

\subsection{Hopping parameter expansion}


To reduce the time complexity of the determinant calculation,
we can use the hopping parameter expansion (HPE).
In this method, we firstly rewrite the determinant as follows:
\begin{align}
	\det \Delta &= \exp \Bigl[ \mathrm{Tr} \log \Delta \Bigr]  \\
		&= \exp \left[ \mathrm{Tr} \log \bigl( 1 - \kappa Q \bigr) \right]  \ .
\end{align}
Then we use the fact that the Wilson fermions form as $\Delta = 1 - \kappa Q$.

Here, if hopping parameter $\kappa$ is small enough,
we can expand the logarithm as
\begin{equation}
	\det \Delta (\mu) = \exp \left[ \textrm{Tr} \sum_{k=1}^{\infty} \frac{1}{k} {\left(\kappa Q(\mu) \right)}^{k} \right]  \ .  \label{HPE}
\end{equation}
This is the hopping parameter expansion.
Using this equation,
we can calculate $\det \Delta(\mu)$ by the polynomial of $Q(\mu)$.
Then, the complexity is $O \left( (N_x N_y N_z N_t)^2\right)$.
This is because the trace of $Q^k$ is calculated by multiplication of $Q$ and a vector.
Such computation costs us $O(N_\textrm{rank}^2)$ operations.
We can reduce the time complexity using this method.

However, the calculation cost is still large.
It is because we have to calculate $\det \Delta(\mu)$ for varying $\mu$.
In order to reduce the cost, we can use the following method.

\subsection{Winding number expansion}


Winding number expansion (WNE)\cite{WNE, WNE_Gatt} is an improved method of HPE.

In HPE, we rewrite $\det \Delta$ to $\exp \left[ \mathrm{Tr} \log \Delta \right]$
and expand $\log \Delta$.
Then, using a method of statistical mechanics, we can rewrite the trace as
\begin{equation}
	\mathrm{Tr} \log \Delta(\mu)
		= \sum_{ \Psi } \bigl\langle \Psi \bigl| \log \Delta(\mu) \bigr| \Psi \bigr\rangle  \  ,
\end{equation}
where $\left| \Psi \right\rangle$ is a state in the system.
The right hand side of this equation means a summation of paths
of which the initial state and the final state are the same (closed loops).
Thus, we can rewrite the trace to a summation of closed loops.

Here, there are many types of size or shape of loops.
We therefore have to distinguish them to calculate in practice.
Now, let us focus on the chemical potential dependence.

From the definition of $\Delta(\mu)$,
we notice that the coefficient $\exp(\mu a N_t) = \exp(\mu/T)$ appears
when a fermion hop $N_t$ times along the time direction.
We hence can write $\det \Delta(\mu)$ formally,
\begin{equation}
	\det \Delta(\mu) = \exp \left[ \sum_{k=-\infty}^{\infty} W_k \left( {e^{\mu/T}} \right)^k \right]  \ .  \label{WNE}
\end{equation}
We can calculate $W_k$ numerically. The algorithm is discussed in Appxs.~\ref{alg_WNE} and \ref{alg_noise}.
(See also Refs.~\cite{Oka-3} and \cite{Oka-4}.)
Using this equation, we can see the $\mu$ dependence on $\det \Delta$ clearly.

Note that Eq.~(\ref{WNE}) can be used at any chemical potential.
That is, we can calculate $Z(\mu)$ from $\det \Delta(\mu)$ via WNE without canonical ensembles.
Such a calculation is discussed in Appx.~\ref{direct}.

Since $W_k$ does not depend on $\mu$, if ONCE we measure $W_k$,
we can calculate $\det \Delta$ with arbitrary $\mu$.
The time complexity is
$O \left( (N_x N_y N_z)^2 \times \right.$ $\left. (N_t)^2 \right)$
in this method.
(See Appxs.~\ref{alg_WNE} and \ref{alg_noise}.)
The complexity of spatial part of this method is smaller than the one of reduction formula.
Therefore, we can calculate $Z_C(n)$ in large lattice using WNE.
In this thesis, we adopt this method when we calculate $Z_C(n)$.

\vspace{2ex}

\section{Problems of canonical approach}


It is known that canonical approach has several numerical problems.
In this section, we will see the problems
using a raw data of canonical partition function as an example.

The following figure shows a raw data of canonical partition function
calculated on one configuration $Z^{(1)}_C(n)$.
We then generate a configuration at $T/T_c = 0.93(5)$ and $m_\pi/m_\rho = 0.80$,
and set the sample size of Fourier transform as $N_{FT} = 1024$.
\begin{figure}[h]
	\centering
 	\includegraphics[width=0.8\hsize, clip]{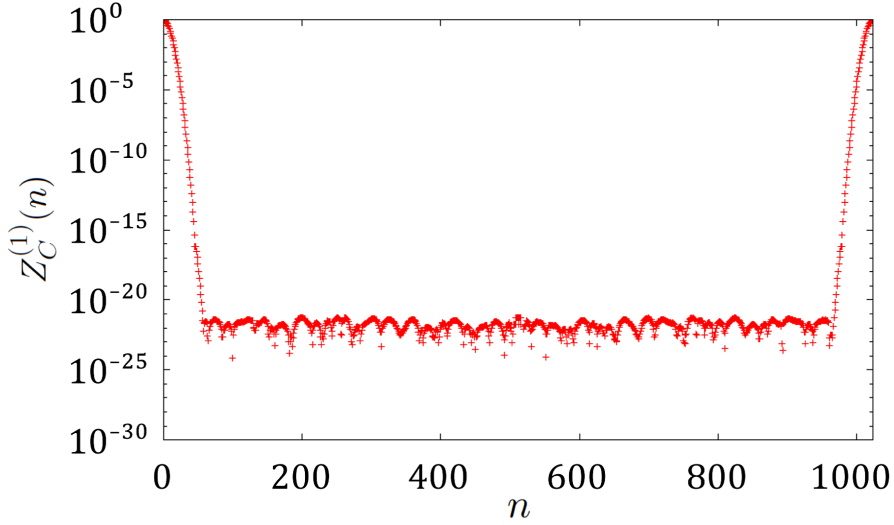}
	\caption{Raw data of canonical partition function with one configuration.}
	\label{Zn_one_orig}
\end{figure}
We find that $Z_C(n)$ is a monotonic decrease function when $n$ is small
and it becomes smaller rapidly with increasing $n$.
However, 
\begin{itemize}
	\item $Z_C(n)$ is saturated at $n = 50$, and
	\item it recovers near $n = 1023 = N_{FT}-1$.
\end{itemize}
As we know, $Z_C(n)$ describes an existence probability of a system
which $n$ particles are in.
Therefore, above figure means that, when $\mu = 0$,
$n=1$ system and $n=1023$ system exists with the same probability!

It is manifestly ill data.
In following sections, we discuss why such data were generated.

\subsection{Property of discrete Fourier transform}


First, we remember a property of discrete Fourier transform (DFT).

DFT is a transformation which is defined as the following equation:
\begin{equation}
	\tilde{F}(k) = \frac{1}{N_{FT}} \sum_{j=0}^{N_{FT}-1} f(j) e^{-i \phi_k j}  \  ,
		\hspace{1em}  \phi_k = \frac{2 \pi k}{N_{FT}}  \  .  \label{DFT}
\end{equation}
Now, we only have $f(j)$ ($j \in \mathbf{Z} , 0 \leq j < N_{FT}$)
which are sampled on $N_{FT}$ points.
In DFT, we construct the finite series $\tilde{F}(k)$ using the finite series $f(j)$.

Then, we note that,
\begin{equation}
	\exp \left( -i \, \phi_k \, j \right)
		= \exp \left( -i \left[\phi_k + 2 \pi n \right] j \right)
		= \exp \left( -i \left[ \phi_{k + N_{FT} \, n} \right] j \right)  \  ,
\end{equation}
with any $n \in \mathbf{Z}$.
That is why we can not distinguish $\tilde{F}(k)$ from $\tilde{F}(k+N_{FT} \, n)$.
This phenomenon is called aliasing.

Aliasing comes from the periodic boundary condition of DFT,
and it makes us equate low negative frequency with high positive frequency.
We can see this considering $n = 1$ and $k \leq 0$ case for example.
Concretely,
\begin{align}
	\tilde{F}(0) &= \tilde{F}(N_{FT}) \\
	\tilde{F}(-1) &= \tilde{F}(N_{FT}-1) \\
	\tilde{F}(-2) &= \tilde{F}(N_{FT}-2) \  .
\end{align}
Thus, Fig.~\ref{Zn_one_orig} shows that
$n=1$ system and ``$n=-1$" system exists with the same probability when $\mu = 0$.
This means that the symmetry of the exchange of particles for anti--particles
(C--symmetry) is satisfied in the QCD system.
Then, the highest baryon number is $n_\mathrm{max} = N_{FT}/2$.

Such a property is seen in the work of H. Nyquist\cite{Nyquist} in 1928
and proved by C.E. Shannon\cite{Shannon} in 1949.
$n_\mathrm{max} = N_{FT}/2$ is called Nyquist frequency
or the folding frequency.

\begin{figure}[h]
	\centering
 	\includegraphics[width=0.8\hsize, clip]{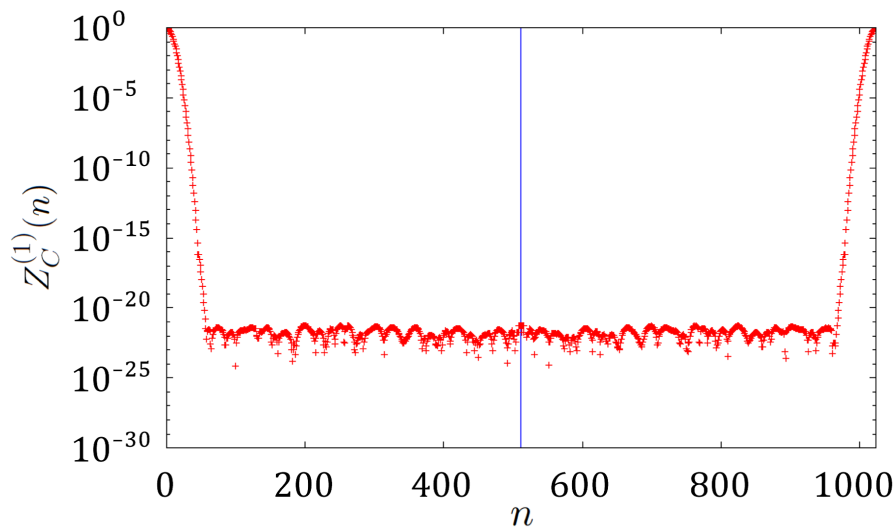}
	\caption{Raw data of canonical partition function with one configuration.
			Blue line denotes half of data.}
	\label{Zn_one_fold}
\end{figure}

We may see this property from Fig.~\ref{Zn_one_orig}.
We noticed that $Z_C(n)$ in Fig.~\ref{Zn_one_orig} is symmetric around $n = N_{FT}/2 = 512$.
Figure~\ref{Zn_one_fold} shows it clearly.
Note that this data is not symmetric completely
because there is the numerical errors which we will discuss below in detail.

Therefore, although we calculate $Z_C(n)$ using DFT with $N_{FT}$, 
we can use only half of data.

\subsection{$^*$An instability of DFT}

\label{err_def}

We do not understand yet why $Z_C(n)$ is saturated at $n = 50$.
Although it had not been investigated for a long time,
the author studied it from a numerical standpoint recently\cite{Oka-1, Oka-2}. \\

First, the author considered that this saturation is caused by the numerical
errors, and if we understand how the errors affect to our calculation, we can reduce this saturation.
He notes that we can classify the numerical errors into four types of errors\cite{Oka-1}
(see also Refs.\cite{TheArt} and \cite{guardbit}).
\begin{enumerate}
	\item Rounding error: for a numerical calculation, numerical values are always rounded when
		they are assigned to each variable because the variables have only finite significant digits.
		Therefore, we can not represent a mathematical value completely by using the numerical variables;
		the difference of mathematical and numerical values is called rounding error.

		For example, this error arises in the following type of calculation,
		\begin{align}
				\frac{1}{3} &= 3.333333\cdots \times 10^{-1}
						\hspace{0.6em} \textrm{(Exact mathematical value)}  \notag \\
				&\xrightarrow{\textrm{assignment}} 3.333333 \times 10^{-1}
						\hspace{0.6em} \textrm{(Numerical value; 7 digits.)}  \  ,
				\label{round_err}
		\end{align}
		In this case, the exact mathematical value has infinite significant digits, 
		but the numerical value has only seven significant digits.
		The numerical result is less than the exact mathematical one as $3.333\cdots \times 10^{-8}$.
		To perform rounding exactly, computers usually have a guard digit\cite{guardbit}.
			
	\item Truncation error:
		when we estimate an infinite series numerically,
		we have to truncate it because we can use only finite computation resource.
		Then, there is a difference between the exact mathematical result and its truncated one.
		This difference is called truncation error.
		We can estimate the effect of this error by varying a truncation point of the series.

	\item Cancellation of significant digits:
		when we calculate a subtraction between two nearly equal values,
		a lot of higher significant digits are canceled.
		Therefore, the result has only a few significant digits
		because the number of significant digits is limited in the numerical calculation.
		This phenomenon is called the cancellation of significant digits.

		It occurs in the following type of calculation:
		\begin{align}
			&1.234567 - 1.234566 \hspace{0.6em} \textrm{(7 digits)}  \notag \\
			&\hspace{2em} = 0.000001 \hspace{0.6em} \textrm{(1 digit.)}  \  .  \label{drop_1}
		\end{align}
		In this case, variables have seven significant digits at first.
		However, six significant digits are lost when we calculate the subtraction.
		If we want to reduce the effect of this cancellation,
		we should increase the number of significant digits.
		For instance, we can consider the following calculation
		instead of above one:
		\begin{align}
			&1.234567444444444444444 - 1.234566111111111111111 \hspace{0.6em} \textrm{(22 digits)}  \notag \\
			&\hspace{2em}= 0.000001333333333333333 \hspace{0.6em} \textrm{(16 digits.)}  \label{drop_16}
		\end{align}
		Then, although six significant digits are similarly lost in this calculation,
		sixteen digits still remain in the final result.

	\item Loss of trailing digits:
		when we calculate an addition or subtraction between a huge number and small one,
		many lower significant digits of the small number are cut off.
		This is because the variables hold only finite significant digits.
		Then, the numerical result is underestimated. 
		This phenomenon is called the loss of trailing digits.

		Specifically, it arises the following type of calculation:
		\begin{align}
			&3 \times 10^{10} + 2 \times 10^{0}  \notag \\
			&\hspace{2em} = 3.000000002 \times 10^{10} \hspace{0.6em} \textrm{(Exact mathematical result)}  \notag \\
			&\hspace{2em} \xrightarrow{\textrm{assignment}} 3.000000 \times 10^{10}
				\hspace{0.6em} \textrm{(Numerical result; 7 digits.)}  \,  ,  \label{loss_7}
		\end{align}
		In this case, variables have seven significant digits,
		and the numerical result is underestimated from the mathematical one by $2 \times 10^{0}$.

		As the cancellation of significant digits, 
		we should increase the number of significant digits in order to reduce this loss.
		\begin{align}
			&3 \times 10^{10} + 2 \times 10^{0}  \notag \\
			&\hspace{2em} = 3.000000002 \times 10^{10} \hspace{0.6em} \textrm{(Exact mathematical result)}  \notag \\
			&\hspace{2em} \xrightarrow{\textrm{assignment}} 3.0000002000000 \times 10^{10}
				\hspace{0.6em} \textrm{(Numerical result; 16 digits.)} \,  ,  \label{loss_16}
		\end{align}
		In this calculation, by increasing significant digits,
		we reduce the error between the mathematical result and numerical one.

\end{enumerate}
Next, the author considered as follows.
\begin{itemize}
	\item Usually, the rounding error is not transmitted to higher significant digits
		since computers have a guard digit\cite{guardbit}.
		Sometimes we may see that this error transmits to higher significant digits,
		but actually, only the cancellation of significant digits
		or loss of trailing digits occurs at the same time.
		In this paper, we consider the transmission of rounding error as an effect of
		the cancellation or loss.

	\item The truncation error may occur in our program.
		When we use HPE or WNE in order to calculate $\Delta(\mu)$,
		we have to truncate the infinite series of these expansion.
		Although the truncation error arises in our calculation,
		we can control it and estimate its effect by changing a truncation point of this series.

\end{itemize}
Thus, we conclude that the instability which we do not control is caused
by the cancellation of significant digits or the loss of trailing digits, or both.

Actually, he monitored the behavior of all variables in a DFT program,
and found that the cancellation of significant digits occurs and cause the instability of DFT.
This cancellation arises almost at the last addition of the summation in Eq.~(\ref{DFT});
specifically,
\begin{align}
	Z_C(n) &= \frac{1}{N} \left[ \left( \sum_{k=0}^{N-2} Z \left( \frac{2 \pi k}{N} \right) e^{-i (2 \pi k / N) n} \right) \right.  \notag  \\
		&\hspace{4em} + Z \left( \frac{2 \pi (N-1)}{N} \right) e^{-i (2 \pi (N-1) / N) n} \left] \phantom{\sum_{k=0}^{N-2}} \right.  \label{dFour_plus}  \  ,  \\  \vspace{-5em}
		&\hspace{4em} \uparrow  \textrm{This addition.} \notag
\end{align}
\begin{figure}[h]
	\centering
 	\includegraphics[width=0.8\hsize, clip]{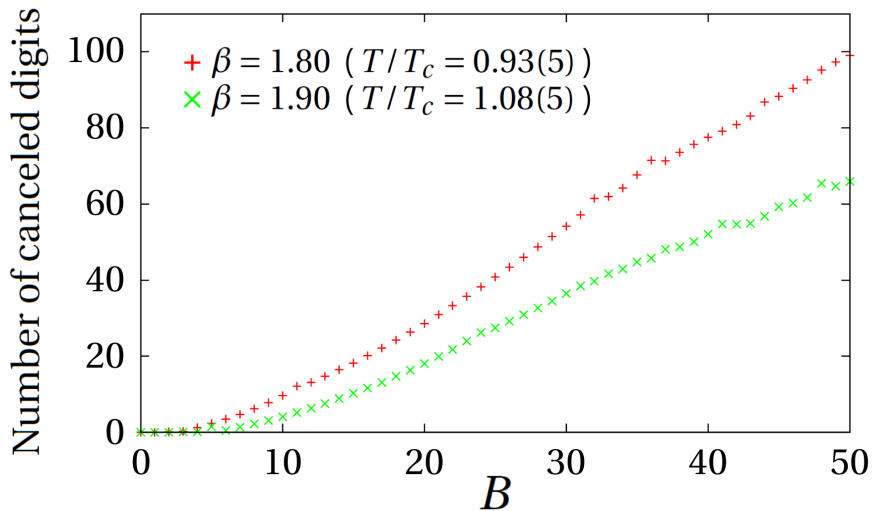}
	\caption{Number of canceled significant digits as a function of baryon number.
		The red and green points are data of low temperature and high temperature.}
	\label{FFT_drop}
\end{figure}
Figure~\ref{FFT_drop} shows how many digits are lost at the summation of Eq.~(\ref{dFour_plus}).
(Note that the author then uses the multi precision calculation which is described in section below.)
The algorithm is discussed in detail in Appx.~\ref{alg_cancel}.

From this figure, we can see that this cancellation becomes serious with increasing $n$.
For example, about 60 digits are canceled at $n=100$, 130 digits are canceled at $n=200$.
Usually, one use the double precision variables in order to calculate the physical quantities,
but it has only 16 significant digits.
Thus, we can not calculate $Z_C(n)$ well using the usual (double precision) variables.

\subsection{$^*$Reduce these problems}

\label{multipre_def}

In order to reduce these problems,
the author suggests to calculate $Z_C(n)$ with large enough data space,
specifically,
\begin{enumerate}
	\item
		\textbf{$N_{FT}$ should be set large enough value for one's purpose.\\}
		For instance, the author aims to study the finite density
		phase transition of QCD, i.e. the confinement--deconfinement phase transition.
		We will use a small lattice in the discussion below,
		whose size is $8^3 \times 4$.
		One considers that QCD critical point may be at roughly $T_c \approx 160$ MeV and
		$\mu^c = \mu^c_B/3 \approx$ 200--400 MeV\cite{Fodor-Katz, Taylor_critical, models}.
		Therefore, at $T = T_c$, lattice spacing $a$ is about $0.3$ fm
		because $N_t a = 1/T$ in finite temperature lattice QCD.
		Then, the lattice size is $(2.4 \textrm{ fm})^3 \times (1.0 \times 10^{-24} \textrm{ s})$.

		If critical chemical potential $\mu^c$ is $200$ MeV,
		it corresponds to length $l = 1.0$ fm.
		We then may consider that $l$ means the distance between quarks.
		Thus, by putting about 10--20 quarks in the system, we may simulate near the critical point.
		Accordingly, we need to set $N_{FT}/2 > 20$.

	\item
		\textbf{The number of significant digits should be set too.\\}
		The magnitude of the cancellation of significant digits
		is related with the net quark number $n$.
		If $N_{FT}$ is large, the cancellation is serious,
		and then we have to increase the number of significant digits.
		(We are going to see how to increase it below.)
		
		The canceled significant digits are roughly proportional to $n$.
		Therefore, one has to estimate or measure how many digits are canceled
		and set the number of significant digits to large enough value.

\end{enumerate}

The author studied how to enhance the accuracy of $Z_C(n)$ using the multi precision calculation.
This calculation extends the number of significant digits.
We can make it by binding several variables.
For instance, a real number $x$ represents as follows in the base--$b$ numerical system, 
\begin{equation}
	x = x_\textrm{sign} \cdot \left( x_1 b^{-1} + x_2 b^{-2} + \cdots + x_f b^{-f} \right) \cdot b^{x_0}  \  ,
\end{equation}
where
\begin{itemize}
	\item $x_\textrm{sign}$ is the signature of $x$ $\left( x_\textrm{sign} \in \{ +1, -1 \} \right)$;

	\item $x_0$ is the exponent of $x$ ($x_0 \in \mathbf{Z}$);
	
	\item $x_1, x_2, \cdots, x_f$ are the mantissas (or significands,
			$x_1 \in \{ 1, 2, \cdots, b-1 \}$,
			$x_i \in \{ 0, 1, 2, \cdots, b-1 \}$ ($i = 2, 3, \cdots, f$));
	
	\item $f$ is the number of significant digits; and
	
	\item $b$ is the base (or radix).
\end{itemize}
Usually, the computers are designed with $b=2$ and fixed $f$.
Then, $x_i$ are just the bits, $x_0$ is a set of several bits.
However, if we program this framework as a software ($x_i$ are constructed 
by integer variables with any $f$, also $b$ is a variable), we can change $b$ and $f$,
and then we can use a original variable with wide enough range and high accuracy.

In this thesis, we use the libraries of multi precision calculation,
MPFR (C++) and FMlib (FORTRAN) for our simulation.\\

\begin{figure}[h]
	\centering
 	\includegraphics[width=\hsize, clip]{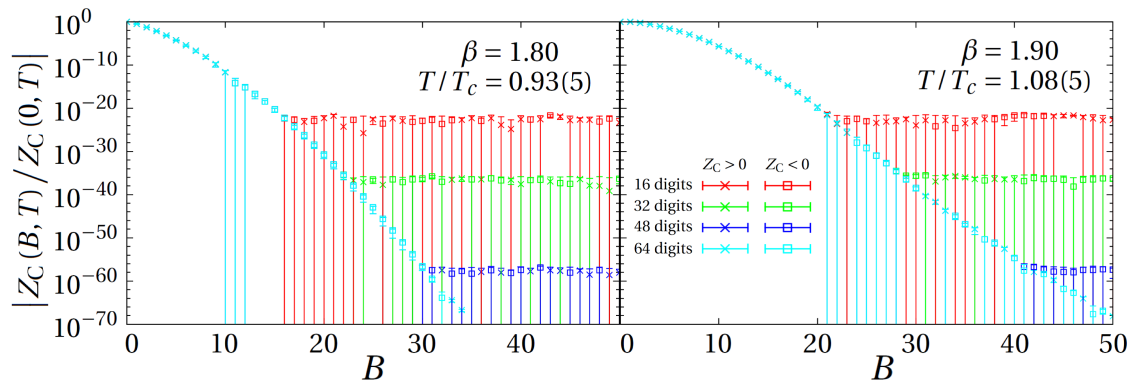}
	\caption{Normalized canonical partition function as a function of baryon number.
		The left and right panels show data of low temperature and high temperature, respectively.
		The red, green, blue and cyan points are the data which is calculated
		16, 32, 48 and 64 significant digits, respectively.}
	\label{Zn_drop}
\end{figure}

We can reduce the cancellation error by increasing the number of significant digits with multi precision.
The author calculated $Z_C(n)$ changing the precision of variables, 
and compared them as Fig.~\ref{Zn_drop} shows.
We find that, by increasing the number of significant digits $f$,
we can calculate $Z_C(n)$ up to large $n$.
This is because $Z_C(n)$ is just a remainder of the calculation of Eq.~(\ref{dFour_plus}).
That is, the number of \textit{canceled} significant digits is directly related with a magnitude of $Z_C(n)$.

He notes that, when we change the lattice volume $V$ or temperature $T$
fixing the number of significant digits, a saturation point (kink) of $Z_C(n)$ moves.

$Z_C(n)$ describes the existence probability of a system which $n$ particles are in.
If only $T$ becomes large, excited states become easy to appear in the system,
then $Z_C(n)$ decreases slowly with increasing $n$.
We can see this actually in Fig.~\ref{Zn_drop}.

In the other situation, if only $V$ becomes large,
also $Z_C(n)$ decreases slowly too
since the net quark number density $n/V$ becomes small.
If volume is changed as $V \rightarrow V^\prime = 2V$,
$Z_C(n)$ corresponds to $Z^\prime_C(n^\prime = 2n)$
because they represent the same density system.
Thus, we can replace $Z_C(1) \rightarrow Z^\prime_C(2),
Z_C(2) \rightarrow Z^\prime_C(4), \cdots$,
therefore $Z_C(n)$ decreases slowly.


\vspace{2ex}

\section{Study of finite density phase transition with canonical approach}

As seen above, by introducing these methods,
we can avoid the sign problem and calculate $Z_C(n)$ with high accuracy.
In this thesis, we calculate several thermodynamic observables using the canonical approach
with the multi precision calculation.
We will see the results below in detail.

\subsection{Thermodynamical observables and partition function}

From grand partition function $Z(\mu)$,
we can calculate thermodynamical observables as follows,
\begin{align}
	\frac{p(\mu_B,T)}{T^4} &= \frac{1}{V T^3} \log Z(\mu_B,T)  \notag \\
		&= \frac{{N_t}^3}{N_x N_y N_z} \log Z(\mu_B,T) \  ,
\end{align}
where $p$ is the pressure, $V = N_x a \times N_y a \times N_z a$ is the spatial volume of the system
and $\mu_B = 3 \mu$ is the baryon chemical potential.
We then use a relation in the finite temperature lattice QCD,
$N_t a = 1/T$, with lattice spacing $a$.

Taking a deviation of the pressure, we can get a useful relation,
\begin{align}
	\frac{\Delta p(\mu_B,T)}{T^4} &= \frac{p(\mu_B,T)}{T^4} - \frac{p(0,T)}{T^4}  \notag \\
		&= \frac{{N_t}^3}{N_x N_y N_z} \log \left( \frac{Z(\mu_B,T)}{Z(0,T)} \right)  \  .
\end{align}
It is useful because the partition function is normalized by $Z(0, T)$
as the expectation values of observables.

Other observables, baryon number density $n_B$ and baryon
number susceptibility $\chi$ are given by the differential of pressure,
\begin{align}
	\frac{n_B(\mu_B,T)}{T^3} &= \frac{\partial}{\partial (\mu_B/T)} \frac{p(\mu_B,T)}{T^4}  \  , \\
	\frac{\chi(\mu_B,T)}{T^2} &= \frac{\partial^2}{\partial (\mu_B/T)^2} \frac{p(\mu_B,T)}{T^4}  \  .
\end{align}\\

Note that, the partition function has the following periodicity as Eq.~(\ref{RW_period}),
\begin{equation}
	Z\left( \frac{\mu}{T} \right) = Z\left( \frac{\mu}{T} + i\frac{2 \pi k}{3} \right)  \  ,
\end{equation}
which is caused by RW symmetry.
Using this equation, we can rewrite the partition function as
\begin{equation}
	Z\left( \frac{\mu}{T} \right) = \frac{1}{3} \left[ \sum_{k=0}^{2} Z\left( \frac{\mu}{T} + i\frac{2 \pi k}{3} \right) \right]  \  .
\end{equation}
Then, the canonical partition function $Z_C(n)$ becomes
\begin{align}
	Z_C(n, T) &= \frac{1}{2 \pi} \int_{0}^{2 \pi} d\left( \frac{\mu_I}{T} \right) \left[ \frac{1}{3} \sum_{k=0}^{2} Z\left( \frac{i \mu_I}{T} + i\frac{2 \pi k}{3} \right) \right] e^{i (\mu_I/T) n}  \notag  \\
		&= \frac{1}{2 \pi} \int_{0}^{2 \pi} d\left( \frac{\mu_I}{T} \right) \left[ \frac{1 + e^{2 \pi i / 3} + e^{4 \pi i / 3}}{3} Z\left( \frac{i \mu_I}{T} \right) \right] e^{i (\mu_I/T) n}  \  .
\end{align}
Therefore, $Z_C$ has the following property,
\begin{equation}
	Z_C\left( n \not= 3 k \right) = 0  \  ,
\end{equation}
for any $k \in \mathbf{Z}$.
This property is called the triality.

Thus, the grand partition function is written as
\begin{equation}
	Z(\mu_B) = \sum_{B=-\infty}^{\infty} Z_C(B) \left( e^{\mu_B/T} \right)^{B}  \  ,
\end{equation}
where $B = n/3$ is the baryon number.
Because of this, we use the baryon number $B$ and baryon chemical potential $\mu_B$
instead of the quark number $n$ and quark chemical potential $\mu$ below.\\

In the following section, we compare our results of the canonical approach
with the results of MPR in order to discuss the validity range of canonical approach.
We use the results of MPR in Ref.~\cite{NN_eos},
and we adopt the same numerical setup as Ref.~\cite{NN_eos}
when we calculate our observables of canonical approach.


In Ref.\cite{NN_eos}, 
the authors also discuss consistency between MPR and Taylor expansion method with Wilson--clover fermions
and concluded that both methods produced consistent results in a small chemical potential region where
errors of both methods could be under control.
Therefore, our work enables us to check consistency among our canonical approach, MPR and Taylor expansion method.

\subsection{Numerical setup}

We adopted the clover improved Wilson fermion action with $N_f = 2$ and
$C_{SW} = (1 - 0.8412/\beta)^{-3/4}$ evaluated by a one--loop perturbation theory\cite{clover_fermions},
and Iwasaki gauge action\cite{Iwasaki}.
All simulations were performed on a
$N_x \times N_y \times N_z \times N_t = 8 \times 8 \times 8 \times 4$ lattice.
Number of flavor is set to $N_f = 2$.
We considered values of $\beta = 2.00$, $1.95$, $1.90$, $1.85$, $1.80$ and $1.70$;
they correspond to $T/T_c = 1.35(7)$, $1.20(6)$, $1.08(5)$, $0.99(5)$, $0.93(5)$ and $0.84(4)$,
respectively.
The values of the hopping parameter $\kappa$ were determined for each value of $\beta$
by following the line of constant physics for the case of $m_\pi / m_\rho = 0.80$,
as in Ref.~\cite{WHOT}.
In Ref.\cite{WHOT}, the lined of constant $T/T_c$ is determined by the relation,
$T_c/m_\rho = A(1+B(m_\pi/m_\rho)^2) / (1+C(m_\pi/m_\rho)^2)$ with
$A = 0.2253(71)$, $B = -0.933(17)$ and $C = -0.820(39)$, which obtained in Ref.~\cite{WHOT_ref}
to evaluate $T_c/m_\rho$ for each $m_\pi/m_\rho$.

We generated gauge configurations at $\mu = 0$ using the HMC method.
We then set the step size $d\tau$ and number of steps $N_\tau$ for HMD part
to $d\tau = 0.2$ and $N_\tau = 50$ so that the simulation time was $N_\tau \times d\tau = 1$.
We neglected the first 2000 trajectories for termalization,
after we extracted 400 configurations every 200 trajectory for each parameter set.

Number of Fourier series $N_{FT}$ is set to $N_{FT} = 1024$ for test in this thesis.
When we calculate the grand partition function $Z(\mu)$ from the canonical partition function $Z_C(n)$,
we have to truncate the series of fugacity expansion.
The truncation point $B_\textrm{max}$ is listed in Tab.~\ref{param_table},
which is discussed below in detail.
Number of significant digits $f$ is set to $f = 400$ in this thesis.

\begin{table}
	\vspace{-5ex}
	\centering
		\begin{tabular}{ccccccc}
			$\beta$ & $C_\textrm{SW}$ & $\kappa$ & $T / T_c$ & $m_\pi / m_\rho$ & \# of Conf. & $B_\textrm{max}$ \\ \hline
			$2.00$ & $1.50579$ & $0.136931$ & $1.35(7)$ & $0.80$ & $400$ & $36$ \\
			$1.95$ & $1.52717$ & $0.137716$ & $1.20(6)$ & $0.80$ & $400$ & $34$ \\
			$1.90$ & $1.55044$ & $0.138817$ & $1.08(5)$ & $0.80$ & $400$ & $28$ \\
			$1.85$ & $1.57589$ & $0.140070$ & $0.99(5)$ & $0.80$ & $400$ & $34$ \\
			$1.80$ & $1.60380$ & $0.141139$ & $0.93(5)$ & $0.80$ & $400$ & $12$ \\
			$1.70$ & $1.66885$ & $0.142871$ & $0.84(4)$ & $0.80$ & $400$ & $30$
		\end{tabular}
	\caption{Parameters of each simulation.}
	\label{param_table}
\end{table}

\subsection{Error estimation}

\label{error_def}

When we calculate the fugacity expansion of the grand partition function $Z(\mu)$,
we must truncate the series as
\begin{equation}
	Z(\mu_B) = \sum_{B=-B_\textrm{max}}^{B_\textrm{max}} Z_C(B) \left( e^{\mu_B/T} \right)^{B}  \  .
\end{equation}
Thus, the truncation error arises, and then we have to estimate how large the error is.

In order to estimate it, we vary the number of series $B_\textrm{max}$
and check the convergence of results.
If the results are stable when $B_\textrm{max}$ is varied,
we can trust them because they should not depend on $B_\textrm{max}$ physically.

For a convergence check, we evaluate the relative error $R$ as follows\cite{Oka-2, Oka-5},
\begin{equation}
	R(\mu_B) := \frac{ \left\langle O(\mu_B) \right\rangle_{B_\textrm{max}} - \left\langle O(\mu_B) \right\rangle_{B_\textrm{max}-1} }{ \left\langle O(\mu_B) \right\rangle_{B_\textrm{max}} }  \  ,
\end{equation}
where $\left\langle O(\mu_B) \right\rangle_{B_\textrm{max}}$ is the expectation value
with the truncation at $B_\textrm{max}$.
In this thesis, we require that the results have at least two significant digits.
Therefore, we trust the results only for
\begin{equation}
	R(\mu_B) < 10^{-3}  \  .  \label{bound}
\end{equation}

\subsection{$^*$Numerical results}

Using the error estimation method,
we study the chemical potential dependence on the thermodynamic observables 
and estimate the validity range of our canonical approach.

\subsubsection{Canonical partition function and analysis of truncation point}

Before we introduce our results of thermodynamic observables,
we show the results of canonical partition function.
\begin{figure}
	\center
	\includegraphics[width=0.8\hsize, clip]{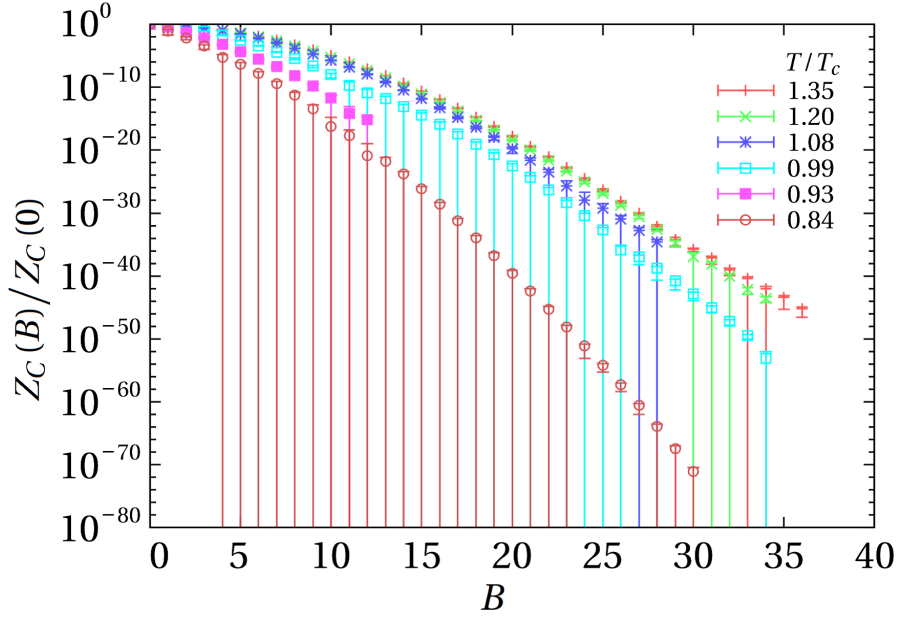}
	\caption{Baryon number dependence on normalized canonical partition function.}
	\label{Zn_graph}
\end{figure}
Figure~\ref{Zn_graph} shows the baryon number dependence on the canonical partition function.
From this figure, we can see that $Z_C(B)$ becomes smaller rapidly with increasing $B$
at both of temperatures below and above $T_c$.
Also, we see that $Z_C(B)$ depends on temperature; it is raised with increasing temperature.
We can consider that they are because $Z_C(B)$ represents the existence probability of the system
which $B$ particles are in\footnote{
	In detail, $Z_C(B) (e^{\mu/T})^B / Z(\mu)$ roles the weight function
	as we saw in Sec.~\ref{review},
	and then it represents the existence probability of the system.
	When $\mu=0$, $Z_C(B)/Z(0)$ represents the probability.
	We now neglect the factor $1/Z(0)$ in the discussion of this section
	because the factor is constant for $B$.
}.
That is, $B=1$ system is more likely than $B=2$ or another system when $\mu=0$,
and $B=2$ system at high temperature is easier to arise
than such systems at low temperature.

In Fig.~\ref{Zn_graph}, the canonical partition function is cut at $B_\textrm{max}$.
We define the value of $B_\textrm{max}$ using the following way.
Because of numerical and statistical errors,
sometimes $Z_C(B)$ becomes negative, although it has to satisfy $Z_C(B) > 0$.
Then, if the error of $Z_C(B)$ is larger than the absolute value of $Z_C(B)$,
we can use the positive value $-Z_C(B)$ instead of the negative value $Z_C(B)$.
However, if $Z_C(B)$ is negative and $\delta \left[ Z_C(B) \right] < \left| Z_C(B) \right|$,
we have to consider that the calculation does not work well at this parameter.
Therefore, when we find such ill data, we reject this and later data.
Then, we set the truncation point $B_\textrm{max}$ from the previous data.

\subsubsection{Pressure, baryon number density and baryon number susceptibility}

Let us introduce our results of thermodynamic observables.

\begin{figure}
	\center
	\includegraphics[width=0.8\hsize, clip]{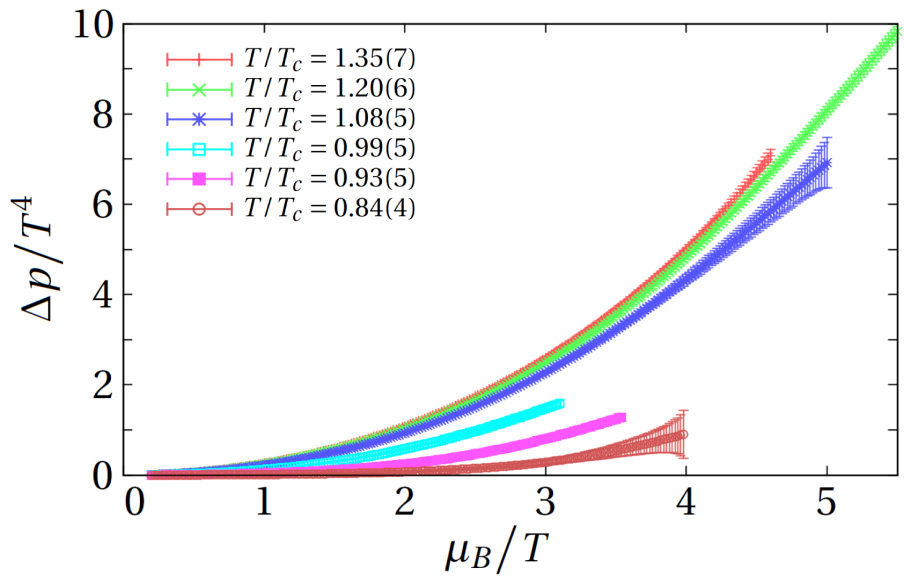}
	\caption{Chemical potential dependence on pressure.
			The red, green, blue, cyan, magenta, and brown points are the results
			at $T/T_c = 1.35(7)$, $1.20(6)$, $1.08(5)$, $0.99(5)$, $0.93(5)$ and $0.84(4)$, respectively.
			The upper bound of the baryon chemical potential is determined by Eq.~(\ref{bound}).
			This figure was already published in Ref.~\cite{Oka-1}}
	\label{pre-Tdep}
\end{figure}

First, we examine the pressure.
Fig.~\ref{pre-Tdep} shows the chemical potential dependence on pressure.
From this figure, it is found that the results of pressure at high temperature $T > T_c$ 
do not suffer from large errors up to $\mu_B/T = 5$, approximately. 
The results at low temperature $T < T_c$ are reliable up to $\mu_B/T = 3.5$--$4$.
Conversely, they are reliable only up to $\mu_B/T \approx 3$.
This may be because we generated configurations at $\mu_0=0$ 
and they suffered from fluctuations caused by the (crossover) phase transition at zero density. 
We may obtain clearer signals if we generate configurations at pure imaginary chemical potentials
because we can leave from the crossover phase transition line.

\begin{figure}
	\center
	\includegraphics[width=0.8\hsize, clip]{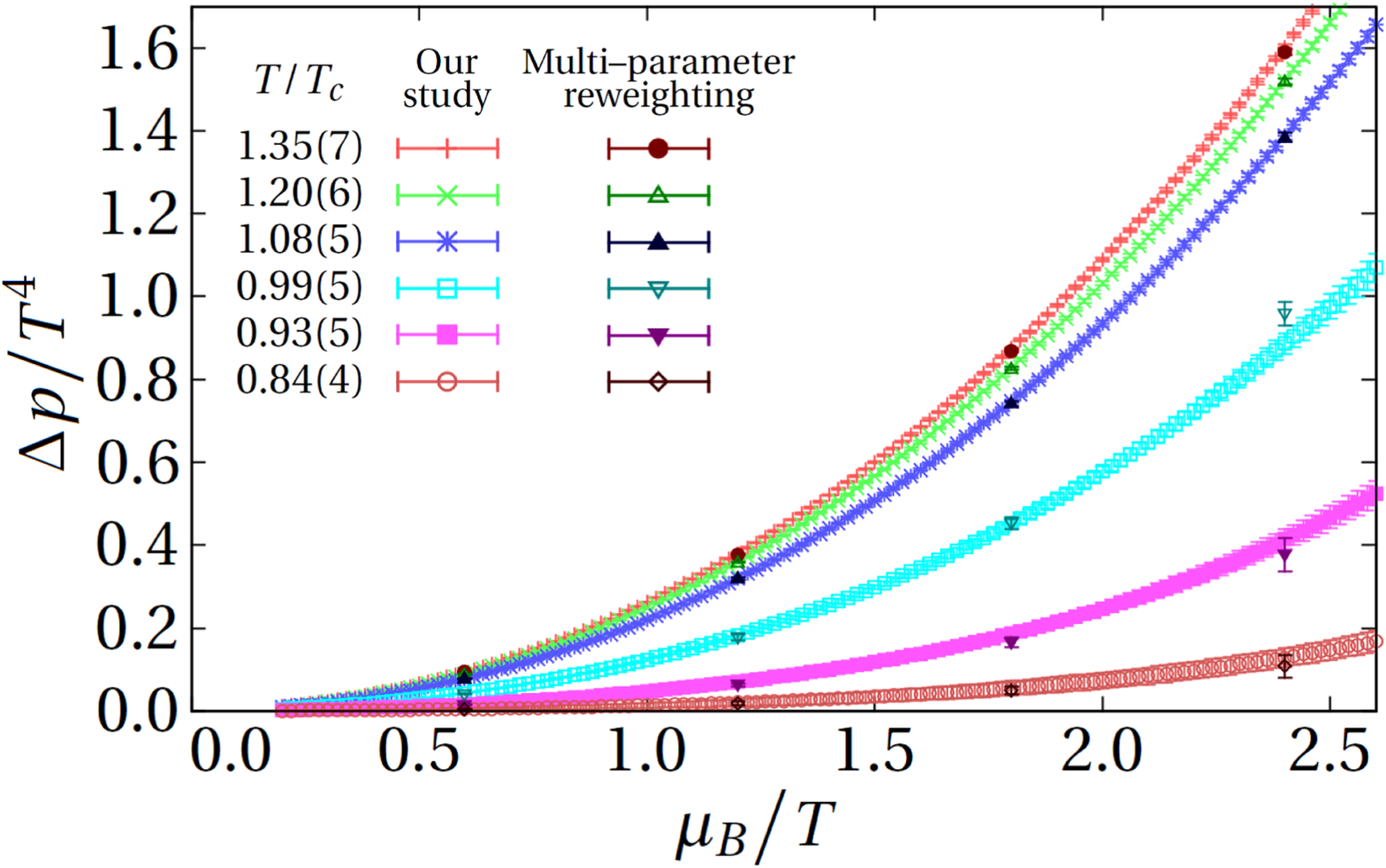}
	\caption{Comparison of pressure calculated
			by the canonical approach and the MPR method.
			The colors of the data points are the same as in Fig.~\ref{pre-Tdep}
			with several additional colors. 
			The data points plotted in the additional colors of dark red, dark green, dark blue,
			dark cyan, dark magenta, and dark brown points
			are the results at $T/T_c = 1.35(7)$, $1.20(6)$, $1.08(5)$, $0.99(5)$, $0.93(5)$ and $0.84(4)$,
			respectively, as calculated by the MPR method.
			This figure was already published in Ref.~\cite{Oka-1}}
	\label{pre-hikaku}
\end{figure}

Fig.~\ref{pre-hikaku} shows the comparison of pressure calculated by the canonical approach
and MPR method.
We can see that the results of canonical approach are consistent with the results
of MPR method.\\
\begin{figure}
	\center
	\includegraphics[width=0.8\hsize, clip]{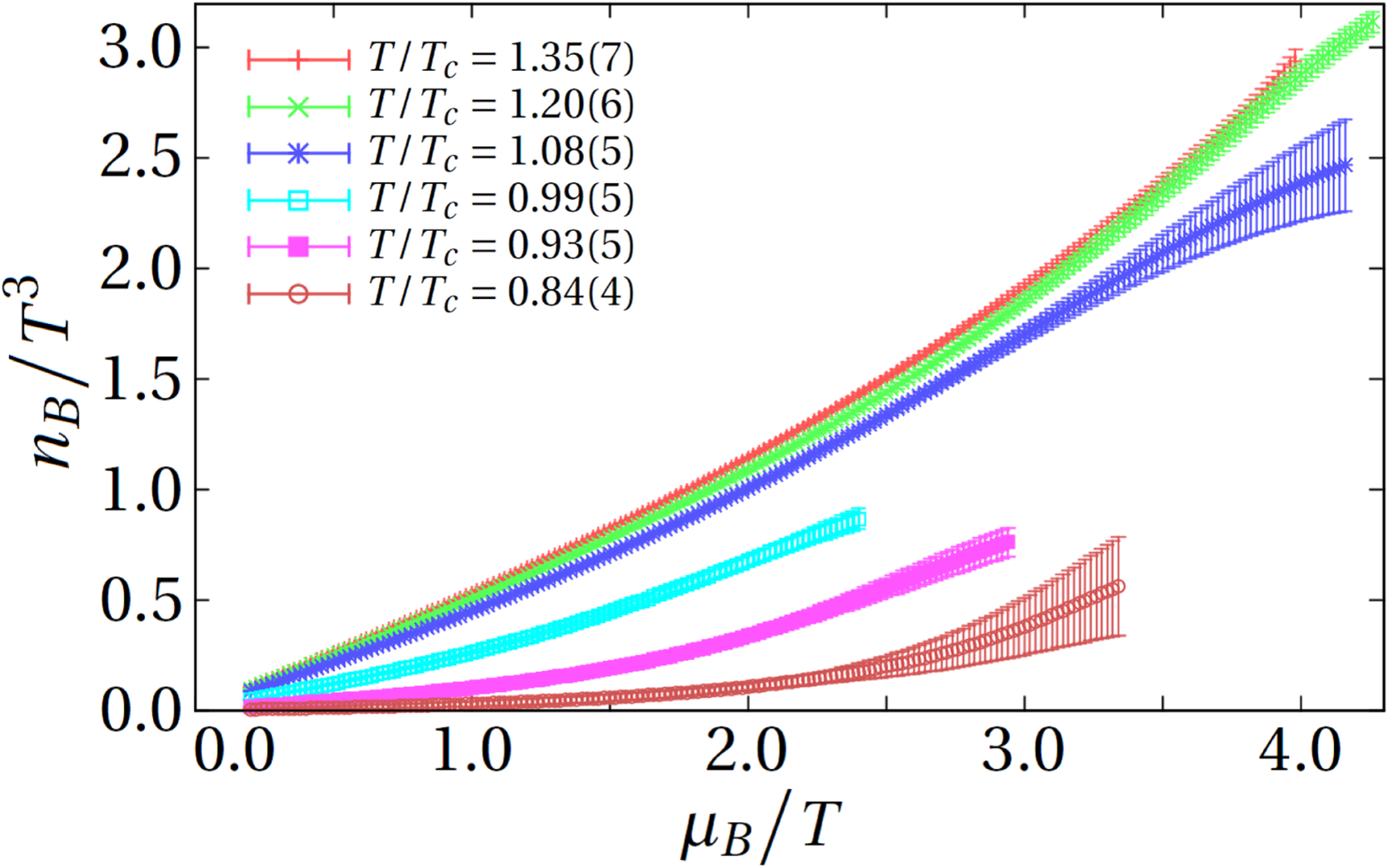}
	\caption{Chemical potential dependence on baryon number density.
			The colors of the data points are the same as in Fig.~\ref{pre-Tdep}. 
			The upper bound of the baryon chemical potential is determined by Eq.~(\ref{bound}).
			This figure was already published in Ref.~\cite{Oka-1}}
	\label{nd-Tdep}
\end{figure}
\begin{figure}
	\center
	\includegraphics[width=0.8\hsize, clip]{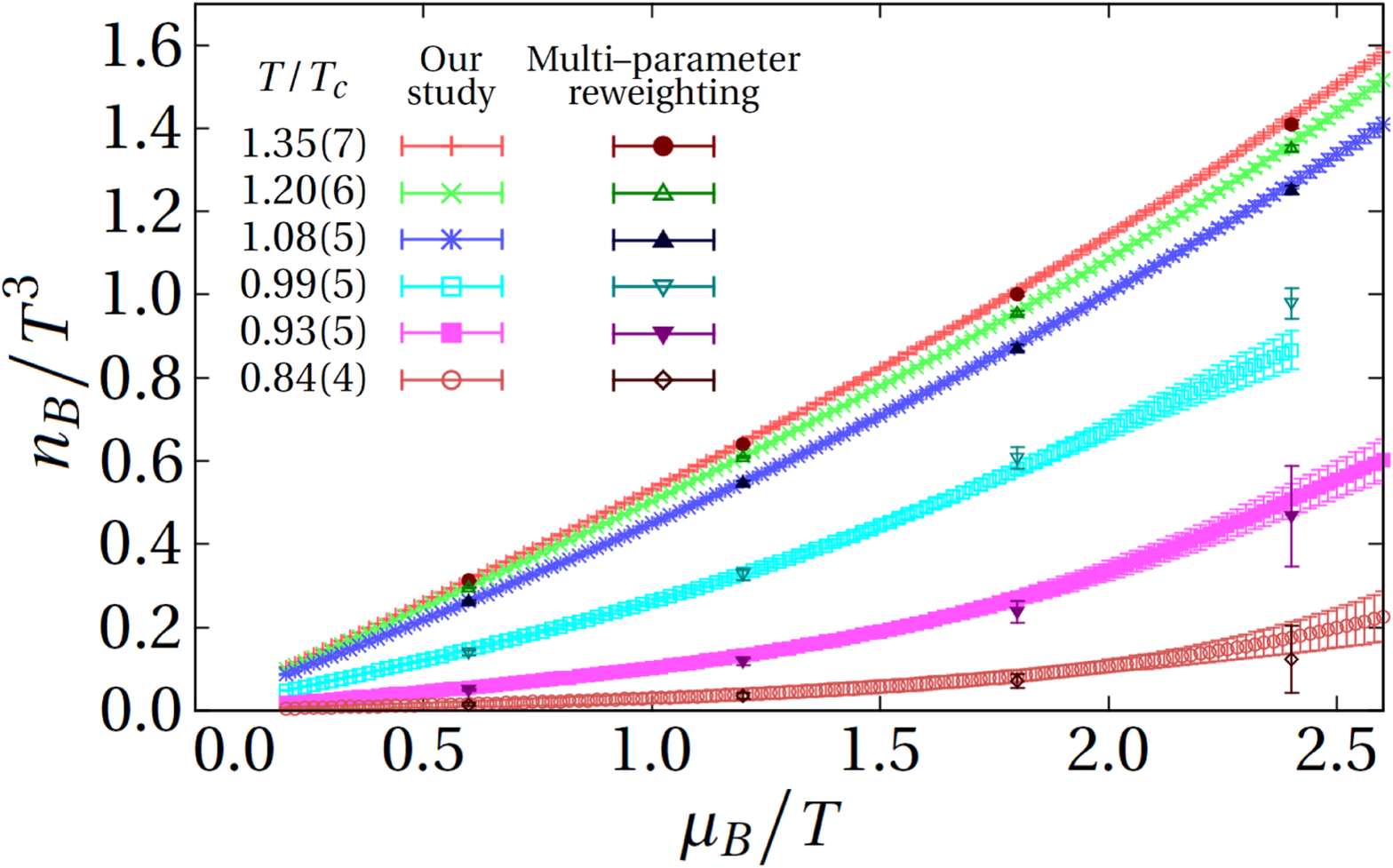}
	\caption{Comparison of the baryon number densities calculated
			by the canonical approach and the MPR method.
			The colors of the data points are the same as in Fig.~\ref{pre-hikaku}.
			This figure was already published in Ref.~\cite{Oka-1}}
	\label{nd-hikaku}
\end{figure}

Next, we consider the expectation value of the baryon number density.
Fig.~\ref{nd-Tdep} shows the chemical potential dependence on baryon number density.
This figure demonstrates that,
for temperatures above and below $T_c$, the results are reliable up to $\mu_B/T = 4$
and $\mu_B/T = 3$--$3.5$, respectively.
However, near $T_c$, the reliable chemical potential range are limited up to $\mu_B/T = 2.4$.
This may be for the same reason described in the pressure analysis.

Figure~\ref{nd-hikaku} represents the comparison of baryon number density between the methods.
It demonstrates good agreement between the results of the canonical approach
and those of MPR method in this case.

Moreover, the gradient of the baryon number density $n_B/T^3$ 
as a function of baryon chemical potential $\mu_B$
becomes small as the temperature decreases.
In a zero temperature case, 
$n_B$ is expected to be zero up to $\mu_B=m_B$ where $m_B$ is the lightest baryon mass of a system,
and becomes a finite value at this point.
This phenomenon is called the silver blaze phenomenon\cite{Gatt-SB}.
The data at $T/T_c=0.84$ in Fig.~\ref{nd-Tdep} does in fact show such a feature.\\

\begin{figure}
	\center
	\includegraphics[width=0.8\hsize, clip]{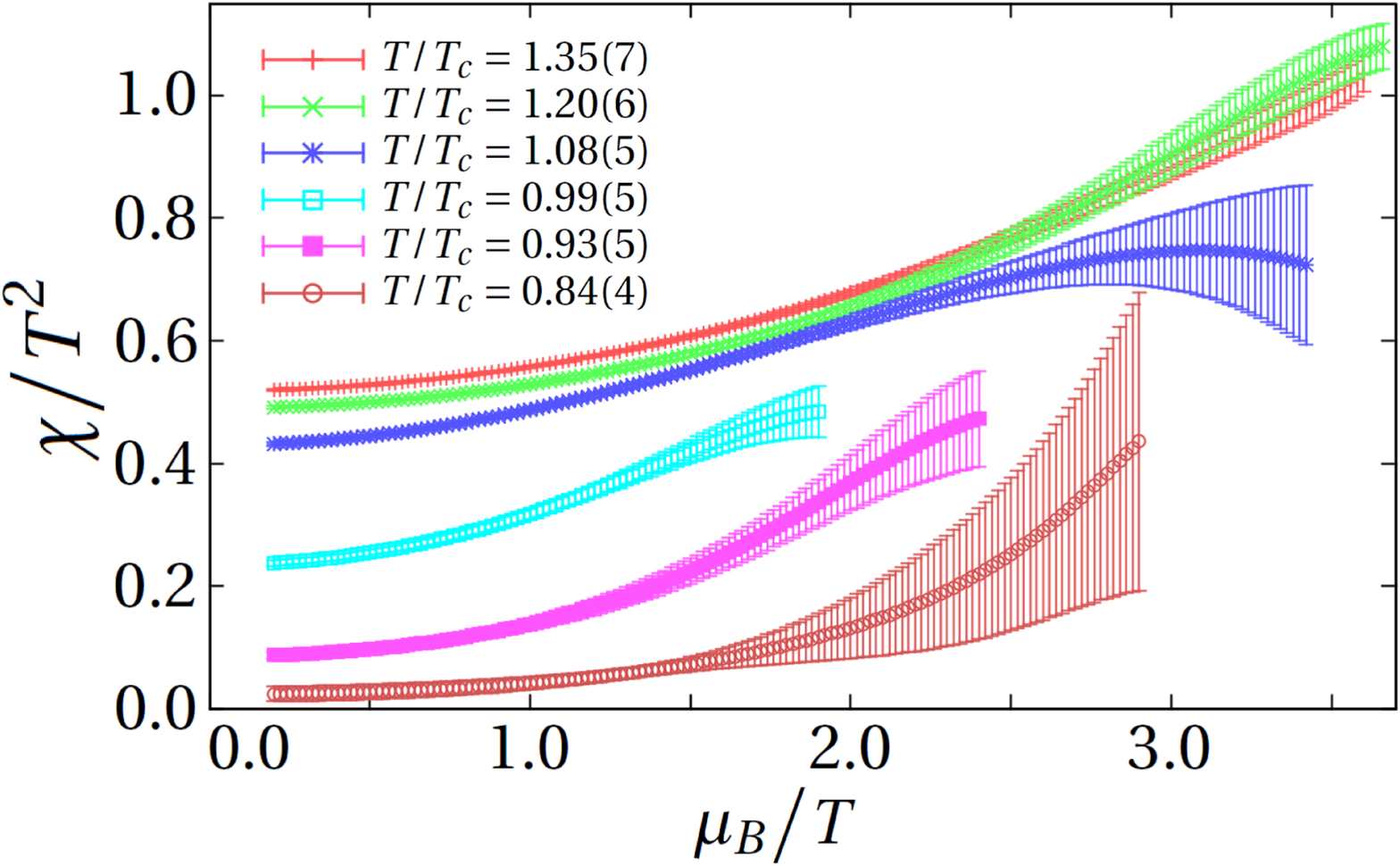}
	\caption{Chemical potential dependence on baryon susceptibility.
			The colors of the data points are the same as in Fig.~\ref{pre-Tdep}. 
			The upper bound of the baryon chemical potential is determined by Eq.~(\ref{bound}).
			This figure was already published in Ref.~\cite{Oka-1}}
	\label{sus-Tdep}
\end{figure}
\begin{figure*}
	\center
	\includegraphics[width=\hsize, clip]{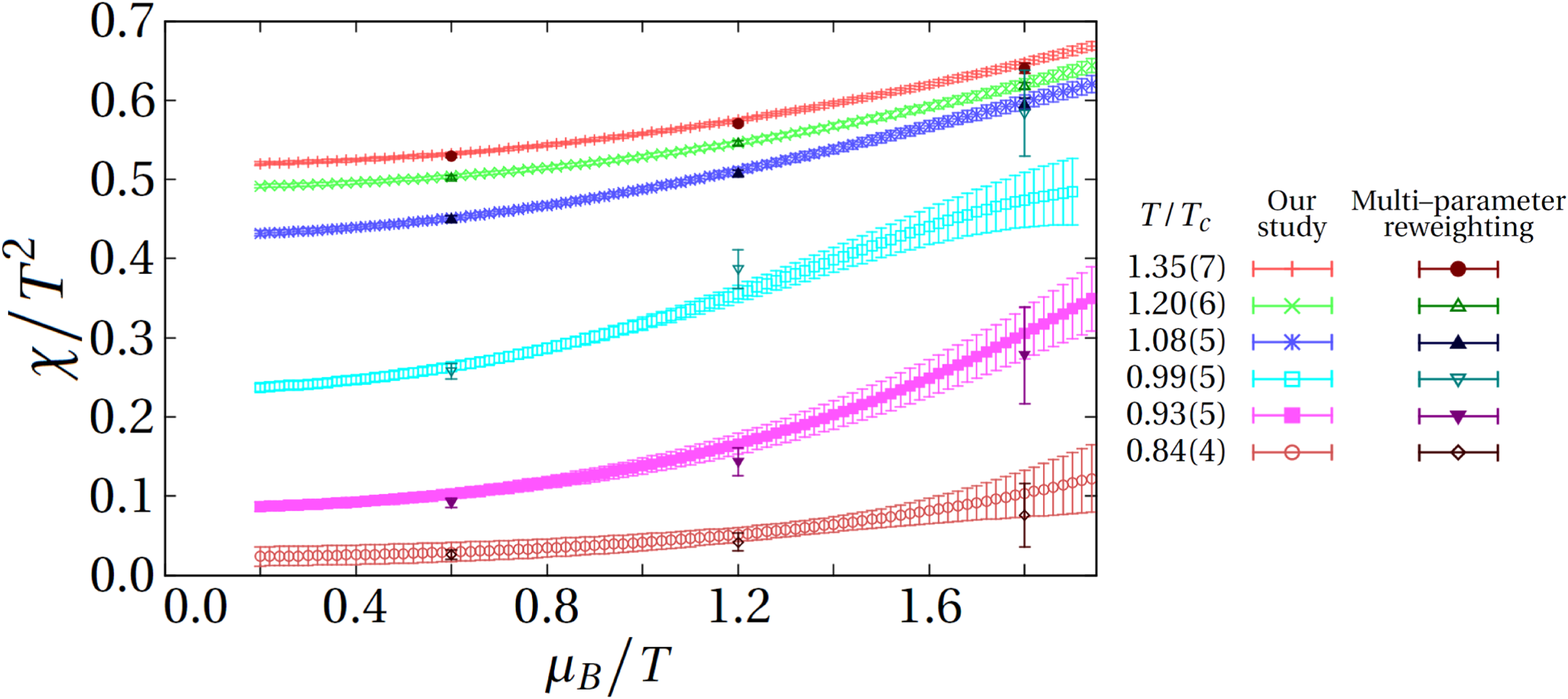}
	\caption{Comparison of baryon susceptibilities calculated
			by the canonical approach and the MPR method. 
			The colors of the data points are the same as in Fig.~\ref{pre-hikaku}.
			This figure was already published in Ref.~\cite{Oka-1}}
	\label{sus-hikaku}
\end{figure*}

Finally, we investigate the baryon susceptibility.
Figure~\ref{sus-Tdep} shows the results of the susceptibility.
From this figure, we find that the results are reliable up to $\mu_B/T = 3.5$ at $T > T_c$,
whereas they are reliable up to $T_c = 2.4$--$2.9$ at $T < T_c$. 

Figure~\ref{sus-hikaku} represents the comparison between the methods in the susceptibility case.
We see that the susceptibility results of canonical approach are in very good agreement
with those of the MPR method.

The baryon susceptibility as a function of $\mu_B/T$ does not show a clear peak.
Thus, signals of the finite density transition between the confinement--deconfinement phases
is not observed yet in this case.

\newpage

\section{Conclusion and discussion}

In this thesis,
we have shown that the canonical approach is consistent with MPR method.
Moreover, the canonical approach provides reliable results 
beyond $\mu_B/T=3$ for almost all observables.
This is very encouraging for the first--principles calculation of
finite density QCD because other methods, such as MPR, Taylor expansion and 
imaginary chemical potential method\cite{imag_chem}, yield reliable information 
in practical situations only up to $\mu_B/T=3$.
Here, the multi precision calculation significantly contributes to this conclusion.

We have investigated the QCD phase structure via the pressure, baryon number density
and baryon number susceptibility.
We calculated these thermodynamic observables in this thesis.
Our results have not shown the QCD phase transition, yet.
However, it is important that we reduced the instability of canonical approach
using the multi precision calculation and calculated the observables
with higher accuracy than previous studies.\\


In order to get more reliable signals of thermodynamic quantities in a large baryon chemical potential region,
we need to calculate the canonical partition functions more accurately at large baryon numbers.
As shown in Fig.~\ref{Zn_drop},
our canonical partition functions currently have unphysical phases at several baryon numbers
although these should have a real positive value in principle.
After the author works, ones have studied
why such phases appear in the canonical partition function $Z_C(n)$\cite{Suzuki-0, Russia-0}.
However, we have not understood it well.

In our opinion, this may be originated from the overlap problem\cite{sign1}.
In this work, we calculate the grand canonical partition functions at pure imaginary chemical potential 
with gauge configurations generated at zero chemical potential through the simplest reweighting method.
However, practically speaking, we have to calculate them with gauge configurations generated 
at each pure imaginary chemical potentials to realize the appropriate importance sampling.
This is a weak point of our work and we need to improve it in future.\\

After the author's study was published, 
the improvement of canonical approach have been applied to the other works.
For example, the author and his collaborators calculated the higher cumulants
with our improvement of canonical approach\cite{Suzuki-1}.
However, the error bars are still large in this study because of the weak point
that we discussed above.

On the other hand, D.L. Boyda, et al. (the lattice QCD group of Vladivostok)
used our improvement and studied how to calculate the canonical partition function
$Z_C(n)$ with more high accuracy\cite{Russia-1, Russia-2}.
Then, they used fitting procedure for the number density at imaginary chemical 
potential and constructed $Z_C(n)$ from it with more high accuracy.

In addition, they calculated the higher cumulants
with their procedure based on our improvement of canonical approach\cite{Russia-3}.
Then, they showed that their results was consistent to experimental results.\\

The author notes that
previous studies of sign problem are the mere resummation techniques
of a series or gauge configurations.
In the reweighting technique, we extract the phase of probability
and resum up gauge configurations with new probability.
In the Taylor expansion method and canonical approach,
we expand the observables using the Taylor, fugacity and winding number expansion.
In addition, recently ones study the Lefschetz thimble to reduce the sign problem.
This is just a resummation technique of gauge configurations.

The author considers that
such methods may include the calculation which causes the cancellation of significant digits.
This is because, in the canonical approach, we can not avoid the cancellation
even though we use and resum the discrete Fourier transformation
or fast Fourier transformation.
Also he consider that
the multi precision calculation is always effectual for such methods.

The author notices that
we can calculate $Z_C(n)$ with high enough accuracy using the reduction formula
even though we do not use the multi precision calculation\footnote{
	Then, we only need to compute the mantissas and exponent part respectively.
	This is because very small eigenvalues appear ($\lambda < 10^{-308}$) in this calculation.
	We do not need to extend the number of significant digits.
}\footnote{
	But, as seen above, the calculation cost of reduction formula is much larger
	than the cost of winding number expansion and Fourier transformation.
}.
This means that, in the reduction formula,
small values result from the multiplication, not from the subtraction between two large values.
From the algorithm of reduction formula, we may study how to reduce
the cancellation of significant digits.\\

The canonical approach has been investigated in the previous studies
\cite{canonical, SNT, Forcrand-1, Forcrand-2, WNE, Gatt, Alex, Li:2010qf, Li:2011ee, Gat1, Gat2}. 
We can also find through our work that the canonical approach is a useful and promising method.
However, our method may be improved further to obtain results under more realistic conditions,
i.e. a lighter quark mass, larger volume, a finer lattice spacing and higher 
density.
Although the hopping parameter expansion yielded very interesting results in this thesis, 
we want to calculate the fermion determinant without the approximation which is from the truncation,
We have learned from this study that the key point is 
how to calculate the determinant at imaginary chemical potential values 
in order to compute the Fourier transformation in Eq.~(\ref{fourier}) with high accuracy.
This requires more computational resources than what has been reported here
but is within the scope of the next--generation high--performance era.


\newpage

\section*{Acknowledgment}



I am deeply indebted to my supervisors,
Prof. Tohru Eguchi and Prof. Hidekazu Tanaka
for insightful comments and beneficial support.
In particular, my laptop which I am using right now was given by Prof. Eguchi.
Also I would like to thank my sub--chief examiners,
Prof. Yu Nakayama and Prof. Jiro Murata for useful advice and comments.

I am greatly thankful to my collaborators,
especially Prof. Atsushi Nakamura and Dr. Ryutaro Fukuda,
for productive discussion, significant support and sincere encouragement.
In particular, my tablet PC which I had used for study had been given by Prof. Nakamura.
I am grateful to Mr. Ryo Sugawara for linguistic support.

I would like to thank everyone who has supported me,
especially those who gave me beer, sake, wine, comics, PCs and money.\\

My calculations were performed
using the supercomputers, SX--9 and SX--ACE at Reserach Center for Nuclear Physics (Osaka Univ.),
PACS--IX and HA--PACS at Center for Computational Sciences (Univ. of Tsukuba),
SR16000 at Yukawa Institute for Theoretical Physics (Kyoto Univ.)
and FX10 at Information Technology Center (Univ. of Tokyo).

My study was partially supported by
the Large Scale Simulation Program in KEK [No.14/15-19 (FY2014)],
Interdisciplinary Computational Science Program in CCS [16a44],
joint research program in JHPCN [EX1631],
HPCI System Research Project [hp170197]
and KAKENHI [26610072, 24340054, 22540265, 15H03663 and 26610072].

\begin{figure}[h]
	\centering
 	\includegraphics[width=\hsize, clip]{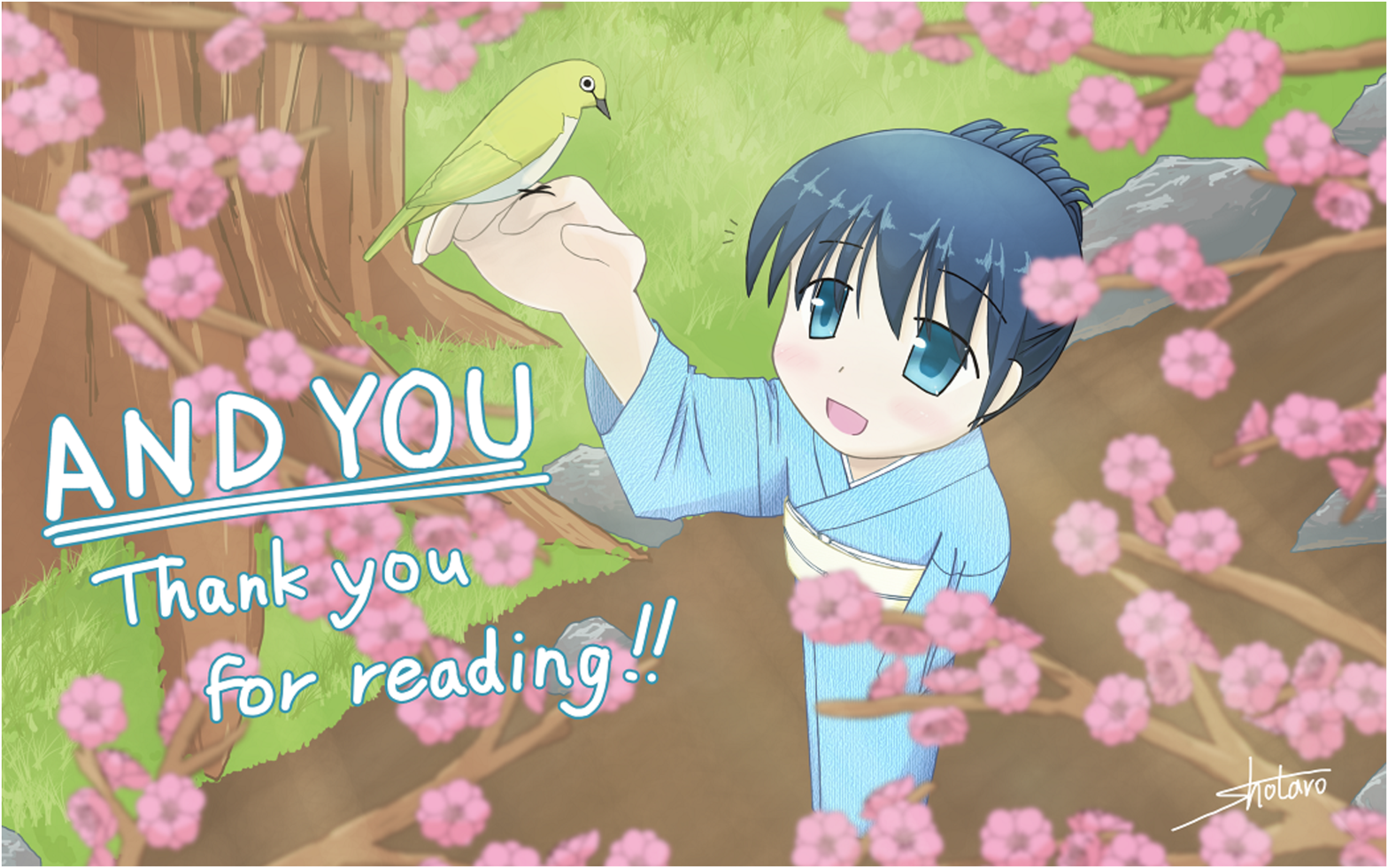}
\end{figure}


\newpage

\appendix

\part*{Appendix}

\section{Grassmann integral of fermion action}

\label{Grassmann_int}

Fermion fields are the Grassmann numbers which satisfy the anticommutation relation,
\begin{align}
	\Psi_i \Psi_j &= -\Psi_j \Psi_i  \  ,  \label{anticommutation}  \\
	\bar{\Psi}_i \bar{\Psi}_j &= -\bar{\Psi}_j \bar{\Psi}_i  \  .  \label{anticommutation_bar}
\end{align}
The Grasmann integral is defined by the following relations,
\begin{align}
	\int d\Psi_i &= \int d\bar{\Psi}_i = 0  \  ,  \label{integral_fermion} \\
	\int d\Psi_i \, \Psi_i &= \int d\bar{\Psi}_i \, \bar{\Psi}_i = 1  \  .  \label{integral_fermion_bar}
\end{align}
Also, we get the following relations from Eq.~(\ref{anticommutation}),
\begin{equation}
	\Psi_i \Psi_i = \bar{\Psi}_i \bar{\Psi}_i = 0  \  .  \label{property_fermion}
\end{equation}
From this equation, we find that the polynomial of Grassmann numbers includes 
the terms of $\Psi_i$ and $\bar{\Psi}_i$ finitely.

We perform the Grassmann integral theoretically using these equation.
For simplification, we consider two flavor fermions $\Psi_i$ ($i=1, 2$)
and their chiral partners $\bar{\Psi}_i$ ($i=1, 2$).
The integral of $\exp(-\bar{\Psi} \mathbf{A} \Psi)$ ($\mathbf{A}$ is a matrix)
is calculated as
\begin{align}
	&\int d\Psi_1 d\bar{\Psi}_1 d\Psi_2 d\bar{\Psi}_2 \, e^{-\sum_{i,j=1}^{2} \bar{\Psi}_i A_{i j} \Psi_j}  \notag  \\
	&\hspace{3em}= \int d\Psi_1 d\bar{\Psi}_1 d\Psi_2 d\bar{\Psi}_2 \, e^{-\bar{\Psi}_1 A_{1 1} \Psi_1} e^{-\bar{\Psi}_1 A_{1 2} \Psi_2} e^{-\bar{\Psi}_2 A_{2 1} \Psi_1} e^{-\bar{\Psi}_2 A_{2 2} \Psi_2}  \notag  \\
	&\hspace{3em}= \int d\Psi_1 d\bar{\Psi}_1 d\Psi_2 d\bar{\Psi}_2 \, \left( 1 - \bar{\Psi}_1 A_{1 1} \Psi_1 \right) \left( 1 - \bar{\Psi}_1 A_{1 2} \Psi_2 \right)  \notag  \\
	&\hspace{3em}\phantom{= d\Psi_1 d\bar{\Psi}_1 d\Psi_2 d\bar{\Psi}_2 \, } \hspace{3em} \times \left( 1 - \bar{\Psi}_2 A_{2 1} \Psi_1 \right) \left( 1 - \bar{\Psi}_2 A_{2 2} \Psi_2 \right)  \notag  \\
	&\hspace{3em}= (-A_{11}) (-A_{22}) + (-A_{21}) (+A_{12}) = \det \mathbf{A}  \  .
\end{align}
We used the Taylor expansion of the exponential in the third line.
Note that the higher powers (e.g. $(\bar{\Psi}_1 A_{1 1} \Psi_1)^2$) do not appear
because of Eq.~(\ref{property_fermion}).
We also use Eq.~(\ref{integral_fermion}) and Eq.~(\ref{integral_fermion_bar})
in the fourth line.

We can generalize this equation to more general formulation.
T. Matthews and A. Salam give us the following formulas\cite{Matthews-Salam},
\begin{align}
	\int D\Psi D\bar{\Psi} \, e^{-\bar{\Psi} \Delta \Psi} &= \det \Delta  \  ,  \\
	\int D\Psi D\bar{\Psi} \, \Psi_i \bar{\Psi}_j e^{-\bar{\Psi} \Delta \Psi} &= -\left( \Delta^{-1} \right)_{i j} \det \Delta  \  ,
\end{align}
where $\Delta$ is the fermion matrix.
These are called Matthews--Salam formula.


\section{$^*$Algorithm of winding number expansion}

\label{alg_WNE}


We can calculate the coefficient $W_k$ of Eq.~(\ref{WNE}) via HPE\cite{Oka-4, Oka-3}.
This is because $\textrm{Tr} Q^m$ is given as follows,
\begin{equation}
	\textrm{Tr} Q^m = \sum_{x, a, \alpha} \left\langle \Psi^{a, \alpha}(x) \left| Q^m \right| \Psi^{a, \alpha}(x) \right\rangle  \  ,
\end{equation}
where $a$ and $\alpha$ denote the Dirac and color index, respectively.
Here, we dissolve $Q$ into the spatial component $Q_i$ ($i=1, 2, 3$),
forward temporal component $Q_4^{(+)}$ and backward temporal component $Q_4^{(-)}$,
\begin{equation}
	Q^m \left| \Psi^{a, \alpha}(x) \right\rangle = \left( \sum_{i=1}^{3} Q_i + e^{+\mu a} Q_4^{(+)} + e^{-\mu a} Q_4^{(-)} \right)^m \left| \Psi^{a, \alpha}(x) \right\rangle  \  .
\end{equation}
Note that $Q_i$ includes the clover term.
Then, we can classify this with chemical potential dependence.
That is,
\begin{align}
	Q^m \left| \Psi^{a, \alpha}(x) \right\rangle &= \sum_{k=-m}^{+m} \left( e^{+\mu a} \right)^k Q^{(k)}(m) \left| \Psi^{a, \alpha}(x) \right\rangle  \notag  \\
		&=: \sum_{k=-m}^{+m} \left( e^{+\mu a} \right)^k \vec{X}^{(k)}(m; x, a, \alpha)  \  ,
\end{align}
where $Q^{(k)}(m)$ is the product of $Q_i$ and $Q_4^{\pm}$.

Thus, we can write $\textrm{Tr} Q^m$ as
\begin{equation}
	\textrm{Tr} Q^m = \sum_{x, a, \alpha} \left\langle \vec{\Psi}(x, a, \alpha), \sum_{k=-m}^{+m} \left( e^{+\mu a} \right)^k \vec{X}^{(k)}(m; x, a, \alpha) \right\rangle  \  ,
\end{equation}
where $\vec{\Psi} := \left| \Psi \right\rangle$ and
$\langle \cdot , \cdot \rangle$ is the inner product.

Here, note that $k$ denotes the time slice because it is the power of $e^{\mu a}$.
Then, we can neglect the variable $t$ of the vectors,
therefore we can reduce the time complexity of the multiplication $Q \vec{\Psi}$
from $O \left( (N_x N_y N_z N_t \times 3 \times 4)^2 \right)$
to $O \left( (N_x N_y N_z \times 3 \times 4)^2 \right)$.
In the practical computing, we do not perform the summation of $m$.
We thus save $\left\langle \vec{\Psi}, (e^{\mu a})^k \vec{X}^{(k)} \right\rangle =: Z^{(k)}$
as an array for $k$.
After we perform the summation of $m$,
we extract the coefficient of $(e^{\mu/T})^k = (e^{\mu a})^{k N_t}$.
The coefficient is just $W_k$.\\

Since the algorithm is intricate, we show it as a pseudo code herein.
Note that the outside loops of $(x,y,z,t,c,\mu)$ for the trace
are performed by the noise method which is discussed below.
\begin{algorithm}
	\caption{Algorithm of WNE via HPE}
	\label{algorithm_WNE}
	\begin{algorithmic}
		\STATE{Initialize: $x,y,z,t,c,\mu,k,m \in \mathbf{Z}$}	
		\STATE{Initialize: $X^{(k)}(x,y,z,c,\mu), Y^{(k)}(x,y,z,c,\mu) \in \mathbf{C}$}
		\STATE{Initialize: $Z^{(k)} \in \mathbf{C}$}
		\STATE{Initialize: $S(x,y,z,c,\mu), T(x,y,z,c,\mu) \in \mathbf{C}$}
		\STATE{Initialize: $W(k) \in \mathbf{C}$}
		\FOR{$(x,y,z,t,c,\mu) = (1,1,1,1,1,1)$ \TO $(N_x,N_y,N_z,N_t,3,4)$}
			\FOR{$k = -k_\textrm{max}$ \TO $+k_\textrm{max}$}
				\STATE{$Z^{(k)} = 0$}
				\STATE{$\vec{X}^{(k)} = \vec{0}$}
			\ENDFOR
			\STATE{$X^{(0)}(x,y,z,c,\mu)$ = 1}
			\FOR{$m = 1$ \TO $m_\textrm{max}$}
				\STATE{$k_\textrm{range} = m$}
				\FOR{$k = -k_\textrm{range}$ \TO $+k_\textrm{range}$}
					\STATE{$\vec{S} = \kappa * (Q_1(t) + Q_2(t) + Q_3(t)) * \vec{X}^{(k)}$}
					\STATE{$\vec{T} = \kappa * (Q_4^{(+)}(t) * \vec{X}^{(k-1)} + Q_4^{(-)}(t) * \vec{X}^{(k+1)})$}
					\STATE{$\vec{Y}^{(k)} = \vec{S} + \vec{T}$}
				\ENDFOR
				\FOR{$k = -k_\textrm{range}$ \TO $+k_\textrm{range}$}
					\STATE{$Z^{(k)} = Z^{(k)} + Y^{(k)}(x,y,z,c,\mu)/m$}
					\STATE{$\vec{X}^{(k)} = \vec{Y}^{(k)}$}
				\ENDFOR
			\ENDFOR
			\FOR{$k = -(k_\textrm{range}-1)$ \TO $+(k_\textrm{range}-1)$}
				\IF{$\textrm{mod}(k,N_t) \not= 0$}
					\STATE{$W(k/N_t) = -Z^{(k)}$}
				\ENDIF
			\ENDFOR
		\ENDFOR
	\end{algorithmic}
\end{algorithm}

\newpage

\section{Noise method}

\label{alg_noise}

The trace of matrix is performed by the loops of summation.
Naively, we perform the loops $N_x N_y N_z N_t \times 3 \times 4$ times.
We can reduce this calculation using the noise method\cite{noise}.

In this method, we start from Eq.~(\ref{QCD_Z}), that is,
\begin{equation}
	\textrm{Tr} \mathbf{A} = \sum_{\phi} \left\langle \phi \left| \mathbf{A} \right| \phi \right\rangle  \  ,
\end{equation}
where $\mathbf{A}$ is a matrix with the rank of $N_\textrm{rank}$.
Then, $\left| \phi \right\rangle$ is a vector.
Here, we use the notation $\vec{\phi} := \left| \phi \right\rangle$ as above section;
we then write as
\begin{equation}
	\textrm{Tr} \mathbf{A} = \sum_{\phi} \left\langle \vec{\phi}, \mathbf{A} \vec{\phi} \right\rangle  \  ,
\end{equation}
where $\langle \cdot, \cdot \rangle$ is the inner product.

We can calculate it choosing $\phi$ as the vector of noises\cite{noise},
\begin{equation}
	\textrm{Tr} \mathbf{A} = \lim_{N \rightarrow \infty} \frac{1}{N} \sum_{i=1}^{N} \left\langle \vec{\eta}^{(i)}, \mathbf{A} \vec{\eta}^{(i)} \right\rangle  \  ,
\end{equation}
where $\vec{\eta}^{(i)}$ is the $i$--th noise vector.
The noise vector satisfies the following property,
\begin{equation}
	\frac{1}{N_\textrm{rank}} \left\langle \vec{\eta}^{(i)}, \vec{\eta}^{(j)} \right\rangle = \delta_{i j}  \  .
\end{equation}
This method corresponds to the Monte Carlo method for each component of $\vec{\phi}$.\\

\begin{figure}[h]
	\centering
 	\includegraphics[width=0.7\hsize, clip]{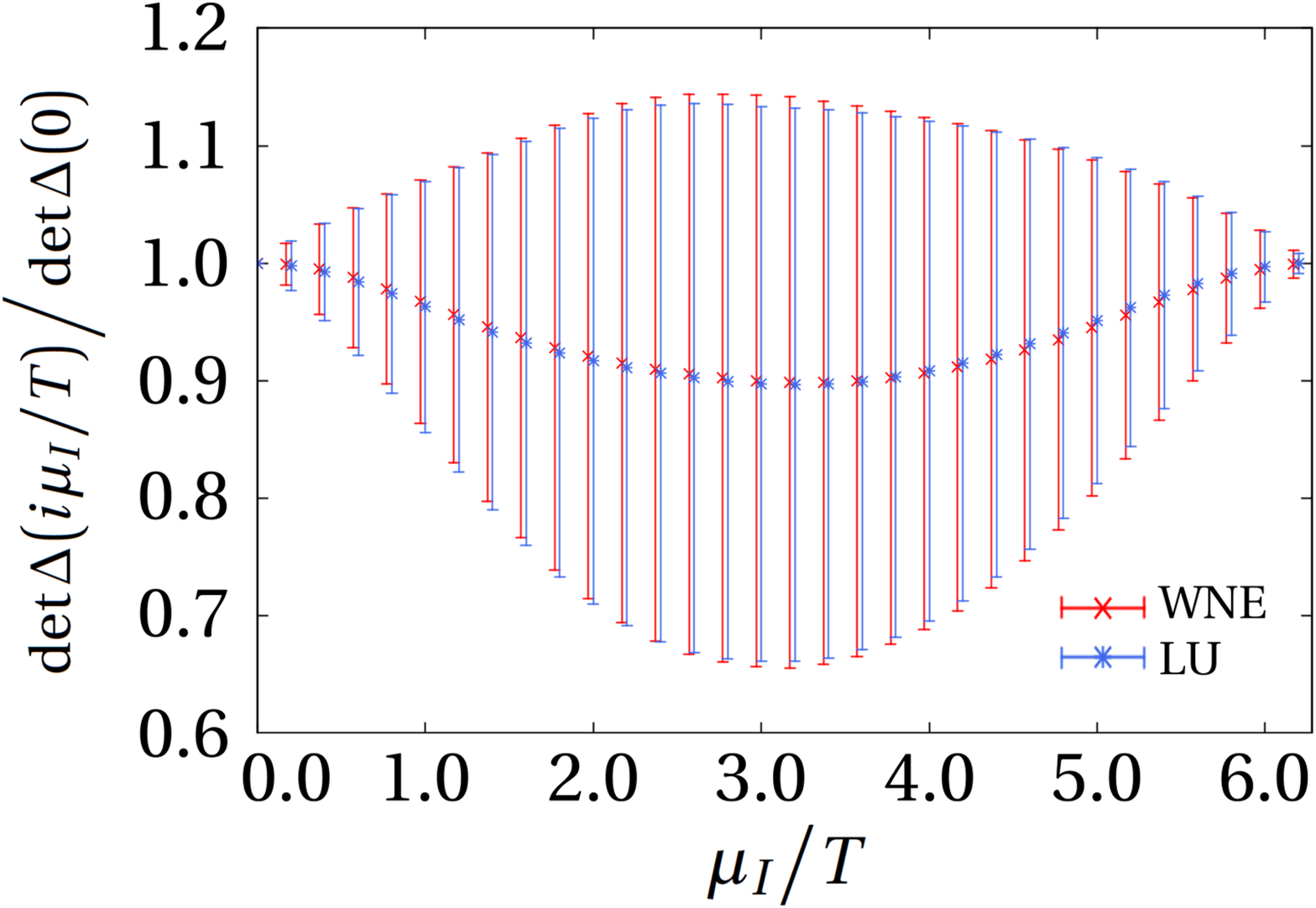}
	\caption{Pure imaginary chemical potential
		dependence on fermion determinants. Red and blue points
		are calculated by the winding number expansion with 16 noise
		vectors and LU decomposition respectively.}
	\label{noise_result}
\end{figure}

J. Foley, et al. showed that we can approximate the trace by 16 noise vectors
with high enough accuracy\cite{noise}.
In this thesis, for safety, we use 64 noise vectors for high temperature $T > T_c$ simulations and
128 noise vector for low temperature $T < T_c$ simulations.
An example calculation is shown as follows. See Fig.~\ref{noise_result}.
This figure shows that the noise method can produce
consistent results with these obtained by LU decomposition
within a range of statistical errors.
We thus use the noise method for the trace of winding number expansion.\\

Note that we can reduce the computation cost of WNE using this method.
This is because the times of loops are reduced as
$N_x N_y N_z N_t \times 3 \times 4 = 24576 \rightarrow N_\textrm{noise} = 128$
when the lattice size is $8^3 \times 4$.
Then, the time complexity of WNE with noise trace is
$O \left( (N_x N_y N_z)^2 \times (N_t)^2 \right)$.

\vspace{2ex}

\section{$^*$Direct method via winding number expansion without canonical ensembles}

\label{direct}


When we calculate thermodynamic observables using the canonical approach with WNE,
we calculate the following quantities in following order,
\begin{equation}
	W_k \rightarrow Z(i \mu_I) \rightarrow Z_C(n) \rightarrow Z(\mu)  \  .
\end{equation}
Then we use WNE, Fourier transform and fugacity expansion.
We can avoid the sign problem using above method.

\begin{figure}[h]
	\centering
 	\includegraphics[width=0.75\hsize, clip]{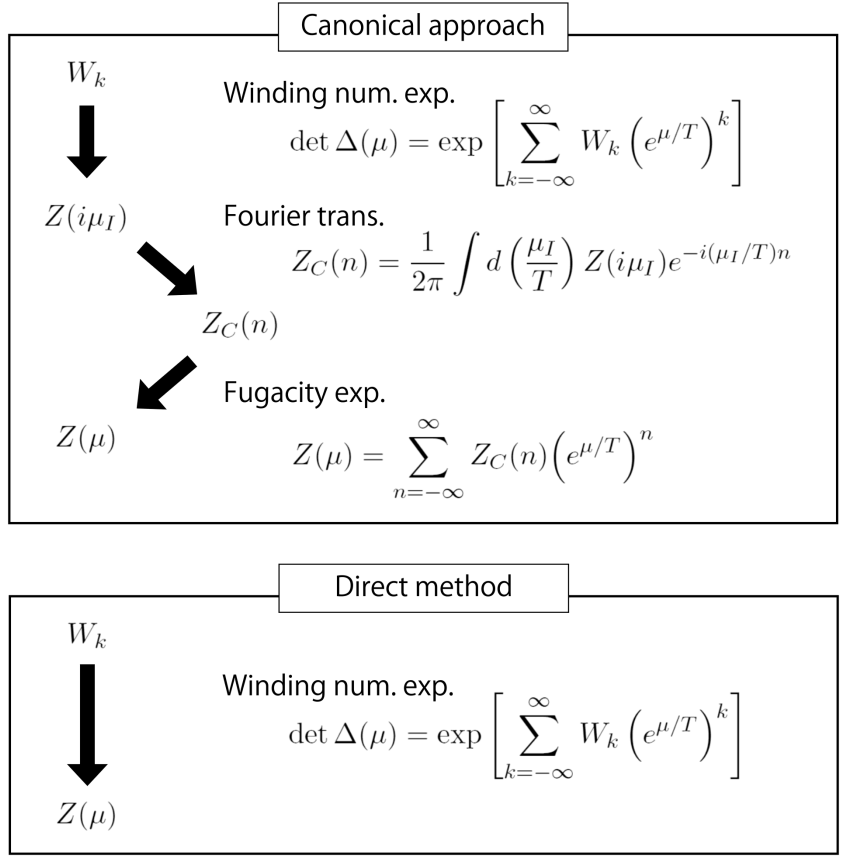}
	\caption{Difference of canonical approach and direct method.}
	\label{canonical_vs_direct}
\end{figure}

On the other hand, we notice that we can calculate $Z(\mu)$ from $W_k$ directly,
\begin{equation}
	W_k \rightarrow Z(\mu)  \  ,
\end{equation}
because we can use the WNE of fermion determinant at any $\mu$ (not only $i \mu_I$).
(See Fig.~\ref{canonical_vs_direct}.)
In this thesis, we call it the direct method.\\

It may seem that the direct method works out well without sign problem.
However, this method differs from the standard canonical approach in the following points.
\begin{itemize}
	\item In the direct method, we calculate $Z(\mu)$ without the canonical ensembles
			unlike the canonical approach. 
	\item In the direct method, we perform the finite density calculation,
			\begin{equation}
				\det \Delta(\mu) = \exp \left[ \sum_k W_k (e^{\mu/T})^k \right] \  ,
			\end{equation}
			\textit{before} we calculate the integral of gauge field.
			On the other hand, in the canonical approach,
			we perform the finite density calculation
			$Z(\mu) = \sum_{n} Z_C(n) (e^{\mu/T})^n$
			\textit{after} we calculate the integral.
\end{itemize}
From the second point,
we expect that the direct method does not work well
because the sign problem occurs when we calculate
$Z(\mu) = \int DU $ $ \left[ \det \Delta(\mu) \right]^{N_f} e^{S_G}$
using complex $\det \Delta(\mu)$.

\begin{figure}
	\centering
 	\includegraphics[width=0.75\hsize, clip]{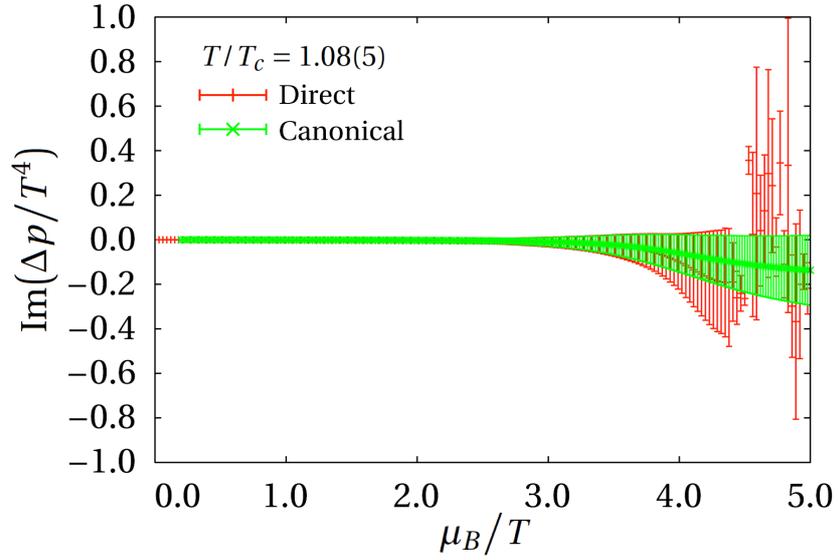}
	\caption{Imaginary part of pressure by canonical approach and direct method.
		The red and green points are the data of direct method and canonical approach
		at $T/T_c = 1.08$, respectively.}
	\label{canonical_vs_direct_pre}
\end{figure}

\begin{figure}
	\centering
 	\includegraphics[width=0.75\hsize, clip]{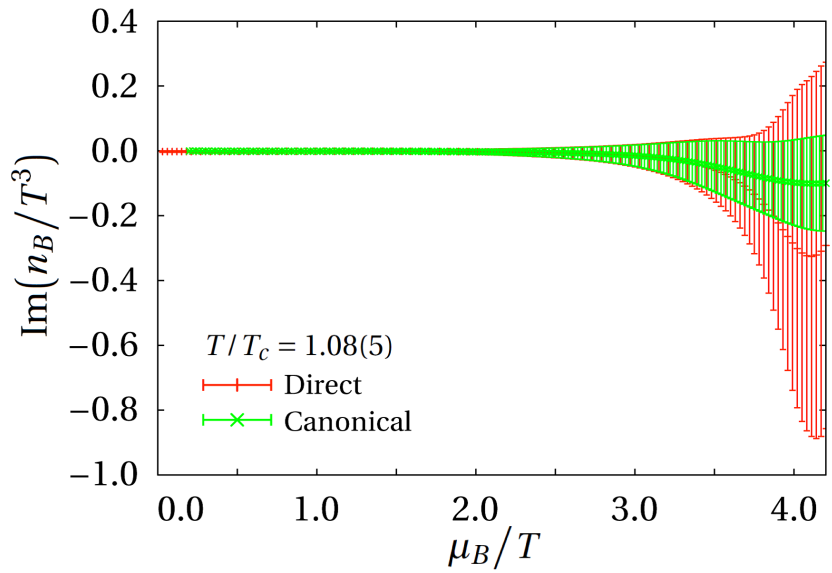}
	\caption{Imaginary part of baryon number density by canonical approach and direct method.
		The red and green points are the data of direct method and canonical approach
		at $T/T_c = 1.08$, respectively.}
	\label{canonical_vs_direct_num}
\end{figure}

\begin{figure}
	\centering
 	\includegraphics[width=0.75\hsize, clip]{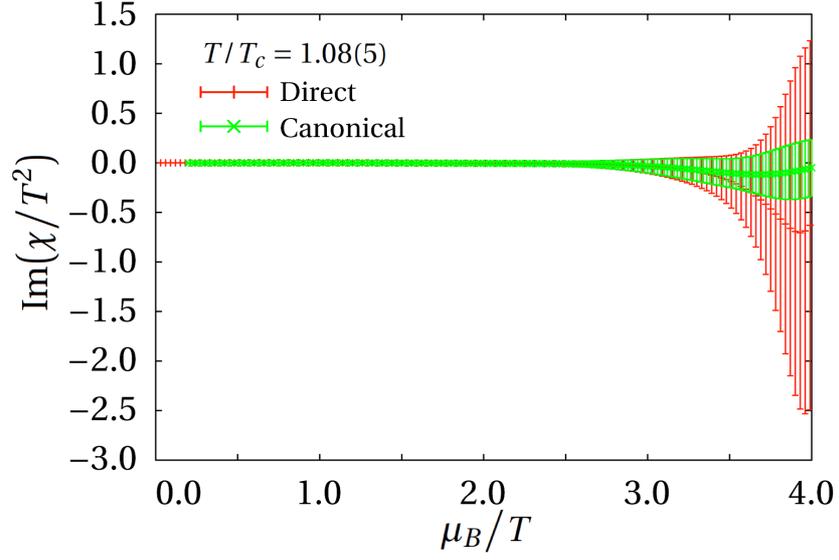}
	\caption{Imaginary part of baryon susceptibility by canonical approach and direct method.
		The red and green points are the data of direct method and canonical approach
		at $T/T_c = 1.08$, respectively.}
	\label{canonical_vs_direct_sus}
\end{figure}

Actually, the canonical approach is more accurate than the direct method.
For example, although the pressure is real,
sometimes it has the imaginary part since there are numerical errors.
Figure~\ref{canonical_vs_direct_pre} shows the imaginary part of pressure
which we calculate using the canonical approach and direct method.
From this figure, we see that the results of canonical approach and direct method
are consistent to zero.
However, the statistical error of canonical approach is smaller than
the one of direct method.
We can see this property from other observables.
For instance, Fig.~\ref{canonical_vs_direct_num} and \ref{canonical_vs_direct_sus}
show the imaginary part of baryon number density and baryon susceptibility.
We find that the error of canonical approach is always smaller than the one of direct method.

\begin{figure}
	\centering
 	\includegraphics[width=0.75\hsize, clip]{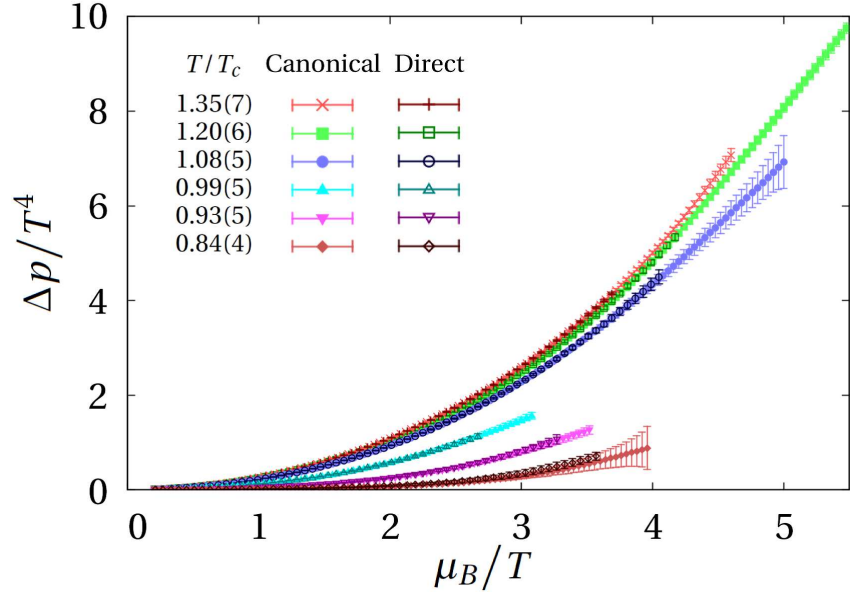}
	\caption{Baryon chemical potential dependence on pressure
		which is calculated by canonical approach and direct method
		at $T/T_c = 1.35$, $1.20$, $1.08$, $0.99$, $0.93$ and $0.84$.}
	\label{direct_pre}
\end{figure}
\begin{figure}
	\centering
 	\includegraphics[width=0.75\hsize, clip]{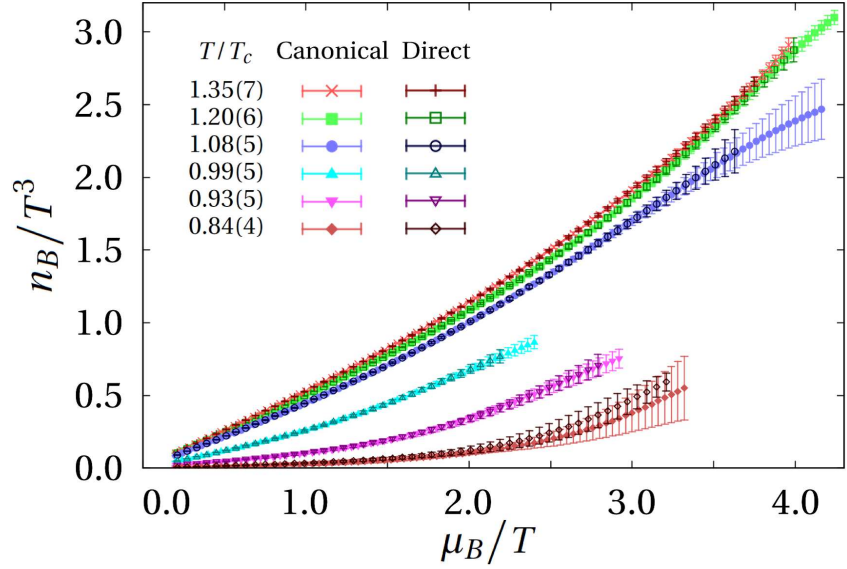}
	\caption{Baryon chemical potential dependence on baryon number density
		which is calculated by canonical approach and direct method
		at $T/T_c = 1.35$, $1.20$, $1.08$, $0.99$, $0.93$ and $0.84$.}
	\label{direct_num}
\end{figure}
\begin{figure}
	\centering
 	\includegraphics[width=0.75\hsize, clip]{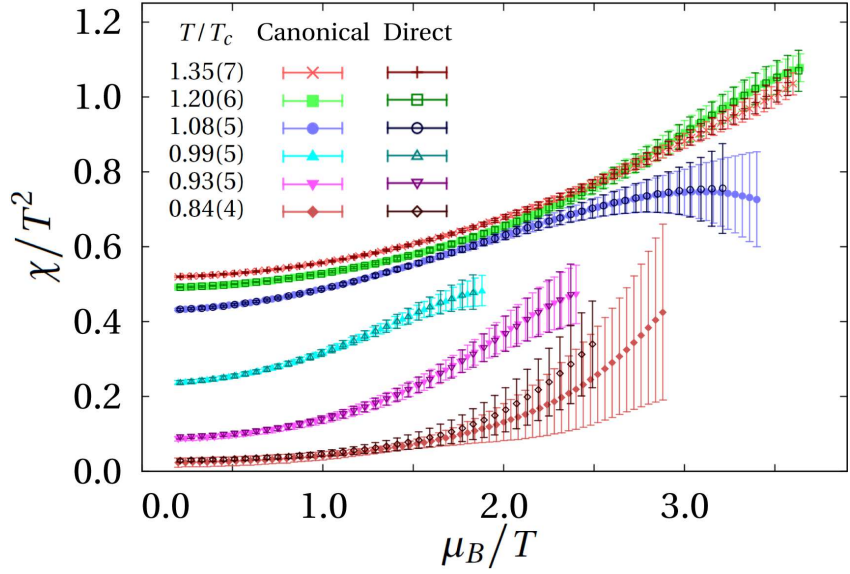}
	\caption{Baryon chemical potential dependence on baryon susceptibility
		which is calculated by canonical approach and direct method
		at $T/T_c = 1.35$, $1.20$, $1.08$, $0.99$, $0.93$ and $0.84$.}
	\label{direct_sus}
\end{figure}

Figure~\ref{direct_pre}, \ref{direct_num} and \ref{direct_sus} show the thermodynamic observables
which we calculate using the canonical approach and direct method.
Then, the validity range of canonical approach is estimated as Sec.~\ref{error_def}.

The range of direct method is estimated
by the magnitude of statistical error of imaginary part.
The procedure is as follows.

\newpage

\begin{enumerate}
	\item Save the error of imaginary part of observables of canonical approach
		at the boundary of validity range, e.g.
		$\delta \left( \textrm{Im} \left[ \Delta p(\mu_\textrm{val(can)}) / T^4 \right]_\textrm{can} \right)$.
	\item Compare the one of direct method
		$\delta \left( \textrm{Im} \left[ \Delta p(\mu) / T^4 \right]_\textrm{dir} \right)$ with\\
		$\delta \left( \textrm{Im} \left[ \Delta p(\mu_\textrm{val(can)}) / T^4 \right]_\textrm{can} \right)$.
	\item Find $\mu_1$ which satisfies,
		\begin{equation}
			\delta \left( \textrm{Im} \left[ \Delta p(\mu_1) / T^4 \right]_\textrm{dir} \right)
			= \delta \left( \textrm{Im} \left[ \Delta p(\mu_\textrm{val(can)}) / T^4 \right]_\textrm{can} \right)  \  .
		\end{equation}
	\item Consider the boundary of validity range of direct method is $\mu_\textrm{val(dir)} = \mu_1$.
\end{enumerate}
That is, we plot the data of canonical approach and direct method with the same accuracy.

From these figure, we find that the validity range of canonical approach
is larger than the one of direct method.
Especially, in the pressure data, canonical approach is valid beyond $\mu_B/T=5$ at $T/T_c=1.20$,
although direct method is valid to $\mu_B/T=4$.

Therefore, when we calculate the thermodynamic observables at finite density,
it is better that we use the canonical approach in order to avoid the sign problem.

\section{$^*$Algorithm for estimation of cancellation of significant digits}

\label{alg_cancel}

In this section, we will see how to estimate the cancellation of significant digits.
The cancellation is caused by the subtraction between two similar numbers.

The logarithm of numbers give us the magnitude of the numbers.
For example, the logarithm of $a = 6.022 \times 10^{23}$ is $\log_{10} a = 23.78$.
Then, ``$23$" corresponds to the power of $a$,
and ``$0.78$" corresponds to the significant digits of $a$.
This is because we can dissolve $\log_{10} a$ as
\begin{equation}
	\log_{10} a = \log_{10} (6.022 \times 10^{23}) = \log_{10} (6.022) + \log_{10} (10^{23})  \  .
\end{equation}
Using this idea, we can estimate how many digits are the same between two numbers.

Let $a$ and $b$ be real numbers.
For simplicity, we consider that $a$ and $b$ are positive and $a > b$.
Here, we define the following quantity,
\begin{align}
	r &:= a - b  \  ,  \\
 	d &:= \log_{10} (a) - \log_{10} (r)  \  .
\end{align}
where ``$r$" means the result and ``$d$" means the difference between the magnitudes of $a$ and $r$.
If $a$ and $b$ are
\begin{align}
	a &= 6.022 \times 10^{23}  \  ,  \\
	b &= 6.021 \times 10^{23}  \  ,
\end{align}
then $r$ and $d$ become the following numbers,
\begin{align}
	r &= \log_{10} (0.001 \times 10^{23}) = 20.00  \  ,  \\
 	d &= 23.78 - 20.00 = 3.78  \  .  \label{drop}
\end{align}

In this case, when we calculate $r = a - b$, three significant digits are lost.
This ``three" appears in Eq.~(\ref{drop}).
That is, the number of canceled digits are given as
\begin{equation}
	\left( \textrm{Number of canceled digits} \right) = \lfloor d \rfloor = \left\lfloor \log_{10} (a) - \log_{10} (r) \right\rfloor  \  ,
\end{equation}
where $\lfloor \cdot \rfloor$ is the floor function.

Here, what does ``$0.78$" of Eq.~(\ref{drop}) mean?
This is the difference between the magnitudes of significant digits.
Let $a_\textrm{sig}$ and $a_\textrm{pow}$ be the significant digits and power of $a$, respectively.
We then can write $d$ as
\begin{align}
	d &= \log_{10} (a) - \log_{10} (r) = \log_{10} (a_\textrm{sig} \times 10^{a_\textrm{pow}}) - \log_{10} (r_\textrm{sig} \times 10^{r_\textrm{pow}})  \notag  \\
		&= \log_{10} \left( \frac{a_\textrm{sig}}{r_\textrm{sig}} \right) + \log_{10} \left( \frac{10^{a_\textrm{pow}}}{10^{r_\textrm{pow}}} \right)  \notag  \\
		&= \log_{10} \left( \frac{a_\textrm{sig}}{r_\textrm{sig}} \right) + \left( a_\textrm{pow}-r_\textrm{pow} \right)  \  ,  \\
	\lfloor d \rfloor &= a_\textrm{pow}-r_\textrm{pow}  \  .
\end{align}
Thus, ``$0.78$" is $\log_{10} \left( a_\textrm{sig}/r_\textrm{sig} \right)$.
(In fact, we can compute it as $\log_{10} (6.022/$ $1.0) = 0.78$.)
Note that $1 \leq a_\textrm{sig}/r_\textrm{sig} < 10$.
This is because $a_\textrm{sig}$ and $r_\textrm{sig}$ satisfy
$1 \leq a_\textrm{sig} < b$ and $1 \leq r_\textrm{sig} < b$ where $b$ is the base of logarithm.
($b=10$; See also Sec.~\ref{multipre_def}.)\\

The Algorithm for estimation of the number of canceled digits is as follows.
Note that we now adopt the multi precision calculation.
\begin{enumerate}
	\item Preparation: adopt the multi precision calculation to the program.

	\item Initialization:
		\begin{enumerate}
			\item Set the number of significant digits is large enough.
			\item Choose $a$ as the minuend and $b$ as subtrahend.
				They are multi precision variables.
		\end{enumerate}

	\item Estimation:
		\begin{enumerate}
			\item Save $\log_{10} (a)$.
			\item Calculate the subtraction $r = a - b$ which may cause the cancellation of significant digits.
			\item Calculate $d = \log_{10} (a) - \log_{10} (r)$.
			\item Print $\lfloor d \rfloor$ which is the number of canceled digits.
		\end{enumerate}
\end{enumerate}




\newpage

\pagestyle{empty}
\newgeometry{top=2.7cm, bottom=0.5cm, left=0.5cm, right=0.5cm}


\newpage

\hspace{1.6cm}{\huge \textbf{Extra}}

\vspace{1em}

\hspace{1.6cm}In this section, there are some illustrations which are drawn by the author.

\vspace{11cm}

\begin{flushright}
 	\includegraphics[width=\hsize, clip]{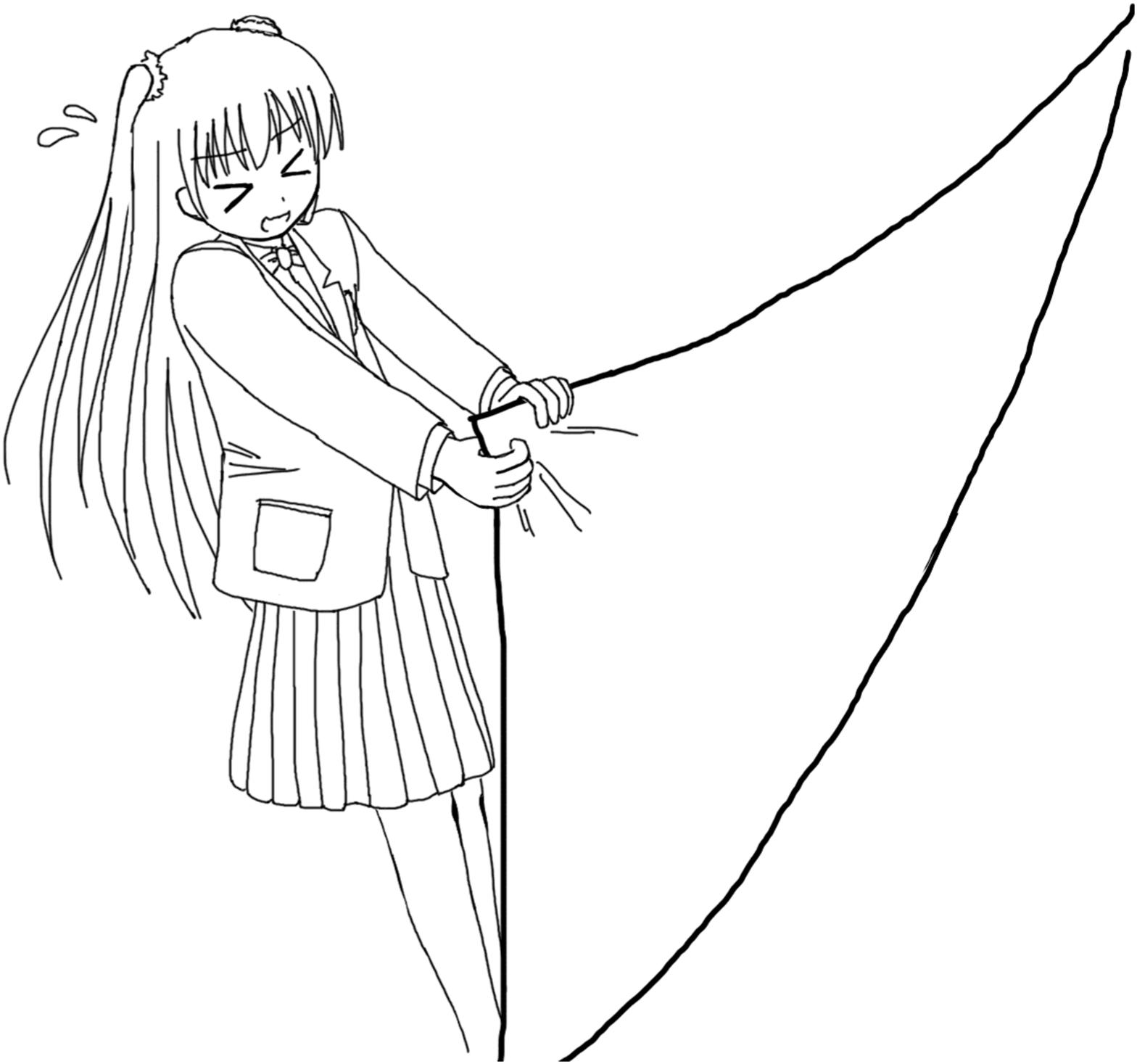}
\end{flushright}

\newpage

\begin{figure}
	\centering
 	\includegraphics[height=0.8\vsize, clip]{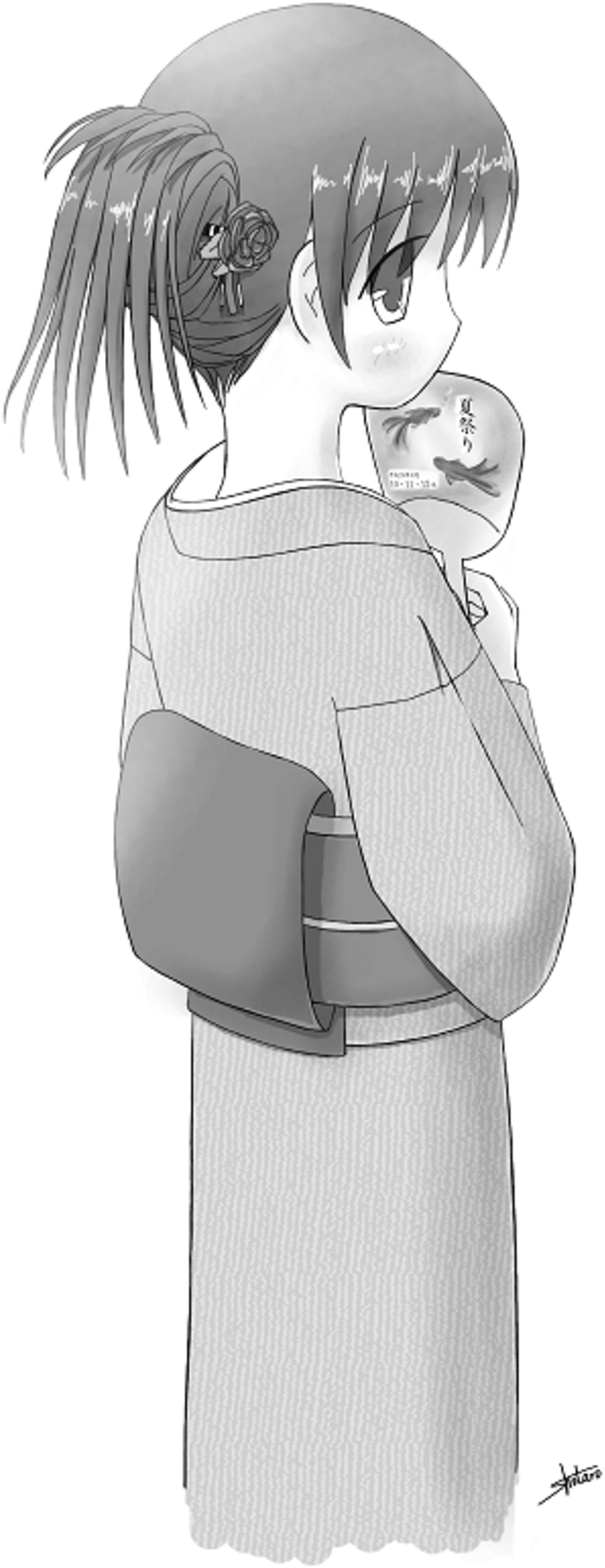}
	{\phantom{.}

	Figure EX1: Summer festival in Japan (Gray scale image)}
\end{figure}

\newpage

\begin{figure}
	\centering
 	\includegraphics[height=0.8\vsize, clip]{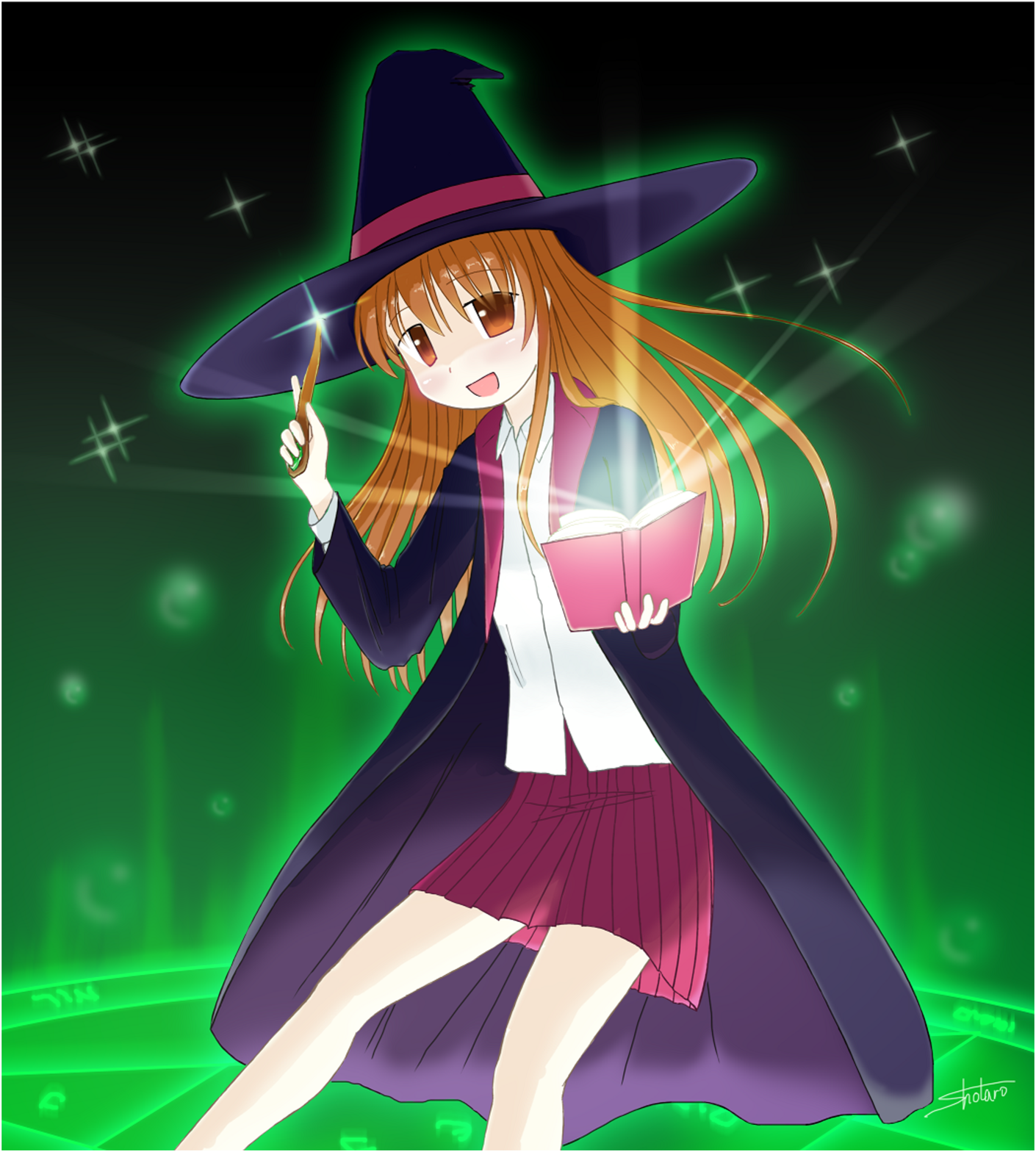}
	{\phantom{.}

	Figure EX2: Happy Halloween!}
\end{figure}

\newpage

\begin{figure}
	\centering
 	\includegraphics[width=0.99\hsize, clip]{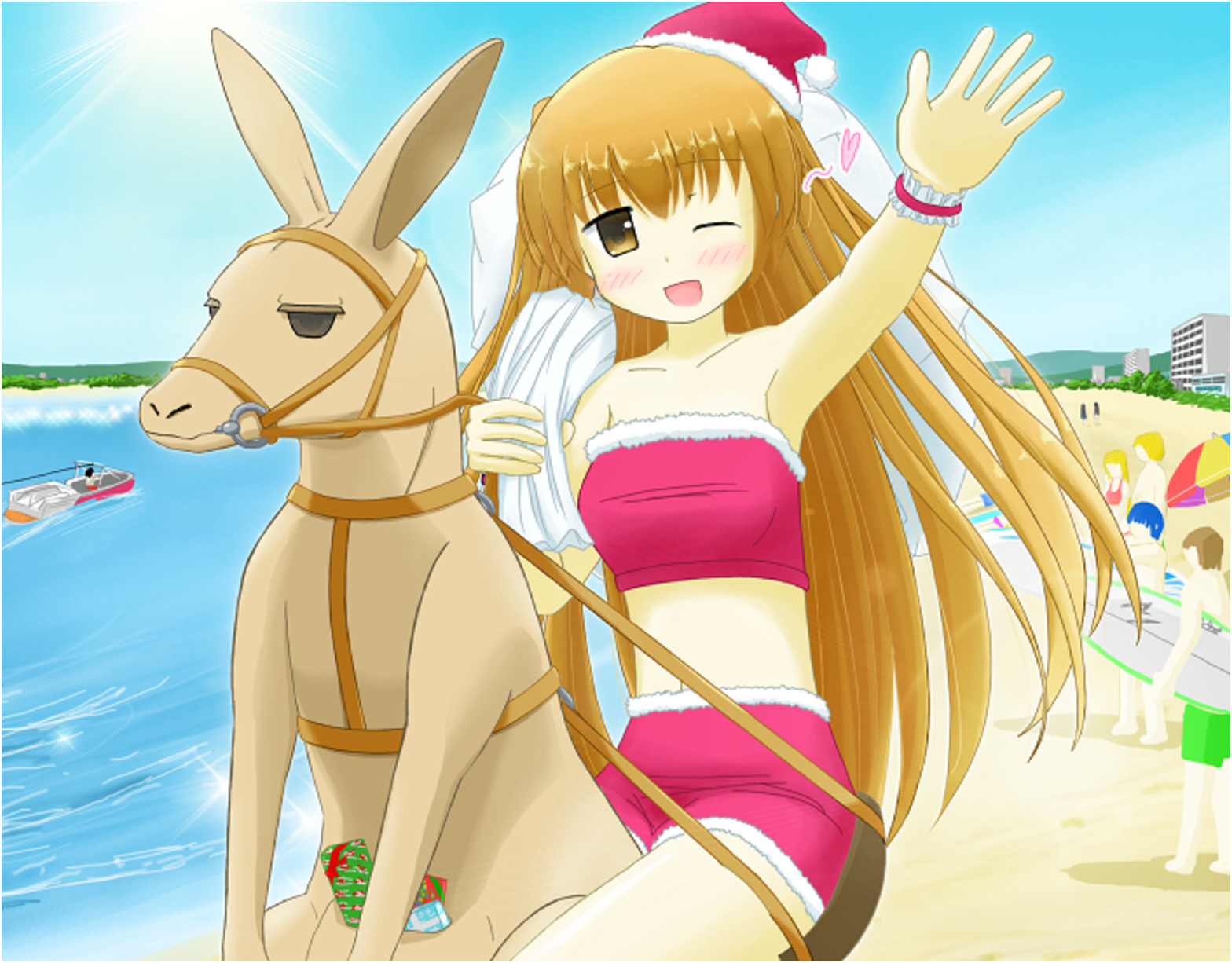}
	{\phantom{.}

	Figure EX3: Christmas in Australia}
\end{figure}

\newpage

\begin{figure}
	\centering
 	\includegraphics[width=0.90\hsize, clip]{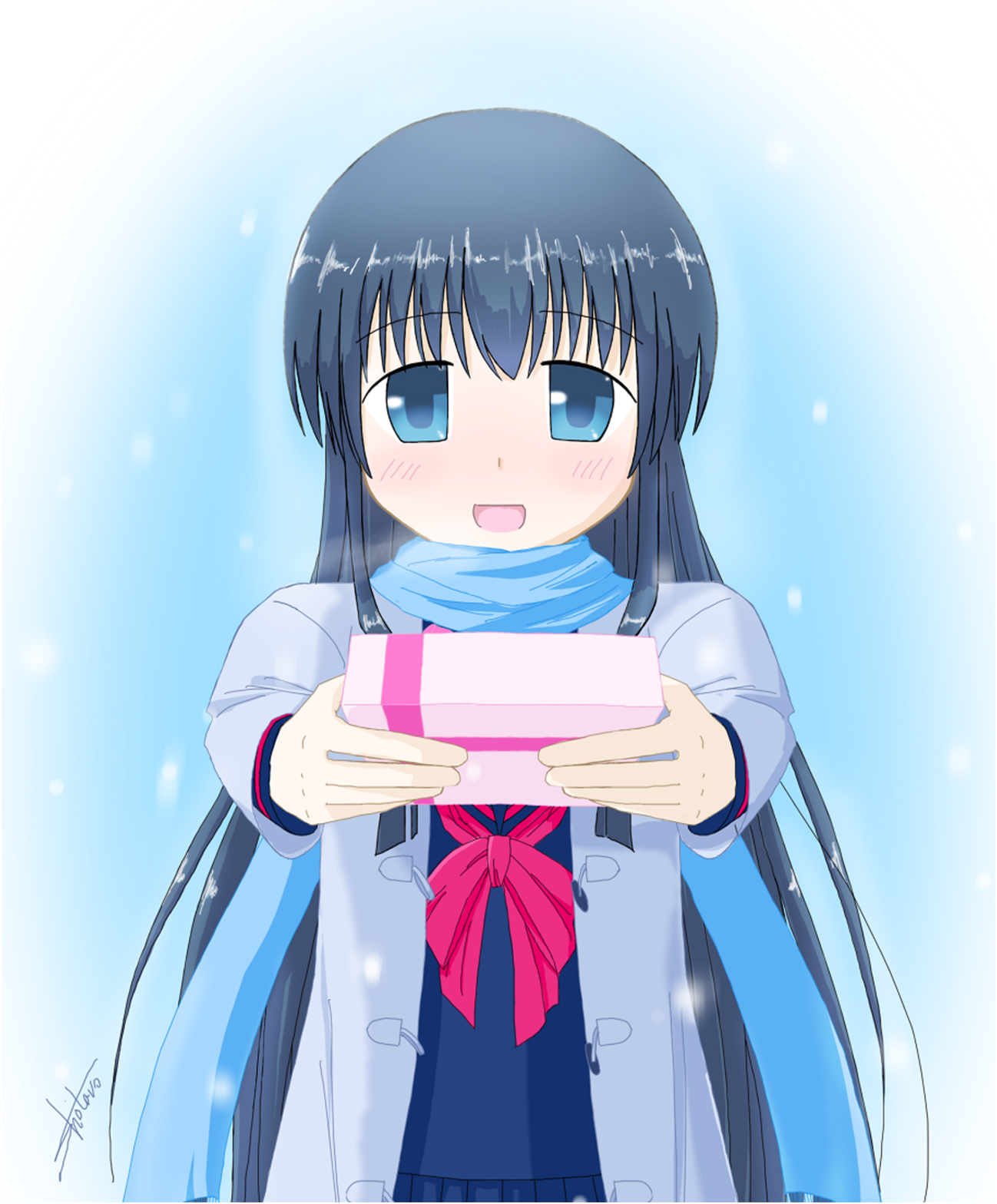}
	{\phantom{.}

	Figure EX4: Happy Valentine's Day! (2018 version)}
\end{figure}


\newpage


\begin{thebibliography}{99}


	
	\bibitem{Fukushima}
		K. Fukushima and T. Hatsuda, ``The phase diagram of dense QCD,"
		 	Rept. Prog. Phys. 74 (2011) 014001.


	\bibitem{QM}
		M. Buballa, and M. Oertel, ``Color--flavor unlocking and phase diagram with
		self--consistently determined strange-quark masses,"
		Nucl. Phys. A 703 (2002) 770.



	\bibitem{beta_function}
		C. Itzykson, and J. B. Zuber, ``Quantum Field Theory," Dover Publications (2006).

	\bibitem{SLAC-exp}
		E.D. Bloom, D.H. Coward, H.C. DeStaebler, J. Drees, G. Miller, L.W. Mo, R.E. Taylor,
		M. Breidenbach, J.I. Friedman, G.C. Hartmann and H.W. Kendall,
		``High-Energy Inelastic $e$--$p$ Scattering at $6^\circ$ and $10^\circ$,"
		Phys. Rev. Lett. 23 (1969) 930.
	
	\bibitem{SLAC-th}
		J.D. Bjorken and E.A. Paschos,
		``Inelastic Electron--Proton and $\gamma$--Proton Scattering and the Structure of the Nucleon,"
		Phys. Rev. 185 (1969) 1975.

	\bibitem{Gross}
		D.J. Gross, F. Wilczek. ``Ultraviolet behavior of non-abelian gauge theories,"
		Phys. Rev. Lett. 30 (1973) 1343. 
	
	\bibitem{Politzer}
		H.D. Politzer. ``Reliable perturbative results for strong interactions,"
		Phys. Rev. Lett. 30 (1973) 1346. 

	\bibitem{Creutz-1}
		M. Creutz, ``Monte Carlo study of quantized SU(2) gauge theory," Phys. Rev. D 21 (1980) 2308.

	\bibitem{Creutz-2}
		M. Creutz, ``Asymptotic-freedom scales," Phys. Rev. Lett. 45 (1980) 313.
		


	\bibitem{finite_temperature_QCD}
		D.J. Gross, R.D. Pisarski and L. G. Yaffe,
		``QCD and instantons at finite temperature,"
		Rev. Mod. Phys. 53 (1981) 43.

	\bibitem{RW}
		A. Roberge and N. Weiss. ``Gauge theories with imaginary chemical potential and the phases of QCD,"
		Nucl. Phys. B 275 (1986) 734.

	\bibitem{Fodor-Katz}
		Z. Fodor and S.D. Katz,
		``Critical point of QCD at finite T and ƒÊ, lattice results for physical quark masses,"
		JHEP 0404 (2004) 050.


	\bibitem{Taylor_critical}
		S. Datta, R.V. Gava and, S. Gupta,
		``The QCD Critical Point : marching towards continuum,"
		Nucl. Phys. A 904 (2013) 883.

	\bibitem{models}
		O. Scavenius, A. Mocsy, I.N. Mishustin and D.H. Rischke,
		``Chiral Phase Transition within Effective Models with Constituent Quarks,"
		Phys. Rev. C 64 (2001) 045202.




	\bibitem{kappa}
		K. Jansen and R. Sommer,
		``O(a) improvement of lattice QCD with two flavors of Wilson quarks,"
		Nucl. Phys. B 530 (1998) 185.

	\bibitem{Iwasaki}
		Y. Iwasaki, ``Renormalization Group Analysis of Lattice Theories and
		Improved Lattice Action. II --- four--dimensional non--abelian SU(N) gauge model ---,"
		UTHEP-118 (1983), preprint.


	\bibitem{clover_fermions}
		B. Sheikholeslami and R. Wohlert,
		``Improved continuum limit lattice action for QCD with Wilson fermions,"
		Nucl. Phys. B259 (1985) 572.




	\bibitem{sign1}
	A. Nakamura, ``Quarks and Gluons at Finite Temperature and Density,"
	Phys. Lett. 149B (1984) 391.

	\bibitem{sign2}
	Ph. de Forcrand, ``Simulating QCD at finite density," PoS (LATTICE2009) 010 (2010).


	\bibitem{sign_orig1}
		J. Engels and H. Satz,
		``Deconfinement at finite baryon number density," Phys. Lett. B 159 (1985) 151.
	
	\bibitem{sign_orig2}
		I. Barbour, N.E. Behilil, E. Dagotto, F. Karsch, A. Moreo, M. Stone, and H.W. Wyld,
		``Problems with finite density simulations of lattice QCD,"
		Nucl. Phys. B 275 (1986) 296.


	\bibitem{reweighting-1}
		Z. Fodor and S.D. Katz, ``A new method to study lattice QCD at finite temperature
		and chemical potential," Phys. Lett. B 534 (2002) 87.

	\bibitem{reweighting-2}
		F. Csikor, G.I. Egri, Z. Fodor, S.D. Katz, K.K. Szabo and A.I. Toth,
		``The QCD equation of state at finite T/ƒÊ on the lattice,"
		Prog. Theor. (Phys. Suppl.) 153 (2004) 93.

	\bibitem{NN_eos}
		K. Nagata and A. Nakamura, ``EoS of finite density QCD with Wilson fermions
		by Multi--Parameter Reweighting and Taylor expansion," JHEP 04 (2012) 092.



	\bibitem{Taylor-1}
		S. Ejiri, C.R. Allton, S.J. Hands, O. Kaczmarek, F. Karsch, E. Laermann and C. Schmidt,
		``Study of QCD thermodynamics at finite density by Taylor expansion,"
		Prog. Theor. (Phys. Suppl.) 153 (2004) 118.


	\bibitem{Taylor-2}
		A. Bazavov, H.T. Ding, P. Hegde, O. Kaczmarek, F. Karsch, E. Laermann, Y. Maezawa,
		S. Mukherjee, H. Ohno, P. Petreczky, H. Sandmeyer, P. Steinbrecher, C. Schmidt,
		S. Sharma, W. Soeldner and M. Wagner,
		``The QCD Equation of State to $\mathcal {O}(\mu_B^ 6) $ from Lattice QCD,"
		Phys. Rev. D 95 (2017) 054504.

	\bibitem{Taylor-3}
		C. Schmidt and S. Sharma, ``The phase structure of QCD," J. Phys. G 44 (2017) 104002.


	\bibitem{WHOT}
		WHOT--QCD Collaboration: S. Ejiri, Y. Maezawa, N. Ukita, S. Aoki, T. Hatsuda,
		N. Ishii, K. Kanaya and T. Umeda,
		``Equation of State and Heavy--Quark Free Energy at Finite Temperature and Density
		in Two Flavor Lattice QCD with Wilson Quark Action,"
		Phys. Rev. D 82 (2010) 014508.



	\bibitem{canonical}
		A. Hasenfratz and D. Toussaint, ``Canonical ensembles and nonzero density quantum chromodynamics,"
		Nucl. Phys. B 371 (1992) 539.



	\bibitem{RHIC}
		M.M. Aggarwal, et al,
		``Higher Moments of Net Proton Multiplicity Distributions at RHIC,"
		Phys. Rev. Lett. 105 (2010) 022302.


	\bibitem{RHIC_fit}
		A. Nakamura, R. Fukuda, S. Oka, S. Sakai, A. Suzuki, Y. Taniguchi, K. Morita and K. Nagata,
		``Net-Baryon Multiplicity and QCD Phase Diagram,"
		PoS (CPOD2014) 021 (2015).


	\bibitem{SNT}
		Y. Sasai, A. Nakamura and T. Takaishi,
		``Phase Fluctuation of Fermion Determinant in Lattice QCD at Finite Density,"
		Nucl. Phys. B (Proc. Suppl.) 129--130 (2004) 539


	\bibitem{Forcrand-1}
		S. Kratochvila and Ph. de Forcrand,
		``The canonical approach to Finite Density QCD,"
		PoS (LATTICE2005) 167 (2006).
		
	\bibitem{Forcrand-2}
		Ph. de Forcrand and S. Kratochvila,
		``Finite density QCD with a canonical approach,"
		Nucl. Phys. B (Proc. Suppl.) 153 (2006) 62.


	\bibitem{WNE}
		X.~Meng, A.~Li, A.~Alexandru and K.F.~Liu,
		``Winding number expansion for the canonical approach to finite density simulations,"
		PoS (LATTICE2008) 032 (2009).


	\bibitem{Gatt}
		J. Danzer and C. Gattringer,
		``Properties of canonical determinants and a test of fugacity expansion
		for finite density lattice QCD with Wilson fermions,"
		Phys. Rev. D 86 (2012) 014502.



	\bibitem{NN-reduction}
		K. Nagata and A. Nakamura, ``Wilson Fermion Determinant in Lattice QCD,"
		Phys. Rev. D 82 (2010) 094027.

	\bibitem{reduction-2}
		A. Alexandru and U. Wenger, ``QCD at non--zero density and canonical partition
		functions with Wilson fermions,"
		Phys. Rev. D 83 (2011) 034502.
		
	\bibitem{reduction_stag}
		P.E. Gibbs,
		``The fermion propagator matrix in lattice QCD,"
		Phys. Lett. B 172 (1986) 53.

  

	\bibitem{WNE_Gatt}
		J. Danzer and C. Gattringer,
		``Winding expansion techniques for lattice QCD with chemical potential,"
		Phys. Rev. D 78 (2008) 114506.
  


	\bibitem{Oka-3}
		A. Nakamura, S. Oka and Y. Taniguchi, 
		``QCD phase transition at real chemical potential with canonical approach,"
		JHEP 02 (2016) 054.


	\bibitem{Oka-4}
		R. Fukuda, A. Nakamura, S. Oka, S. Sakai, A. Suzuki and Y. Taniguchi, 
		``Beating the sign problem in finite density lattice QCD,"
		PoS (LATTICE2015) 208 (2016).





	\bibitem{Nyquist}
		H. Nyquist, ``Certain topics in telegraph transmission theory,"
		Trans. AIEE. 47 (1928) 617.

	\bibitem{Shannon}
		C. E. Shannon, ``Communication in the Presence of Noise,"
		Proceedings of the IRE 37 (1949) 10.



	\bibitem{Oka-1}
		S. Oka, ``Exploring finite density QCD phase transition with	canonical approach
		---Power of multiple precision computation---,"
		PoS (LATTICE2015) 067 (2016).

	\bibitem{Oka-2}
		R. Fukuda, A. Nakamura and S. Oka, ``Canonical approach to finite density QCD
		with multiple precision computation,"
		Phys. Rev. D 93, 094508 (2016).


	\bibitem{TheArt}
		D. E. Knuth,
		``The Art of Computer Programming, Volume 2 / Seminumerical Algorithms,"
		Second Edition, Addison--Wesley publishing (1969).

	\bibitem{guardbit}
		D. Goldberg, ``What every computer scientist should know about floating-point arithmetic,"
		ACM Computing Surveys (CSUR) 23 (1991) 5.






	
	\bibitem{WHOT_ref}
		A.A. Khan et al. (CP-PACS Collaboration),
		``Equation of state in finite temperature QCD with two flavors of improved Wilson quarks,"
		Phys. Rev. D 64 (2001) 074510.

	\bibitem{Oka-5}
		R. Fukuda, A. Nakamura and S. Oka,
		``Validity range of canonical approach to finite density QCD,"
		PoS (LATTICE2015) 167 (2016).



	\bibitem{Gatt-SB}
		C. Gattringer and T. Kloiber,
		``Lattice study of the Silver Blaze phenomenon for a charged scalar $\phi^4$ field,"
		Nucl. Phys. B 869 (2013) 56.





	\bibitem{imag_chem}
		Ph. de Forcrand and O. Philipsen,
		``The QCD phase diagram for small densities from imaginary chemical potential,"
		Nucl. Phys. B 642 (2011) 290.




	\bibitem{Suzuki-0}
		R. Fukuda, A. Nakamura, S. Oka, A. Suzuki and Y. Taniguchi,
		``Study of the sign problem in canonical approach,"
		PoS (LATTICE2016) 060 (2017).

	\bibitem{Russia-0}
		V.A. Goy, V. Bornyakov, D. Boyda, A. Molochkov, A. Nakamura, A. Nikolaev and V. Zakharov,
		``Sign problem in finite density lattice QCD,"
		PTEP 2017 (2017) 031D01.

	\bibitem{Suzuki-1}
		A. Nakamura, S. Oka, A. Suzuki and Y. Taniguchi,
		``Calculation of high--order cumulants with canonical ensemble method in lattice QCD,"
		PoS (LATTICE2015) 168 (2016).

	\bibitem{Russia-1}
		D.L. Boyda, V.G. Bornyakov, V.A. Goy, V.I. Zakharov, A.V. Molochkov, A. Nakamura and A.A. Nikolaev,
		``Novel approach to deriving the canonical generating functional in 
		lattice QCD at a finite chemical potential,"
		JETP Lett. 104 (2016) 657.

	\bibitem{Russia-2}
		V.G. Bornyakov, D.L. Boyda, V.A. Goy, A.V. Molochkov, A. Nakamura, A.A. Nikolaev and V.I. Zakharov,
		``New approach to canonical partition functions computation in $N_f=2$ 
		lattice QCD at finite baryon density,"
		Phys. Rev. D 95 (2017) 094506.
	
	\bibitem{Russia-3}
		D. Boyda, V.G. Bornyakov, V. Goy, A. Molochkov, A. Nakamura, A. Nikolaev and V.I. Zakharov,
		``Lattice Study of QCD Phase Structure by Canonical Approach,"
		arXiv:1704.03980 [hep-lat], preprint.




	\bibitem{Alex}
		A. Alexandru, M. Faber, I. Horvath and K.F. Liu,
		``Lattice QCD at finite density via a new canonical approach,"
		Phys. Rev. D 72 (2005) 114513.

	\bibitem{Li:2010qf} 
		A. Li, A. Alexandru, K.F. Liu and X. Meng,
		``Finite density phase transition of QCD with $N_f=4$ and $N_f=2$ using canonical ensemble method,"
		Phys. Rev. D 82 (2010) 054502.

	\bibitem{Li:2011ee} 
		A. Li, A. Alexandru and K.F. Liu,
		``Critical point of $N_f = 3$ QCD from lattice simulations in the canonical ensemble,"
		Phys. Rev. D 84 (2011) 071503.

	\bibitem{Gat1} 
		A. Alexandru, C. Gattringer, H.P. Schadler, K. Splittorff and J.J. M. Verbaarschot, 
		``Distribution of canonical determinants in QCD,"
		Phys. Rev. D 91 (2015) 074501.
  
	\bibitem{Gat2} 
		C. Gattringer, H.P. Schadler, 
		``Generalized quark number susceptibilities from fugacity expansion at finite chemical
		potential for $N_f = 2$ Wilson fermions,"
		Phys. Rev. D 91 (2015) 074511.



	\bibitem{Matthews-Salam}
		T. Matthews and A. Salam, ``Propagators of Quantized Field," Nuovo Cimento 2 (1955) 120.



	\bibitem{noise}
		J. Foley, K.J. Juge, A. O'Cais, M. Peardon, S.M. Ryan and J.I. Skullerud,
		``Practical all--to--all propagators for lattice QCD,"
		Comput. Phys. Commun. 172 (2005) 145.



\end{thebibliography}
\end{document}